\documentclass[11pt]{article}
\usepackage{graphicx} 

\usepackage{amsmath}    
\usepackage{graphicx}   
\usepackage[unicode]{hyperref}   
\usepackage[superscript]{cite}
\usepackage{tcolorbox}
\tcbuselibrary{most}
\usepackage{parskip}
\usepackage{amssymb}
\numberwithin{equation}{section}
\usepackage{chngcntr}      
\counterwithin{figure}{section}
\counterwithin{table}{section}
\usepackage{physics}  
\usepackage{colortbl}  
\usepackage{enumitem}
\usepackage{longtable, booktabs, array}

 \usepackage{pdflscape}
\usepackage[utf8]{inputenc}
 \usepackage{amsmath}
\usepackage{listings}
\usepackage{color}
\usepackage{hyperref}
\usepackage{geometry}
\usepackage{caption}
\usepackage{subcaption}
\usepackage{float}
\usepackage{titlesec}
\usepackage{comment}

\usepackage{fancyhdr}
\usepackage{parskip}
\usepackage{xcolor}
\usepackage{tikz}
\usetikzlibrary{arrows.meta, positioning, calc, decorations.pathreplacing, decorations.pathmorphing}
\usepackage{booktabs}
\usepackage{array}
\usepackage{tcolorbox}
\definecolor{navyblue}{RGB}{15, 77, 196}
\tikzset{arrow/.style={-latex, line width=1.2pt}}
\definecolor{lightblue}{RGB}{173,216,230}
\definecolor{darknavy}{RGB}{0,0,128}
 \definecolor{grayline}{gray}{0.4}

\usepackage{booktabs,tabularx,array,makecell}
\newcolumntype{Y}{>{\raggedright\arraybackslash}X}
\newcolumntype{C}{>{\centering\arraybackslash}X}

\hypersetup{
    colorlinks=true,
    linkcolor=navyblue,
    citecolor=navyblue,
    urlcolor=magenta,
    pdftitle={NE TNXXXX},    
    pdfauthor={G. Ruffini},             
}
\definecolor{codegray}{gray}{0.95}
\definecolor{keywordcolor}{RGB}{0,0,180}
\definecolor{stringcolor}{RGB}{128,0,0}
\definecolor{commentcolor}{RGB}{0,128,0}

\definecolor{coral}{HTML}{F57558}
\definecolor{softblue}{HTML}{668EF2}
\definecolor{slate}{HTML}{76819D}

\titleformat{\section}
  {\normalfont\Large\bfseries}{\thesection}{1em}{}
  
\titleformat{\subsection}
  {\normalfont\large\bfseries}{\thesubsection}{1em}{}

\title{Rosetta Stone of Neural Mass Models \\ }
\author{Francesca Castaldo\thanks{francesca.castaldo@neuroelectrics.com, giulio.ruffini@neuroelectrics.com (equal contribution)}$^*$, Raul de Palma Aristides, \\Pau Clusella, Jordi Garcia-Ojalvo, Giulio Ruffini$^*$}
\date{July 2025}

\usepackage{lineno} 

\usepackage{color}

\newcommand{\rev}[1]{\begingroup\color{black}#1\endgroup}
\begin{document}

\maketitle

\begin{abstract}
    Brain dynamics dominate every level of neural organization—from single-neuron spiking to the macroscopic waves captured by functional magnetic resonance imaging (fMRI), magnetoencephalography (MEG), and electroencephalography (EEG) —yet the mathematical tools used to interrogate those dynamics remain scattered across a patchwork of traditions.
    Neural mass models (NMMs) (aggregate neural models) provide one of the most popular gateways into this landscape, but their sheer variety---spanning lumped parameter models, firing‐rate equations, and multi‐layer generators---demands a unifying framework that situates diverse architectures along a continuum of abstraction and biological detail. Here, we start from the idea that oscillations originate from a simple push-pull interaction between two or more neural populations. We build from the undamped harmonic oscillator and, guided by a simple push–pull motif between excitatory and inhibitory populations, climb a systematic ladder of detail.
    Each rung is presented first in isolation, next under forcing, and then within a coupled network, reflecting the progression from single‐node  to whole‐brain modeling. By transforming a repertoire of disparate formalisms into a navigable ladder, we hope to turn NMM choice from a subjective act into a principled design decision, helping both theorists and experimentalists translate between scales, modalities, and interventions. In doing so, we offer a \emph{Rosetta Stone} for brain oscillation models—one that lets the field speak a common dynamical language while preserving the dialectical richness that fuels discovery.
      
\end{abstract}

\clearpage

\section*{List of Abbreviations}
\addcontentsline{toc}{section}{List of Abbreviations}
\begingroup
\setlength{\columnsep}{18pt}
\renewcommand{\arraystretch}{1.05}
\small
\begin{description}[leftmargin=2.6cm,style=nextline,labelwidth=2.4cm,labelsep=0.2cm,itemsep=0pt,parsep=0pt]
\setlength{\itemsep}{1pt}
\item[AIT] Algorithmic information theory
\item[AMPA] $\alpha$-amino-3-hydroxy-5-methyl-4-isoxazolepropionic acid (receptor)
\item[BOLD] Blood-oxygen-level-dependent (signal)
\item[CSD] Current source density
\item[DBS] Deep-brain stimulation
\item[DCM] Dynamic causal modelling
\item[DDE] Delay differential equation
\item[DHO] Damped harmonic oscillator
\item[dMRI] Diffusion MRI
\item[DMF] Dynamic mean-field (model, Wong--Wang/Deco)
\item[E--I] Excitatory--inhibitory (population motif)
\item[EC] Effective connectivity
\item[EEG] Electroencephalography
\item[EIF] Exponential integrate-and-fire (neuron)
\item[ERP] Event-related potential
\item[FC] Functional connectivity
\item[FCD] Functional connectivity dynamics
\item[fMRI] Functional magnetic resonance imaging
\item[GABA] $\gamma$-aminobutyric acid (receptor)
\item[HFO] High-frequency oscillation
\item[HH] Hodgkin--Huxley (model)
\item[HWHM] Half-width at half-maximum
\item[ING] Interneuron gamma (mechanism)
\item[JR] Jansen--Rit (neural mass model)
\item[LaNMM] Laminar neural mass model
\item[LFP] Local field potential
\item[LIF] Leaky integrate-and-fire (neuron)
\item[MEG] Magnetoencephalography
\item[M/EEG] Magneto- and electroencephalography
\item[MOM] Metastable oscillatory mode
\item[MOU] Multivariate Ornstein--Uhlenbeck (process)
\item[MPR] Montbri\'o--Paz\'o--Roxin (mean-field reduction)
\item[MRI] Magnetic resonance imaging
\item[MSF] Master stability function
\item[NMDA] N-methyl-D-aspartate (receptor)
\item[NMM] Neural mass model
\item[NMM1] First-generation neural mass model (second-order synapses, static sigmoid)
\item[NMM2] Next-generation neural mass model (exact mean-field, dynamic $(r,v)$ relation)
\item[ODE] Ordinary differential equation
\item[OU] Ornstein--Uhlenbeck (process)
\item[PDE] Partial differential equation
\item[PING] Pyramidal--interneuron gamma (mechanism)
\item[PSD] Power spectral density
\item[PSP] Postsynaptic potential
\item[PV] Parvalbumin (interneuron type)
\item[QIF] Quadratic integrate-and-fire (neuron)
\item[RG] Renormalization group
\item[RNG] Random number generator
\item[SC] Structural connectivity
\item[SDDE] Stochastic delay differential equation
\item[SDE] Stochastic differential equation
\item[SEEG] Stereo-electroencephalography
\item[SL] Stuart--Landau (oscillator / Hopf normal form)
\item[SOM] Somatostatin (interneuron type)
\item[tACS] Transcranial alternating-current stimulation
\item[tDCS] Transcranial direct-current stimulation
\item[tES] Transcranial electrical stimulation
\item[TMS] Transcranial magnetic stimulation
\item[TVB] The Virtual Brain (simulator)
\item[VIP] Vasoactive intestinal peptide (interneuron type)
\item[WILCO] Wilson--Cowan (model)
\end{description}
\endgroup
\clearpage

\setcounter{tocdepth}{2}
\tableofcontents

\clearpage

\section{Introduction} 

\begin{quote}
\emph{``Understanding is the ability to see one thing in many ways.''}\\[4pt]
\hfill ---R.~P.~Feynman
\end{quote}

  \rev{Neural oscillations span every measurable scale, from subthreshold membrane resonances to the rhythms recorded by MEG/EEG and fMRI. While these dynamics may reflect critical computational roles or pathological signatures, bridging such empirical observations with a heterogeneous theoretical landscape remains a practical challenge.
  
Mathematical neural mass models (NMMs) provide a coarse-grained, biophysically informed description of population dynamics that bridges microcircuit mechanisms with macroscopic signals and enables principled inference and prediction. In this context, a ``neural mass" refers to a ``lumped", zero-dimensional representation in space: each region or circuit is summarized by a small set of population-averaged state variables, often (but not always) expressed in a firing-rate format. 

A mass is associated with a single point in space — or, more commonly in whole-brain modeling, with a parcel of a brain parcellation that stands in for an extended cortical or subcortical region. The whole brain is then represented as a discrete network of such point-like masses, with anatomical connectivity encoded in their pairwise couplings. When continuous spatial variation is required — for example, traveling waves, cortical patterning, or laminar gradients — this discrete description generalizes to neural field models \cite{coombes2014}. The discrete parcel indices are replaced by a continuous spatial coordinate, and the finite system of ordinary differential equations (ODEs) becomes a partial differential equation (PDE; or integro-differential equation) in space and time.
  
Earlier syntheses have shaped the field from complementary angles: Ermentrout (1998) \cite{ermentrout1998neural} emphasized spatiotemporal pattern formation; Ashwin et al. (2016) \cite{ashwin2016mathematical} surveyed oscillatory network dynamics and phase-reduction perspectives. In parallel, however, connectome-based whole-brain modelling has become a standard way to link anatomy, delays, and neuroimaging observables~\cite{deco_dynamic_2008,cabral_role_2011,DavidFriston2003,castaldo2023multi,sanchez-todoPhysicalNeuralMass2023,breakspearDynamicModelsLargescale2017}, and recent exact mean-field reductions have begun to connect spiking-network models directly to low-dimensional mass equations. As a result, many scientists now move between oscillator normal forms, rate-based neural masses with explicit synaptic filtering/transfer functions, and exact spiking-to-mass reductions — without a shared notation in which the moves compose.}

\rev{Links between these formalisms exist across the literature, but they are scattered and often use different conventions for coupling, forcing, and biological interpretation. The goal of this review is to provide a practical Rosetta Stone: a scoped, step-by-step mapping that starts from canonical oscillator theory and shows how successive modelling choices—damping, forcing, nonlinear saturation, synaptic filtering, and transfer functions—lead to the neural mass models most commonly used in large-scale brain modelling. Our aim is to provide a coherent “ladder” that (i) makes assumptions explicit, (ii) standardizes how exogenous inputs and inter-areal coupling enter at the node level, and (iii) offers concrete recipes for assembling and interpreting connectome-coupled network models and their perturbational responses.

We emphasize the limited scope of this exercise. The mappings we provide are \emph{local} bridges: explicit reductions valid near specific bifurcations (typically Hopf), under specific limits (weak coupling, strong attraction, exact mean-field), or under specific structural assumptions (current-based coupling, identical populations). This review is not an exhaustive survey, and we do not propose a single global theory unifying all neural mass formalisms across all parameter regimes. Local bridges between models already exist in the literature but are obscured by mismatched conventions; a step-by-step walk across model space---in a single shared notation, with the assumptions of each bridge made explicit---turns these scattered equivalences into practical translation rules.}

To this end, the paper's structure mirrors a spiral curriculum: each incremental step in model complexity is introduced first in its isolated, single-mass form and subsequently generalized to the networked or coupled scenario essential for whole-brain modeling.  

We begin with the undamped harmonic oscillator—the archetype of pure phase dynamics—and sequentially introduce dissipation, external forcing, and nonlinearities. We emphasize how a basic push-pull motif underlies its oscillatory dynamics.  This systematically leads us to the Stuart–Landau oscillator (SL), whose characteristic cubic nonlinearity delivers amplitude regulation with robust limit-cycle behavior. From this pivotal point, we pause to discuss some basic elements needed to establish firm connections with biology: synapses, which transform and delay signals arriving at populations, and transfer functions, which shape the response of accumulated synaptic perturbations into output firing rates. 
With this at hand, we jump to the Wilson-Cowan (WILCO) model, which provides a limit cycle linking back to the Hopf bifurcation in the SL model. Here we (typically) interpret abstract amplitude and phase coordinates as firing rates with biologically meaningful excitatory–inhibitory (E–I) population interactions, with the transfer function being the dynamical element. 
Second-order synaptic filters with static transfer functions naturally yield more nuanced models focusing on the dynamics of post-synaptic potentials, such as the Jansen–Rit and laminar neural-mass families (NMM1). This class of models can reproduce empirically observed alpha–gamma oscillatory interactions and respond realistically to physiological and pharmacological interventions.  \rev{At each step a shared push--pull dynamical structure recurs, recognizing it across formalisms is what makes the cross-walk possible.  Finally, we discuss next-generation neural mass models that can be derived exactly from spiking-network limits under mean-field assumptions—building on earlier exact macroscopic descriptions for theta-neuron networks and exemplified by the Montbrió–Pazó–Roxin reduction for quadratic integrate-and-fire (QIF) neuron  populations—and we use this class (NMM2) to show how ``dynamic'' transfer functions arise from microscopic spiking mechanisms~\cite{Montbrio:2015aa,so2014networks,luke2013complete}. Beyond this lineage, the conductance-based, propagation-speed, decision-circuit, and dynamic causal modelling families listed in block~(B) of Table~\ref{tab:roadmap} each sit at a one-step extension of the rate/PSP lineage; we credit them but do not re-derive them in detail.}


\subsection{Key Concepts}\label{sec:keyconcepts}

\rev{We collect here the definitions used throughout the paper.}

\rev{\textbf{Neurons, synapses, mesoscale.}
We focus throughout on the \textit{mesoscale}, or ``lumped,'' description. Motivated by experimental measures like Local Field Potentials (LFPs) which aggregate activity over space, we replace the state of individual cells with variables representing the statistical mean of the population \cite{trappenbergFundamentalsComputationalNeuroscience2023,dayan_theoretical_2001}. In this framework: a) input spikes become a continuous \emph{average firing rate} input (spikes per second).
    b) synaptic dynamics act as a linear low-pass filter, smoothing average inputs into an \emph{average membrane potential} perturbation. c)
     the sharp firing threshold of single neurons is smoothed by population heterogeneity and timing (variance in thresholds and states, event timing), resulting in a continuous, non-linear \emph{sigmoid transfer function} that maps average membrane potential to an output average firing rate.} 
     
\begin{figure}[t!]
    \centering
    \includegraphics[width=0.7\linewidth]{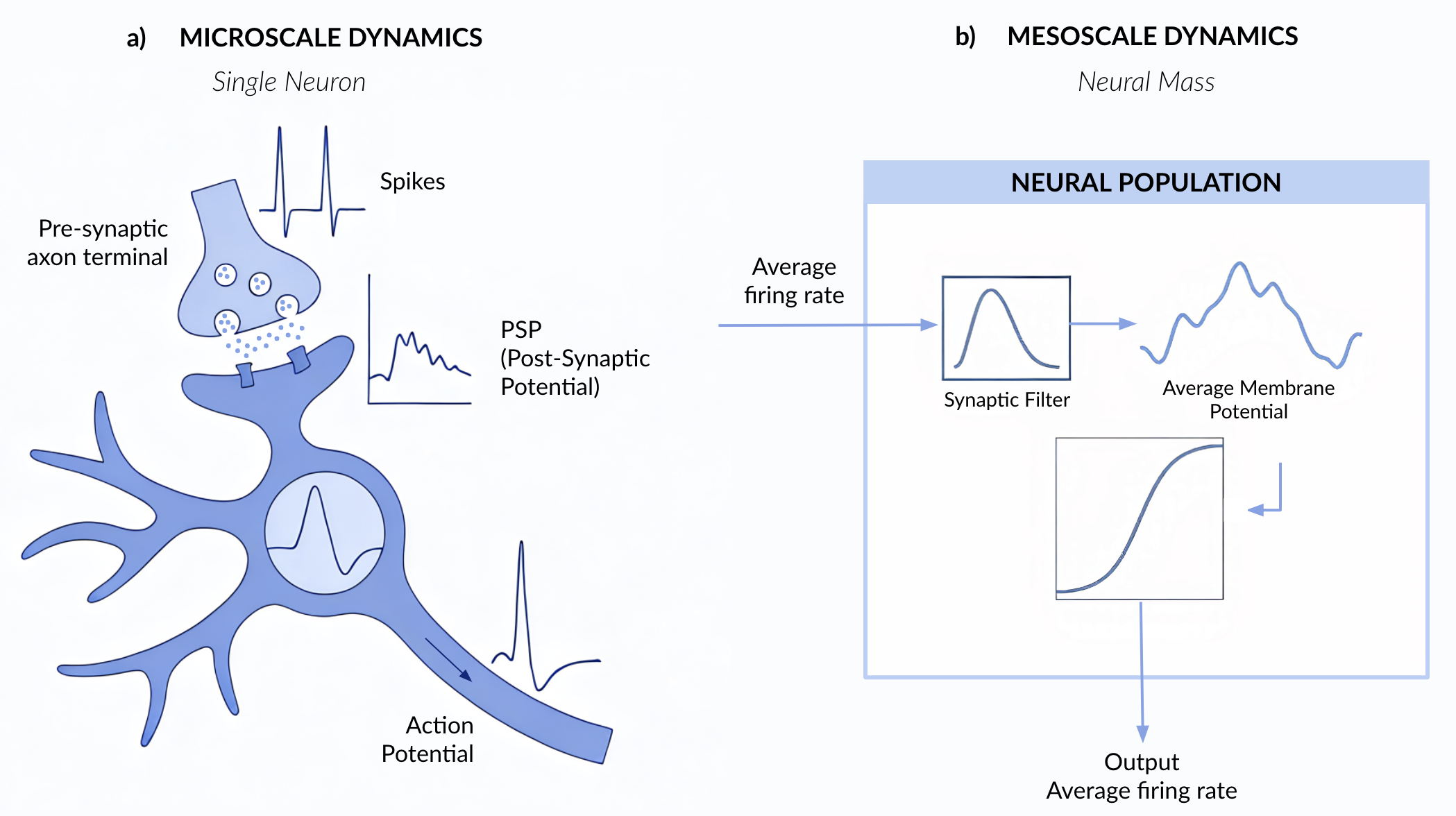}
    \caption{\rev{ The transition from microscopic physiology to mesoscale modeling. Left: Discrete spiking dynamics of a single neuron. Right: Continuous rate-based dynamics of a neural mass, where the sigmoid function replaces the hard firing threshold.}}
    \label{fig:micro_meso}
\end{figure}

 \textbf{Computational modeling concepts} 
\begin{enumerate} 
\item  \rev{\textbf{Neural mass vs.\ neural field.}
A \emph{neural mass model} is a \emph{lumped} (point) description of a neuronal
population (e.g.\ a cortical column, parcel, or nucleus), in which the collective activity
of many neurons is summarized by a finite set of population-averaged state variables.
These variables may represent firing rates, mean membrane potentials, synaptic currents,
or other mesoscopic quantities, and are typically governed by low-dimensional ordinary differential equations (ODEs), stochastic differential equation (SDEs) or,
when propagation delays are retained,  delay differential equations (DDEs).
A \emph{neural field model} is the spatially continuous counterpart, in which the same
variables depend on position and interact through spatial kernels or PDE/integral operators.
Both masses and fields can be rate-based, voltage-based, conductance-based, or derived
as mean-field limits of spiking networks; thus, \emph{mass} should be understood as
\emph{lumped vs.\ spatially extended}, not as a synonym for \emph{rate}.
\smallskip
We use a two-variable $(x,y)$ representation as the minimal autonomous system that supports oscillation. Real neural mass models are typically higher-dimensional; the $(x,y)$ pair is the recurring rotational sub-block, not a model in itself.}

\rev{In this review, the concept of ``neural mass model'' is used in two related but distinguishable ways in the literature. In the Wilson--Cowan / Freeman / Lopes da Silva / Jansen--Rit / Wendling / laminar neural mass model (LaNMM) lineage, and in its exact next-generation extension by Montbri\'o, Paz\'o \& Roxin (MPR / NMM2), ``neural mass'' is the family name of a specific modelling tradition: lumped population descriptions written in terms of firing rates and/or post-synaptic potentials, with current-based synaptic interactions through linear filters, and a population-level transfer relation linking input current to output firing rate. Within this tradition the transfer relation may be a static sigmoid (Jansen--Rit, Wendling, LaNMM), a dynamically filtered sigmoid (Wilson--Cowan, where the firing rate evolves through $\tau_m \dot r = -r + \varphi(I)$), or an exact dynamical $(r,v)$ relation derived from the spiking microscale (MPR / NMM2). In the dynamical-systems and mathematical-neuroscience traditions, by contrast, ``neural mass'' simply means \emph{lumped} (as opposed to \emph{neural field}, spatially extended), and the term covers any zero-dimensional population description regardless of whether its dynamical variables are firing rates, mean voltages, conductances, or order parameters of an exact spiking-network reduction, and regardless of whether synaptic interactions are current- or conductance-based. Under this broader definition, the Liley conductance-based mean-field~\cite{Dafilis2013}, the Robinson corticothalamic propagation model~\cite{Robinson2001,Rennie2002}, the Wong--Wang / Dynamic-Mean-Field reduction~\cite{WongWang2006,Deco2013}, and the David--Friston canonical microcircuit~\cite{DavidFriston2003} are all neural mass models. We adopt this broader (lumped-vs.-field) definition throughout the review. The derivations and cross-walks that follow, however, concentrate on the current-based rate/PSP lineage and its exact next-generation reduction, because that is the sub-family within which the chain of dynamical equivalences --- phase oscillator $\leftrightarrow$ damped oscillator $\leftrightarrow$ Stuart--Landau $\leftrightarrow$ Wilson--Cowan $\leftrightarrow$ NMM1 $\leftrightarrow$ NMM2 --- is tightest, and where a unified notation yields the most leverage. The conductance-based, propagation-speed, decision-circuit, and dynamic causal modelling (DCM) lineages are credited in block~(B) of the roadmap Table and revisited in the synthesis of \S\ref{sec:Closing}, but they are not re-derived in detail here.}

\item 
\rev{\textbf{Oscillators and oscillations.}
We use ``oscillator'' and ``oscillation'' in two related but non-identical senses.
In the \emph{dynamical-systems} sense, an oscillator is an autonomous system that supports
persistent rhythmic motion, most commonly through an attracting periodic orbit (a stable
limit cycle), but also through neutrally stable centers in idealized conservative models.
In the \emph{data-analysis} sense, an oscillation refers to a signal component with a
dominant timescale (often visible as a narrow-band spectral peak).
These notions overlap but need not coincide: for example, a damped linear focus driven by
noise can produce \emph{noise-sustained quasi-cycles} (spectral peaks without a deterministic
limit cycle), whereas chaotic systems are \emph{aperiodic} yet may still show prominent
spectral structure.
In algorithmic information theory terms, we would say that a signal is an oscillation if it can be efficiently compressed by exploiting its approximate periodicity. This reflects the scientific observer's perspective on modeling the phenomenon (see the Appendix~\ref{app:limitcycles} for a more in-depth discussion)}.


\emph{Harmonic oscillators} are prototypical oscillatory systems,  linear and conservative models of   a mass-spring system with an angular frequency and a sinusoid whose amplitude is fixed by initial conditions \cite{Strogatz1994}. 

\emph{Limit-cycle oscillators} are nonlinear and dissipative; after transients, they settle onto a stable closed orbit.

\item 
A \emph{node} is one neural mass; a \emph{network} is a set of nodes connected by synaptic links, which can encode axonal delays and gains.  
Coupling just two nodes is enough for symmetry-breaking, phase locking, and collective bifurcations.

\item 
\rev{\textbf{Push--pull motif as a local dynamical decomposition.}
Across the models reviewed here, oscillations typically arise from an interplay between
(i) a variable that tends to drive activity away from a reference state (``push'') and
(ii) a variable (or process) that provides a restoring or damping effect (``pull'').
In a harmonic oscillator these are displacement and momentum (or two quadrature variables);
in E--I population models they correspond to excitatory and inhibitory components of the loop;
in complex normal forms they appear as the real and imaginary parts of $z$.
Importantly, in biophysically detailed \emph{conductance-based} synapses, synaptic currents
depend on the state through reversal potentials (``shunting''): the effective sign and strength
of an interaction can therefore change with membrane potential.
Throughout, the push--pull language should be read as an \emph{operating-point-dependent}
effective interaction (often after linearization and/or under current-based approximations),
rather than as a universal, state-independent statement that ``excitation always pushes'' and
``inhibition always pulls''.}

\item \emph{Forcing} is any external drive that breaks the autonomy of a node: input from another node in the network (which we call \textit{coupling} below),  electrical stimulation, pharmacological modulation, sensory pulses, or broadband synaptic noise from sources not explicitly modelled.
\rev{When the unforced system has a \emph{strongly attracting} limit cycle and forcing is \emph{weak},
its leading-order effect is typically a modulation of phase captured by a phase response curve \cite{Ermentrout1991}.
When attraction is weak and/or forcing is strong, amplitude deviations and qualitative changes
of the attractor can no longer be neglected, and a phase-only description may fail.\cite{Izhikevich2007}}

\item \emph{Coupling} \rev{is forcing generated by other nodes: one neural mass serves as a
time-dependent drive for another.
Coupling may be instantaneous or delayed, linear or nonlinear, and weak or strong.
In the weak-coupling/strong-attraction regime, coupling often reduces to an effective
interaction between phases (yielding generalized Kuramoto-type descriptions); outside that
regime, amplitude effects (and possibly multistability) become central.
External stimulation can be viewed as a degenerate form of coupling in which the ``partner''
signal is prescribed by the experimenter or device.}\cite{Pikovsky2003,Abrams2004}.

\end{enumerate}

  \textbf{Mathematical tools}
\begin{enumerate}
\item 
\rev{\textbf{Fixed points and \emph{linear} stability.}
Given a dynamical system $\dot{\mathbf{x}}=\mathbf{f}(\mathbf{x})$ with
$\mathbf{x}(t)\in\mathbb{R}^n$ and $\mathbf{f}:\mathbb{R}^n\to\mathbb{R}^n$ at least
continuously differentiable, a fixed point $\mathbf{x}^\ast$ satisfies
$\mathbf{f}(\mathbf{x}^\ast)=\mathbf{0}$.
Unless stated otherwise, ``stability'' in this review refers to \emph{linear stability}:
the local behavior is determined by the eigenvalues of the Jacobian
$D\mathbf{f}(\mathbf{x}^\ast)$ (and, in delay systems, by the corresponding characteristic
spectrum).}
A stable fixed point attracts nearby states, while an unstable one repels nearby states \cite{strogatz2018nonlinear}.
\item \textit{Hopf bifurcation (also \textit{Hopf-Andronov}, HA)}. \rev{A two-dimensional system can begin or cease to oscillate in four ways ; we focus on HA bifurcation throughout, and refer the reader to \cite{strogatz2018nonlinear,hoppensteadt2012weakly,kuznetsov2023elements} for the others.
A Hopf bifurcation occurs when a complex-conjugate pair of eigenvalues of $Df$ crosses the imaginary axis, exchanging the stability of a fixed point.}

\item \emph{Differential Linear operator (\(\hat L\)):} Every synapse in the model behaves like a linear filter: it receives an incoming firing-rate trace \(r(t)\) and converts it into a post-synaptic potential (PSP) \(x(t)\).  Written explicitly, this is a linear operation (convolution, see Appendix~\ref{app:L-op})
\begin{equation}
x(t)=\hat K[r(t)] ,
\end{equation}
or, equivalently,
\begin{equation}
r(t)=\hat L[x(t)] ,
\end{equation}
where \(\hat K\) is the synapse’s impulse-response kernel and \(\hat L\) the inverse operator of \(\hat K\), $\hat L^{-1} = \hat K $.  Framing the dynamics in terms of \(\hat L\) makes the filter’s gain, decay time and delay visible in a handful of coefficients and shows how the population amplifies, attenuates or phase-shifts small perturbations \cite{strogatz2018nonlinear}.  In the simplest push–pull oscillator, there are two such operators, one excitatory and one inhibitory; extending the network adds one \(\hat L\) row per additional connections between populations.

\item \emph{Nonlinearity:} \rev{Models beyond the harmonic oscillator include nonlinear terms.} They reflect the transformation of synaptic inputs into firing rates by the neuronal population. This is necessarily a nonlinear function because firing rates are bounded above and below.

\item \rev{\textbf{Phase reduction and phase--amplitude reductions.}
When a system possesses a stable limit cycle with strong attraction and is subject to weak
forcing/coupling, its dynamics can often be reduced to a phase equation
$\dot{\theta}=\omega+\text{(interaction terms)}$, leading to phase-oscillator networks.
If amplitude excursions matter (e.g.\ near weak attraction, near bifurcations, or under
stronger perturbations), one can extend this to phase--amplitude descriptions (e.g.\
phase--isostable reductions) that retain the leading amplitude modes in addition to phase.
We will point out, as models become more biophysical, when phase-only reductions are
appropriate and when full-state (amplitude-including) descriptions are required.}

\end{enumerate}


\begin{landscape}
\begin{tcolorbox}[
  colback=gray!5,
  colframe=slate,
  boxrule=0.4pt,
  enhanced,
  breakable,
  sharp corners,
  left=4pt,right=4pt,top=3pt,bottom=3pt,
  boxsep=2pt,
]

\scriptsize
\renewcommand{\arraystretch}{0.9}
\begin{tabular}
{p{0.28\textwidth} >{\centering\arraybackslash}p{0.68\textwidth}}

\textbf{Model} & \textbf{Coupled node equation} \\
\hline
\multicolumn{2}{l}{\textit{(A) Normal-form ladder used as a mathematical scaffold in this review}} \\
\hline

\textbf{Phase-only (Kuramoto / undamped HO)} &
$\displaystyle
\dot{\theta}_i(t) = \omega_i
+ G\sum_{j=1}^{N} C_{ij}\,
\sin\!\big(\theta_j(t-\tau_{ij})-\theta_i(t)-\varphi_{ij}\big)
+ \hat{F}_{e;i}(t),\quad i=1,\dots,N.$
\\
\hline

\textbf{Linear damped oscillator network} &
$\displaystyle
\dot{z}_i(t) = (\alpha_i+i\omega_i)\,z_i(t)
+ G\sum_{j=1}^{N} C_{ij}\big[z_j(t-\tau_{ij})-z_i(t)\big]
+ \hat{F}_{e;i}(t),\quad i=1,\dots,N.$
\\
\hline

\textbf{Stuart--Landau (Hopf) network} &
$\displaystyle
\dot{z}_i(t) = (\alpha+i\omega_i)\,z_i(t) - (\gamma+i\beta)\,|z_i(t)|^2 z_i(t)
+ G\sum_{j=1}^{N} C_{ij}\big[z_j(t-\tau_{ij})-z_i(t)\big]
+ \hat{F}_{e;i}(t),\quad i=1,\dots,N.$
\\
\hline

\textbf{Wilson--Cowan (WILCO) E--I rate network} &
$\displaystyle
\tau_x \dot{x}_i + x_i = \sigma_x\!\Big(w_{xx}x_i-w_{xy}y_i+P_{x,i}+\hat{F}_{e;i}(t)+\sum_{j\neq i} C_{ij}x_j\Big),$
\\[-1mm]
& $\displaystyle
\tau_y \dot{y}_i + y_i = \sigma_y\!\big(w_{yx}x_i-w_{yy}y_i\big),\quad i=1,\dots,N.$
\\
\hline

\textbf{NMM1 (second-order synapses, E--I motif)} &
$\displaystyle
\hat{L}_x[x_i] = \sigma_x\!\Big(-w_{xy}y_i+\hat{F}_{e;i}(t)+\sum_{j\neq i} C_{ij}x_j(t-\tau_{ij})\Big),$
\\[-1mm]
& $\displaystyle
\hat{L}_y[y_i] = \sigma_y\!\big(w_{yx}x_i\big),\quad i=1,\dots,N,$
\\[-1mm]
& $\displaystyle
\text{where}\quad \hat{L}_\alpha=\frac{1}{\gamma_\alpha}\Big(\tau_\alpha^2\frac{d^2}{dt^2}+2\tau_\alpha\frac{d}{dt}+1\Big).$
\\
\hline

\textbf{NMM2 (next-generation / QIF-based E--I motif)} &
$\displaystyle
r_x^{(i)}=\hat{\Phi}_x\!\Big(C_{xx}s_x^{(i)}-C_{xy}s_y^{(i)}+\hat{F}_e^{(i)}(t)+\sum_{j\neq i} C_{ij}s_x^{(j)}\Big),\quad
s_x^{(i)}=\hat{K}_x[r_x^{(i)}],$
\\[-1mm]
& $\displaystyle
r_y^{(i)}=\hat{\Phi}_y\!\Big(-C_{yy}s_y^{(i)}+C_{yx}s_x^{(i)}\Big),\quad
s_y^{(i)}=\hat{K}_y[r_y^{(i)}],\quad i=1,\dots,N.$
\\
\hline

\multicolumn{2}{l}{\textit{(B) Other canonical whole-brain node models (listed for completeness; not derived here)}} \\
\hline

 \textbf{Conductance-based / shunting mass (Liley-type) \cite{LileyCaduschWright1999,LileyCaduschDafilis2002,Dafilis2013}} &
$\displaystyle
\tau_a\,\dot{h}_{a,i} = -(h_{a,i} - h_a^{\rm rest}) + \sum_b \psi_{ab}(h_{a,i})\,I_{ab,i},
\qquad
\psi_{ab}(h_a) = \frac{h_a^{\rm eq,b} - h_a}{|h_a^{\rm eq,b} - h_a^{\rm rest}|},$
\\[-1mm]
& $\displaystyle
(\partial_t + \gamma_{ab})^2\,I_{ab,i} = G_{ab}\sum_{j=1}^{N} C_{ij}\,\rho_{b,j}(t-\tau_{ij}) + p_{ab,i},
\qquad
\rho_{a,i} = S(h_{a,i}),\quad a,b\in\{e,i\}.$
\\
\hline

\textbf{Corticothalamic / propagation-speed model (Robinson-type)\cite{Robinson2001,Rennie2002}} &
$\displaystyle
\ddot{\phi}_i + 2\gamma \dot{\phi}_i + \gamma^2\phi_i
= \gamma^2\Big(Q_i(t)+G\sum_{j\neq i}C_{ij}Q_j(t-\tau_{ij})\Big),\quad
\tau_{ij}=d_{ij}/v.$
\\
\hline

\textbf{Reduced Wong--Wang / dynamic mean-field (DMF)  node \cite{WongWang2006,Deco2013}} &
$\displaystyle
\tau_s \dot{S}_i = -S_i + (1-S_i)\gamma r_i,\quad
r_i=\phi(I_i),\quad
I_i = I_0 + wJ S_i + G\sum_{j\neq i} C_{ij}S_j(t-\tau_{ij}) + \eta_i(t).$
\\
\hline

\textbf{David--Friston / DCM neural mass (canonical microcircuit family)\cite{DavidFriston2003}} &
$\displaystyle
\mathcal{D}_{a_p}[x_{i,p}]
= \sum_q g_{pq}\,\sigma(x_{i,q})
+ \sum_{j\neq i} C_{ij}\,\sigma(x_{j,q}(t-\tau_{ij})),
\qquad
\mathcal{D}_{a}=\frac{d^2}{dt^2}+2a\frac{d}{dt}+a^2.$
\\
\hline
\end{tabular}

\captionof{table}{
\scriptsize
\textbf{Coupled whole-brain node equations.} \textbf{Block~(A)} lists the canonical normal-form ladder used as the mathematical scaffold throughout this review. Block~(B) lists representative whole-brain node models that are widely used in the literature but lie at one-step extensions of the rate/PSP lineage and are not re-derived here. \emph{Shared notation:} $C_{ij}$ structural-connectivity weights; $\tau_{ij}=d_{ij}/v$ conduction delays from tract length $d_{ij}$ and effective conduction velocity $v\in[5,10]$\,m/s; $\hat F_{e;i}$ exogenous drive (see Eq.~\ref{eq:forcing}). \emph{Block~(B) notation:} \emph{Liley} --- $h_{a,i}$ mean cell-body potential of population $a\in\{e,i\}$ at node $i$, $h_a^{\rm rest}$ resting potential, $h_a^{\rm eq,b}$ reversal potential at $a$ for synapses from $b$, $\psi_{ab}(h_a)$ normalised shunting weight (1 at rest, 0 at reversal), $I_{ab,i}$ mean post-synaptic potential amplitude from $b{\to}a$, $\gamma_{ab}$ inverse synaptic time constant, $G_{ab}$ synaptic gain, $\rho_{a,i}=S(h_{a,i})$ population firing rate via sigmoid $S$, $p_{ab,i}$ external drive. \emph{Robinson} --- $\phi_i$ axonal pulse field, $Q_i$ population firing rate, $\gamma$ temporal damping rate of the propagation field ($\gamma\sim v/r$ with $v$ axonal speed and $r$ characteristic spatial scale), $G$ global gain. \emph{Reduced Wong--Wang / DMF} --- $S_i$ mean NMDA gating fraction at node $i$, $r_i=\phi(I_i)$ firing rate via the LIF mean-field transfer function $\phi$, $I_0$ background current, $w,J$ local recurrent self-coupling parameters, $\eta_i$ noise. \emph{David--Friston / DCM} --- $x_{i,p}$ state variable of intra-circuit population $p$ at node $i$, $\mathcal{D}_a$ a second-order $\alpha$-synaptic operator with kinetic rate $a$, $g_{pq}$ intra-circuit gain matrix, $\sigma$ sigmoidal voltage-to-rate map. 
}
\label{tab:roadmap}
\end{tcolorbox}

\end{landscape}


\begin{figure}[ht!]
    \centering
    \includegraphics[width=1\linewidth]{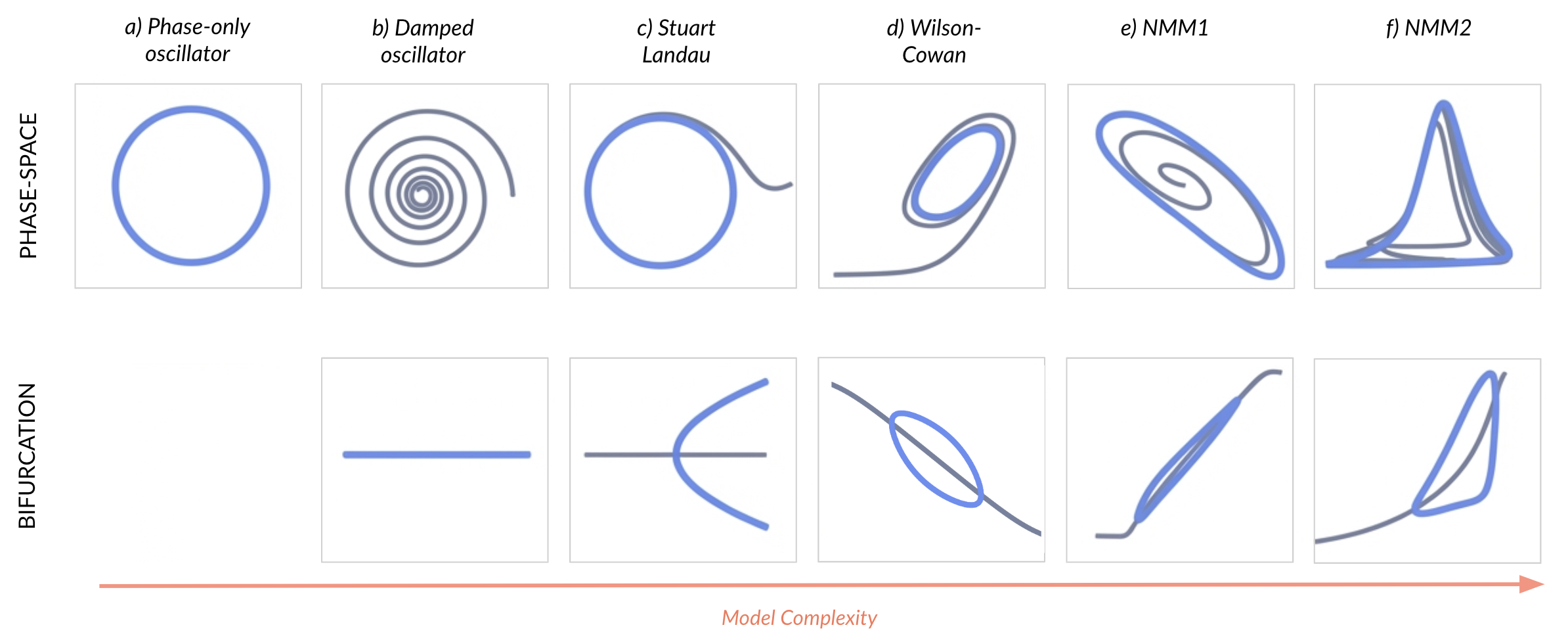}
    \caption{\textbf{Oscillatory dynamics across increasing model complexity}. Top row: phase-space portraits. Bottom row: qualitative one-parameter bifurcation diagrams. The horizontal arrow indicates increasing model complexity from left to right. Blue traces highlight attracting (or neutrally stable) periodic orbits and their stable cycle branches; gray traces denote unstable invariant sets and illustrative transients. Panels: (a) phase-only oscillator ( no amplitude dynamics); (b) damped oscillator with a globally attracting fixed point; (c) Stuart–Landau oscillator (SL) where a Hopf bifurcation creates a stable limit cycle; (d) Wilson–Cowan (WILCO) $E$--$I$ model with coexistence of equilibria and oscillations; (e–f) two neural-mass models (NMM1, NMM2) showing parameter-dependent onset and growth of oscillations and possible bistability. Sketches are schematic and not to scale.}
    \label{fig:portraits}
\end{figure}

\section{Linear Response and Phase-Only Limits}
\label{sec:linear-phase}
\rev{Linear and phase-only descriptions are a natural starting point in any physics-oriented review of neural mass models, for three concrete reasons. First, networks of linear (or linearised) damped oscillators admit closed-form expressions for stationary covariances, lagged covariances, and power spectra in terms of the Jacobian and the connectome, yielding analytic predictions for empirical observables---functional connectivity (FC), MEG/EEG power spectral densities---that are routinely fitted to data~\cite{PonceAlvarez2024HopfLinear,Gilson2016}. Second, every nonlinear neural mass model introduced later in this review (Stuart--Landau, Wilson--Cowan, NMM1, NMM2) reduces near a stable focus to a linear damped oscillator, so the present section supplies the universal local description that subsequent models inherit. Third, the linear network on a connectome is the natural setting for graph-spectral and connectome-harmonic decompositions, in which the dynamics diagonalise in the eigenbasis of the connectivity Laplacian and link directly to large-scale brain modes observed in MEG/EEG~\cite{atasoyConnectomeharmonicDecompositionHuman2017a}. At its core, an oscillation requires (i) at least two dynamical variables---or one complex variable whose real and imaginary parts exchange ``energy'' or activity, pushing or pulling on the another---and (ii) a mechanism to rotate or cycle continuously through phase space. The simplest realisation of these requirements is the undamped, phase-only oscillator, in which a single phase variable advances uniformly while the amplitude remains fixed. This phase reduction is itself an explicit modelling choice---valid under weak forcing and strong attraction to a limit cycle---and serves throughout this review as the pedagogical starting point onto which damping, external forcing, and nonlinear amplitude dynamics are progressively grafted~\cite{Strogatz2018,Winfree2001}.}

\subsection{Undamped (Phase‐Only) Oscillator}

The basal mechanism leading to oscillations in the activity of neuronal populations relies on the interplay between two subpopulations of neurons, one of them inhibitory and the other excitatory.
We can represent the activity of these two populations by two variables $x(t)$ and $y(t)$, respectively.
In the simplest description, we can assume that $x$ decreases $y$ linearly, whereas $y$ increases $x$ also linearly.
Assuming that the proportionality coefficient is the same in the two cases, we obtain a simple set of two coupled linear differential equations:
\begin{align}
    \dot{x} &= -\,\omega\,y,
    \label{eq:undamped_cart_x}\\
    \dot{y} &= \;\omega\,x.
    \label{eq:undamped_cart_y}
\end{align}
Defining
\begin{equation}
    x = r \cos\theta,\qquad
    y = r \sin\theta,
    \label{eq:undamped_xy_def}
\end{equation}
one obtains directly a rescaled model in \textit{polar form}:
\begin{align}
    \dot{\theta} &= \omega, 
    \label{eq:undamped_polar_theta}\\
    \dot{r}      &= 0.
    \label{eq:undamped_polar_r}
\end{align}
This is simply a phase (undamped) oscillator with natural angular frequency $\omega>0$, which in our case determines the synaptic/membrane time constants of the populations.
From~\eqref{eq:undamped_polar_theta}-\eqref{eq:undamped_polar_r} we immediately obtain $\theta(t) = \omega\,t + \theta(0)$, while $r(t) = r(0)$. Thus this \rev{system} is characterized by a constant \rev{(conserved)} amplitude $r(t)\ge0$ and a linearly advancing phase
$\theta(t)\in\mathbb{R}$.  

  Equations \eqref{eq:undamped_cart_x}-\eqref{eq:undamped_cart_y} provide the fundamental \textit{push-pull motif} underlying oscillations in all our models: whenever \(y\) is positive, it drives \(\dot{x}\) negative, acting like an inhibitory force on \(x\); when \(y\) becomes negative, the sign flips and \(x\) is driven upward, as if released from inhibition into excitation.  In turn, a positive \(x\) pushes \(\dot{y}\) upward—exciting \(y\)—while a negative \(x\) pulls \(\dot{y}\) downward, inhibiting \(y\).  This continuous alternation of “push” and “pull” ensures that neither variable drifts off balance: instead, they chase each other around in a perfect circle of constant amplitude.

Introducing the complex variable
\begin{equation}
    z = x + i\,y = r\,e^{\,i\,\theta},
    \label{eq:undamped_complex_z}
\end{equation}
one finds
\begin{equation}
    \dot{z} = \dot{x} + i\,\dot{y}
            = (-\,\omega\,y) + i\,(\omega\,x)
            = i\,\omega\,(x + i\,y)
            = i\,\omega\,z,    
\end{equation}
so that
\begin{equation}
    \dot{z} = i\,\omega\,z.
    \label{eq:undamped_complex}
\end{equation}
Its solution is $z(t) = z(0)\,e^{\,i\,\omega\,t}$, from which $r$ and $\theta$ can be determined.

\subsubsection{Effect of forcing}

To illustrate the effect of a tonic drive (deterministic or stochastic), we augment the undamped oscillator \eqref{eq:undamped_complex}
with a complex forcing term \(F(t)\). First, consider a constant \(F\in\mathbb{C}=F_x+iF_y\):
\begin{align}
\dot x &= -\,\omega\,y + F_x,
\label{eq:forced_cart_x}\\
\dot y &= \;\omega\,x + F_y.
\label{eq:forced_cart_y}
\end{align}
A purely real \(F\) produces a DC offset \((x^*,y^*) = (-F_x/\omega,0)\), around which the trajectory circulates (see Appendix section~\ref{app:shiftedoscillators} for more details). When \(F(t)\) varies in time, Eqs.~\eqref{eq:forced_cart_x}–\eqref{eq:forced_cart_y} describe how the instantaneous bias modulates both amplitude and phase.
To see this, we note that in the complex representation,
\begin{equation}
\dot z = i\,\omega\,z + F(t),
\label{eq:forced_time_complex}
\end{equation}
the term \(F(t)\) gently shifts the oscillator’s center and can transiently entrain or phase‐shift the trajectory without altering its underlying circular geometry (see Appendix~\ref{app:shiftedoscillators} for further details on the effects of forcing).

\subsubsection{Internal and external contributions to forcing; electric fields}
\rev{Forcing appears in all models, internal (from other nodes, i.e., \textit{coupling}) or external (unmodelled nodes, neuromodulation, electric fields).}  We allow for the possibility of forcing to be of a stochastic nature, indicating this with a hat notation  ($\hat F$).  Next, we divide forcing contributions between internal contributions from the network model in which the node lies ($g_i$, which we include in models typically as \textit{coupling}), and external to it ($\hat F_e$). We also separate the latter into external driving of physiological origin ($f$) and perturbations from an external electric field  $E$,
\begin{align} \label{eq:forcing}
    \hat F(t) &= g_{i}(t) +\hat F_e(t)\\
    \hat F_e(t)& = f(t) +   \Lambda [ \vec E(t)] + \hat \eta(t)
\end{align}
where $\Lambda$ is some operator or function of the electric field.
Thus, we will normally model $\hat F_e$ as a sum of a deterministic component  $f$ --- typically from inputs from external populations and an electric field contribution from transcranial electrical stimulation (tES), transcranial magnetic stimulation (TMS) or deep brain stimulation (DBS), for example ---, and an additive zero-mean stochastic contribution to forcing $\eta(t)$. 

For example, in the case of weak electric fields at low frequencies (\textit{transcranial electrical stimulation} or tES), the contribution from the electric field is of the form,
\begin{equation} \label{eq:lambda-E}
    \Lambda[\vec E] = \vec \lambda \cdot \vec E(t) 
\end{equation}
Here,  the electric field effect is linear through a vectorial coupling constant ($\vec \lambda$) and captures ensuing membrane perturbations \cite{ruffiniTranscranialCurrentBrain2013,Ruffini:2014aa,galan-gadeaSphericalHarmonicsRepresentation2023,aberraBiophysicallyRealisticNeuron2018,aberraSimulationTranscranialMagnetic2020}.  

We will use the notation in Eq.~\eqref{eq:forcing} in the following sections, adding a node index if needed (i.e., $\hat F_{e;i}$ for the forcing on node $i$).

\subsubsection{Network of  undamped harmonic oscillators}

Here, we provide the general coupled network model and show how it connects to the Kuramoto model presented below. Following the additive forcing model for coupling, the equation for a network of oscillators is 
\begin{equation}
\dot{z}_i(t)
=
  i\,\omega_i\,z_i(t)
\;+\;
 \sum_{j\neq i}^{N} C_{ij} z_j(t),
\qquad i=1,\dots,N.
\label{eq:linear_undamped_network}
\end{equation}
 
To connect the coupled linear oscillator network in~\eqref{eq:linear_undamped_network} with a phase‐only description, we write each state in polar form
\begin{equation}
    z_i(t) = r_i(t)\,e^{i\theta_i(t)}.
\end{equation}

\rev{Substituting into~\eqref{eq:linear_undamped_network} and separating real and imaginary parts yields coupled equations for amplitudes and phases. With $C_{ij}=|C_{ij}|\,e^{i\varphi_{ij}}$, the amplitude dynamics read
\begin{equation}
\dot r_i \;=\; \sum_{j\ne i} |C_{ij}|\, r_j\, \cos\!\left(\theta_j - \theta_i + \varphi_{ij}\right),
\label{eq:linear-network-amplitude}
\end{equation}
and the phase dynamics
\begin{equation}
\dot\theta_i \;=\; \omega_i \;+\; \frac{1}{r_i}\sum_{j\ne i} |C_{ij}|\, r_j\, \sin\!\left(\theta_j - \theta_i + \varphi_{ij}\right),
\label{eq:linear-network-phase}
\end{equation}
which already has Kuramoto--Sakaguchi structure, but with time-dependent amplitudes $r_i(t)$. A genuine phase-only model is recovered as an explicit \emph{reduction choice}: when the amplitudes relax to a fixed spatial profile $r_i^{*}>0$ (a slow manifold approached when the linear spectrum has a single marginal mode and all others decay), the amplitude equation~\eqref{eq:linear-network-amplitude} sets $\dot r_i \to 0$ and~\eqref{eq:linear-network-phase} reduces to the constant-coupling form of Eq.~\eqref{eq:kuramoto-effective-coupling} below.}

A convenient way to make this connection more concrete is to add a weak uniform decay term, $\dot z_i = (i\omega_i - \gamma) z_i + \sum_{j\neq i} C_{ij}z_j$, and to interpret the resulting dynamics in terms of the spectrum of the linear operator~\cite{Roberts2008,Conteville2013}. For appropriate choices of $\gamma$ and $C_{ij}$, all but one eigenmode decay, while a single collective mode remains marginal and rotates at a common angular frequency. In this \emph{collective oscillation} regime, the amplitudes $r_i(t)$ relax to a fixed spatial profile $r_i^\ast>0$, so that~\eqref{eq:linear-network-phase} reduces asymptotically to a genuine phase model with constant effective couplings,
\begin{equation}
    \dot\theta_i
    =
    \omega_i
    +
    \sum_{j\neq i}
      \tilde K_{ij}\,
      \sin\!\bigl(\theta_j-\theta_i+\varphi_{ij}\bigr),
    \qquad
    \tilde K_{ij} = |C_{ij}|\,\frac{r_j^\ast}{r_i^\ast}.
    \label{eq:kuramoto-effective-coupling}
\end{equation}
This construction shows how a linear complex network with global $U(1)$ phase symmetry naturally generates a low‑dimensional manifold of phase dynamics that is well captured by Kuramoto–type equations~\cite{Kuramoto1984,Nakao2016,Pietras2019}. At the same time, the Kuramoto model is usually taken in the opposite, more general direction: as a phenomenological phase description for weakly coupled limit cycles with $U(1)$ symmetry, where the coupling matrix and coupling function are not constrained to arise from any particular underlying linear operator.

\subsubsection{The Kuramoto model}

The canonical model for studying collective phase dynamics is the Kuramoto model ---  “the hydrogen atom of synchronisation” due to its simplicity, analytical tractability, and extraordinary reach across disciplines 
\cite{kuramotoSelfentrainmentPopulationCoupled1975,Winfree2001,strogatz_nonlinear_2015,acebronKuramotoModelSimple2005}. 

Starting from the undamped limit cycle in Eq.~\eqref{eq:undamped_polar_theta}, Kuramoto assumed that each oscillator interacts with all others only through the difference of their phases.  For a population of $N$ units this yields  
\begin{equation}
    \dot{\theta}_{i}
    \;=\;
    \omega_{i}
    \;+\;
    \frac{G}{N}
    \sum_{j=1}^{N}
    \sin\!\bigl(\theta_{j}-\theta_{i}\bigr),
    \qquad
    i=1,\dots,N,
    \label{eq:kuramoto_original}
\end{equation}
where $\omega_{i}$ is the natural frequency of oscillator $i$ and $G$ is a global coupling constant.

As additional motivation of this model, the sine term is the first (and usually dominant) Fourier component of any $2\pi$‑periodic interaction function\cite{Ermentrout2010}, and retaining only this term might capture most of the essential physics.
In the classic all‑to‑all case with a unimodal, symmetric frequency distribution and pure sine coupling (no phase lag), increasing $K$ past a critical value produces a continuous (second‑order) transition to synchrony that is solvable exactly in the thermodynamic limit ($N\to\infty$) \cite{strogatzKuramotoCrawfordExploring2000}.
More generally—depending on the frequency distribution, the presence of phase‑lag or higher harmonics in the coupling, and the network structure—the transition can be abrupt and hysteretic (first‑order) or continuous.
No other phase‑only coupling law has proved as analytically transparent or universally useful.

\noindent
For whole-brain simulations, we couple phases over a weighted network and include drive and noise:

\begin{tcolorbox}[
  colback=gray!5,
  colframe=gray!5,   
  boxrule=0pt,       
  enhanced, sharp corners,
  left=4pt,right=4pt,top=3pt,bottom=3pt,
  boxsep=2pt,
  before skip=6pt, after skip=6pt
]
\begin{equation}
\dot{\theta}_i(t)
=
\omega_i
+
G\sum_{j=1}^{N} C_{ij}\,
\sin\!\Big(\theta_j\!\big(t-t_{ij}\big)-\theta_i(t)-\varphi_{ij}\Big)
+
\hat F_{e;i}(t),
\qquad i=1,\dots,N.
\label{eq:kuramoto_connectome}
\end{equation}
\end{tcolorbox}

Here $C=[C_{ij}]\in\mathbb{R}_{\ge 0}^{N\times N}$ is the connectivity matrix, $G\in\mathbb{R}_{\ge 0}$ a global coupling gain, $\omega_i\in\mathbb{R}$ the intrinsic angular frequencies, $\tau_{ij}\ge 0$ the delays, $\varphi_{ij}\in(-\pi,\pi]$ fixed phase lags, and $\hat F_{e;i}(t)\in\mathbb{R}$ the external forcing.
\rev{In whole-brain applications, $C_{ij}$ is most commonly estimated from diffusion magnetic resonance imaging (dMRI; tractography from white-matter streamlines) and $\tau_{ij} = d_{ij}/v$ from tract lengths $d_{ij}$ and an effective conduction velocity $v\in[5,10]\,\mathrm{m/s}$.}


\definecolor{coral}{HTML}{F57558} \definecolor{softblue}{HTML}{668EF2} \definecolor{slate}{HTML}{76819D}

The box below summarises the parameters of the phase-oscillator description and gives guidelines for using it to model neural populations.

\begin{tcolorbox}[
  colback=gray!5,
  colframe=slate,
  colbacktitle=softblue!25,
  coltitle=softblue!60!black,
  fonttitle=\bfseries\small,
 title={\small  \textit{Coupled Undamped Harmonic Oscillators (Eq.~\ref{eq:kuramoto_connectome})}\\ Whole-brain Simulations Parameters \& Physiological Meaning },
  boxrule=0.4pt,
  enhanced, sharp corners,
  left=4pt,right=4pt,top=3pt,bottom=3pt,
  boxsep=2pt,
  before skip=6pt, after skip=6pt
]
\scriptsize
\begin{tabular}{@{}l p{0.9\linewidth}@{}}
Node $i$ & Coarse‑grained neural population (e.g., cortical/subcortical parcel or column) treated as a single phase unit.\\
\textbf{$N$} & \emph{Network size}: number of regions/nodes (parcels) modeled.\\
\textbf{$x,y$} & Quadrature \emph{push–pull} components of a mesoscopic E/I loop. A positive $y$ suppresses $x$ (inhibitory‑like), negative $y$ releases $x$ (excitatory‑like), and vice versa, sustaining rotation.\\
\textbf{$\hat\eta_i(t)$} & Independent standard noise sources.\\
\textbf{$G$} & \emph{Global coupling gain} scaling long‑range synaptic influence; proxies neuromodulatory gain/arousal effects on inter‑areal drive.\\
\textbf{$C_{ij}$} & \emph{Connectivity weight} from region $j$ to $i$. Typically structural (white‑matter tract strength or streamline count); can also be functional (FC) or effective (EC) matrices when modeling phenomenology or directed influence; synthetic graphs (all‑to‑all/modular/distance‑decay) if data are absent.\\
\textbf{$\tau_{ij}$} & Propagation delay along $j\!\to i$ (conduction + synaptic). Estimate from tract length/velocity; typical few–tens of ms.\\
\textbf{$\omega,\;\omega_i$} & \emph{Natural angular frequency} ($f_i=\omega_i/2\pi$); reflects effective synaptic/membrane time constants of the local E/I loop.\\
\textbf{$\theta_i(t)$} & Local \emph{phase} (population timing / excitability window) of node $i$.\\
\textbf{$r$}(const.) & Baseline \emph{amplitude} / power envelope of the population rhythm; held fixed in the phase‑only reduction. $\dot r=0$ (Eq.~\ref{eq:undamped_polar_r}).\\
\textbf{$\varphi_{ij}$} & \emph{Fixed phase‑lag} (Sakaguchi offset) capturing delays/filtering at a carrier frequency; $\varphi_{ij}=0$ recovers pure Kuramoto coupling. If using explicit $\tau_{ij}$, set $\varphi_{ij}{=}0$ or use $\varphi_{ij}\approx\bar\omega t_{ij}$.\\
\textbf{$\hat F_{e;i}(t)$}& Exogenous drive  (external forcing from nodes or elements outside the network or an electric field, see Eq.~\ref{eq:forcing}).

\end{tabular}
\end{tcolorbox}

\begin{tcolorbox}[
  colback=gray!5,
  colframe=slate,
  colbacktitle=softblue!25,
  coltitle=softblue!60!black,
  fonttitle=\bfseries\small,
  title={\small When to Use It},
  boxrule=0.4pt,
  enhanced, sharp corners,
  left=4pt,right=4pt,top=3pt,bottom=3pt,
  boxsep=2pt,
  before skip=6pt, after skip=6pt
]
\scriptsize
\textbf{Use this when} phase relations matter more than amplitudes: synchrony, phase locking, entrainment.\\
\textbf{Assumptions} near a stable limit cycle with approximately constant power.\\
\textbf{Best for} large networks needing analytic clarity and light compute (order parameter, clustering, locking bandwidths).\\
\textbf{Inputs} act mainly as phase biases/periodic drives.\\
\textbf{Avoid} if envelope dynamics, amplitude quenching, or bursty power are essential—use an amplitude–phase model.
\end{tcolorbox}

\begin{tcolorbox}[
  colback=gray!5,
  colframe=slate,
  colbacktitle=softblue!25,
  coltitle=softblue!60!black,
  fonttitle=\bfseries\small,
  title={\small How to Use It},
  boxrule=0.4pt,
  enhanced, sharp corners,
  left=4pt,right=4pt,top=3pt,bottom=3pt,
  boxsep=2pt,
  before skip=6pt, after skip=6pt
]
\scriptsize
\textbf{State} phases only: $\theta_i(t)$; fix amplitude $r_i=r_0$.\\
\textbf{Provide} $C$ (connectome): \emph{structural} (dMRI/tractography), \emph{functional} (empirical FC weights), or \emph{effective} (model‑based directed EC); if unavailable, use all‑to‑all, modular, random, or distance‑decay graphs. Also set $G$.\\
\textbf{Frequencies} $\omega_i\!\approx\!2\pi f_0$ with narrow spread (or heterogeneous if desired).\\
\textbf{Optional} delays $\tau_{ij}$ (or static phase‑lags $\varphi_{ij}$), drive $\hat F_{e;i}(t)$ (e.g., $A_i\sin(\Omega t+\phi_i)$).\\
\textbf{Defaults} normalize $C\leftarrow C/\lambda_{\max}$; set $\varphi_{ij}=0$ initially; $\theta_i(0)\sim\mathcal U(0,2\pi)$; $\Delta t\le 1/(50\,f_{\max})$.\\
\rev{\textbf{Integration \& reproducibility ($\Delta t$).}
\\
With non-zero delays $\tau_{ij}>0$ and stochastic forcing, Eq.~\eqref{eq:kuramoto_connectome} is a stochastic delay differential equation (SDDE), \emph{not} an ODE/SDE. In the deterministic limit ($\eta_i\equiv 0$), we recommend a method-of-steps DDE solver with adaptive step size and continuous interpolation of the delayed history (e.g.\ MATLAB \texttt{dde23}, Julia \texttt{DelayDiffEq.jl}). When stochastic forcing is included, no fully standard SDDE solver is universally adopted; in practice we use a fixed-step Euler--Maruyama scheme on the delayed system as a controlled approximation, with the following caveats: (i) the discretisation is piecewise-constant (or piecewise-linear) noise on each step; (ii) delayed states are recovered from a circular history buffer with linear interpolation when $\tau_{ij}/\Delta t \notin \mathbb{Z}$; (iii) the step size must satisfy both the oscillation-resolution bound $\Delta t \leq 1/(50\,f_{\max})$ \emph{and} the delay-resolution bound $\Delta t \leq \min_{ij}\tau_{ij}/10$, with convergence verified by halving $\Delta t$ and checking statistical observables.
\\
\textbf{Reporting.} For reproducibility, fix and report: solver name and version, the SDDE discretisation scheme (noise model and interpolation), $\Delta t$, total simulated time, discarded transient length, and the seed of the random number generator.}
\textbf{Readouts} order parameter $Z(t)=\frac{1}{N}\sum_j e^{i\theta_j(t)}$, report $\langle|Z|\rangle$; synthesize $x_i(t)=r_0\cos\theta_i(t)$ if needed.
\end{tcolorbox}

\begin{tcolorbox}[colback=gray!5, colframe=gray!50, title=\textbf{Box 2.1: EEG/MEG frequency bands}, fonttitle=\bfseries\small,
  fontupper=\small]
Throughout this review, references to ``alpha,'' ``gamma,'' and other named rhythms follow the standard clinical electroencephalography (EEG) / magnetoencephalography (MEG) band conventions:
\begin{description}\setlength{\itemsep}{1pt}\setlength{\parskip}{0pt}
  \item[$\delta$ (1--4\,Hz)] Slow-wave activity; dominant during deep (non-REM stage 3) sleep and unconsciousness; cortico-thalamic origin.
  \item[$\theta$ (4--8\,Hz)] Drowsiness, memory encoding, navigation; prominent in hippocampus and medial temporal lobe.
  \item[$\alpha$ (8--13\,Hz)] Relaxed wakefulness with eyes closed, posterior dominance; thalamo-cortical generator (Adrian--Berger rhythm).
  \item[$\beta$ (13--30\,Hz)] Active cognition, sensorimotor processing; modulated by movement preparation.
  \item[$\gamma$ ($\gtrsim 30$\,Hz, often 30--80\,Hz)] Local cortical computation, feature binding; pyramidal--interneuron gamma (PING) / interneuron gamma (ING) interneuronal mechanisms.
\end{description}
The boundaries are conventional and vary slightly across the literature. We refer back to this box whenever a specific band is invoked in subsequent sections.
\end{tcolorbox}

\subsubsection{Applications}
Historically, the phase‑oscillator framework grew out of Winfree’s program in theoretical biology, which showed how large populations of weakly coupled limit‑cycle oscillators with heterogeneous natural frequencies could synchronize and be reduced to phase dynamics \cite{Winfree1967}. Kuramoto then analysed an idealized continuum limit with sinusoidal coupling, deriving a closed mean‑field description via a complex order parameter and predicting a continuous transition from desynchrony to partial synchrony \cite{Kuramoto1975}. His 1984 monograph consolidated the theory and notation used today, and later expositions placed the model as a paradigm for collective synchronization across physics, chemistry, and neuroscience \cite{Kuramoto1984,Strogatz2000,Acebron2005}.

Kuramoto‑type networks have become ubiquitous in computational neuroscience, appearing in everything from large‑scale \textit{cortex‑as‑a‑graph} models that reproduce zero‑lag synchrony, realistic  blood-oxygen-level-dependent (BOLD) fluctuations, and structured MEG amplitude envelopes\cite{Breakspear2010,cabral_role_2011,Cabral2014MEGDelays,Cabral2014} to theories of how rhythmic sensory inputs entrain cortical circuits at the level of individual columns\cite{Buzsaki2006}. This popularity stems from the model’s tight mapping to neural ingredients: each oscillator’s intrinsic frequency $\omega_i$ can represent the diverse frequency content of cortical rhythms, and the coupling constant $G$ collects the net gain of synaptic (as well as ephaptic or subcortical) interactions among neurons or brain regions. At the macroscopic scale, the emergence of phase coherence provides a principled analogue of the global signals captured by EEG/MEG\cite{Breakspear2010,Cabral2011}.

As coupling increases, heterogeneous oscillators undergo a transition from desynchrony to partial synchrony\cite{Kuramoto1975,Acebron2005}, offering a mechanistic lens on how large‑scale neural oscillations (e.g., alpha) can arise gradually as effective interactions strengthen. Within connectome‑constrained implementations, this framework has clarified how zero‑lag synchrony can occur across distant cortical areas, how hubs facilitate inter‑modular coordination, and how network activity explores metastable, clustering regimes that wax and wane over time\cite{Schmidt2015KuramotoHubs,Shanahan2010Chimera,Hizanidis2016SciRep,Cabral2014PNB}.

The same phase‑only formalism is well suited to external drive: because inputs enter as phase biases, periodic stimulation can lock phases over predictable bandwidths, accounting for stimulus‑locked entrainment and phase‑reset phenomena in EEG/MEG\cite{Buzsaki2006}. This logic extends beyond the cortex—for example, circadian networks in the suprachiasmatic nucleus are naturally captured as forced phase‑oscillator ensembles\cite{Gu2012SCN}.

To bridge toward biological realism, neuroscience variants relax idealizations by introducing sparse/weighted structural connectomes, heterogeneous propagation delays, stochasticity, and higher‑harmonic couplings. These “realism knobs” generate chimeras (coexistence of synchronized and desynchronized populations), metastable clustering, and frequency‑dependent phase‑lag structure—while retaining a tractable phase core\cite{Shanahan2010Chimera,Hizanidis2016SciRep,Petkoski2016PRE,Petkoski2018PhaseLags,Petkoski2019PhilTransA}. Importantly, when these extras are set to zero, the models reduce to the canonical Kuramoto equation, preserving interpretability and analytical leverage.

Finally, the phase‑only reduction is biologically insightful because many cortical E/I loops operate near a limit cycle, so the theory cleanly separates \emph{when} a population fires (phase) from \emph{how strongly} it fires (amplitude). This turns questions about perception, attention, communication‑through‑coherence\cite{fries_rhythms_2015}, and intervention into questions about phase alignment and effective coupling—precisely the levers neuromodulators and stimulation can adjust. In practice, Kuramoto‑type reductions have informed control strategies in deep‑brain stimulation and desynchronization therapies\cite{Tass2003CR,Popovych2018Multisite,Weerasinghe2019DBS,Weerasinghe2021MultiContactDBS}, and physiologically motivated variants link coupling and phase shifts to transmitters and hemodynamics\cite{Sadilek2015MultiplexKuramoto}; related approaches have begun to quantify coupling abnormalities in clinical cohorts\cite{Bauer2022KuramotoMDD}.

\subsection{Damped Harmonic Oscillator (DHO)}
Introducing a real damping coefficient \( \alpha \) (\( \alpha<0 \) for decay, \( \alpha>0 \) for growth) into Eq.~\eqref{eq:undamped_polar_r} 
(\emph{unforced} case) the equations of motion become
\begin{align}
    \dot{r}      &= \alpha\,r,  \label{eq:damped_polar_r}\\
    \dot{\theta} &= \omega.     \label{eq:damped_polar_theta}
\end{align}
\noindent
which, in Cartesian coordinates, reads
\begin{align}
    \dot{x} &= -\omega\,y + \alpha\,x, \label{eq:damped_cart_x}\\
    \dot{y} &=  \;\;\;\omega\,x + \alpha\,y. \label{eq:damped_cart_y}
\end{align}

The fundamental “push–pull” interaction between \(x\) and \(y\) persists in the damped case: the term \(-\omega\,y\) continues to inhibit (or disinhibit) \(x\) exactly as in the undamped case, while \(\omega\,x\) drives \(y\).  On top of this reciprocal coupling, each variable now experiences its own leakage (if \(\alpha <0\)) or intrinsic amplification (if \(\alpha>0\)) through the terms \(\alpha x\) and \(\alpha y\). As a result, the two‐node loop still chases its 90° phase offset, but gradually spirals inward under leak or outward under growth.  This interplay—orthogonal E/I feedback combined with uniform decay or gain—captures how real neural circuits blend balanced excitation and inhibition with membrane‐ or synaptic‐level dissipation (or recruitment), producing damped (or growing) oscillations rather than perfect, constant‐amplitude rotations.

Letting \( z = x + i\,y \) yields the compact complex form
\begin{align}
    \dot{z} &= (\alpha + i\omega)\,z. \label{eq:damped_complex}
\end{align}

\subsubsection{Effect of Forcing}
 To introduce a constant (or very slowly varying) drive, one again writes
\begin{equation}
    \dot z \;=\; (\alpha + i\,\omega)\,z \;+\; F, \quad  (F\in\mathbb{C})
\end{equation}
This moves the only fixed point of the system from \(z^*=0\) to 
\begin{equation}
    z^{*} \;=\; -\,\frac{F}{\alpha + i\,\omega}.
\end{equation}
From the linear stability analysis, trajectories spiral into (for \(\alpha < 0\)) or out of (for \(\alpha>0\)) the off‐center equilibrium \(z^{*}\) (see details in Appendix \ref{sec:bifurcations}).  In polar coordinates (see Eqs.~\eqref{eq:damped_polar_r}–\eqref{eq:damped_polar_theta}), \(\Re[F\,e^{-\,i\theta}]\) can exactly balance the leakage term \(\alpha r\). A constant real \(F_x\) ensures that \(x(t)\) does not decay to zero, mirroring how a steady current injection holds a neuron at a depolarized potential.  More generally, time‐dependent \(F(t)\) encodes transient pushes and pulls that can transiently boost amplitude, shift phase, or perturb the equilibrium away from its natural damped focus—preparing the oscillator for subsequent coupling effects in network models.

\subsubsection{Coupling: Network of  damped harmonic oscillators}

Consider \(N\) oscillators that, when uncoupled, each satisfy  
\[
    \dot{z} = (\alpha + i\,\omega)\,z + \hat F(t),
\]
where \(\alpha\) (real) is the common damping (\(\alpha<0\)) or growth (\(\alpha>0\)) rate, \(\omega\) is the natural frequency, and \(\hat F(t)\) is any external (complex‐valued) input.  Introducing all‐to‐all diffusive coupling among these units yields the network equation  
\begin{equation}
    \dot{z}_{i} 
    = (\alpha + i\,\omega_{i})\,z_{i} 
    + \frac{G}{N}\sum_{j=1}^{N}\bigl(z_{j} - z_{i}\bigr) 
    +  \hat F_{e;i}(t),
    \quad
    i = 1,\dots, N.
    \label{eq:coupled_linear_damped_complex}
\end{equation}
Here, each \(\omega_{i}\) denotes the intrinsic frequency of node \(i\), \(G\ge0\) is the global coupling strength, and the term \(\tfrac{G}{N}\sum_{j}(z_{j} - z_{i})\) represents a diffusive interaction that pulls each oscillator toward the network centroid \(\tfrac{1}{N}\sum_{j}z_{j}\).    Consequently, this all‐to‐all linear network is precisely the \emph{linearized} limit of the Stuart–Landau network (introduced in the next section).

\noindent
For whole‑brain simulations in the damped regime, we place a linear resonator at each node and couple them through a weighted connectome with optional delays, additive drive, and noise:
\begin{tcolorbox}[
  colback=gray!5,
  colframe=gray!5,   
  boxrule=0pt,       
  enhanced, sharp corners,
  left=4pt,right=4pt,top=3pt,bottom=3pt,
  boxsep=2pt,
  before skip=6pt, after skip=6pt
]
\begin{equation}
\dot{z}_i(t)
=
\bigl(\alpha_i + i\,\omega_i\bigr)\,z_i(t)
\;+\;
G\sum_{j=1}^{N} C_{ij}\,\bigl[z_j\!\bigl(t-t_{ij}\bigr)-z_i(t)\bigr]
+  \hat F_{e;i}(t),
\qquad i=1,\dots,N.
\label{eq:linear_damped_connectome}
\end{equation}
\end{tcolorbox}

Here $z_i=x_i+i\,y_i$ is the complex state of node $i$; $\alpha_i<0$ sets linear damping (decay time $-\!1/\alpha_i$) and $\omega_i$ is the intrinsic angular frequency; $G$ is an optional global coupling gain sometimes used to fit models; $C_{ij}\!\ge\!0$ are (possibly directed) connectome weights; $\tau_{ij}\!\ge\!0$ are propagation delays (set $\tau_{ij}\!\equiv\!0$ if ignored) and $ \hat F_{e;i}(t)$ is the external forcing; Setting $\tau_{ij}\!\equiv\!0$ and taking $C_{ij}\!=\!1/N$ recovers the all‑to‑all form in Eq.~\eqref{eq:coupled_linear_damped_complex}.

\rev{With linear coupling  and real $C_{ij}$, the network coupling term $C_{ij}(z_j-z_i)$ in Eq.~\eqref{eq:linear_damped_connectome} unpacks as
\[
\dot x_i = \cdots + C_{ij}(x_j-x_i),\qquad \dot y_i = \cdots + C_{ij}(y_j-y_i),
\]
i.e., the connection links push to push and pull to pull at equal strength, with no cross-talk between the push of one node and the pull of another. Allowing $C_{ij}=|C_{ij}|e^{i\iota_{ij}}$ to be complex rotates the coupling by $\iota_{ij}$ and mixes the two channels across nodes --- the Sakaguchi--Kuramoto phase-lag structure that emerges in the polar reduction~\eqref{eq:linear-network-phase}, with $\iota_{ij}\approx\bar\omega\,\tau_{ij}$ at a carrier frequency $\bar\omega$ identifying the offset with a conduction delay~\cite{sakaguchi1986soluble}. The interpretation is mode-agnostic: $(x,y)$ may be a pair of firing rates, a post-synaptic potential and its time derivative, or any pair of variables that play the push--pull role of the local rhythm. Linear coupling in $z$ thus corresponds to current-based, frequency-independent inter-areal drive linearised around a stable operating point.}

\begin{tcolorbox}[
  colback=gray!5,
  colframe=slate,
  colbacktitle=softblue!25,
  coltitle=softblue!60!black,
  fonttitle=\bfseries\small,
  title={\small  \textit{Coupled Damped oscillator (Eq.~\ref{eq:linear_damped_connectome})}\\ Whole-brain Simulations Parameters \& Physiological Meaning },
  boxrule=0.4pt, enhanced, sharp corners,
  left=4pt,right=4pt,top=3pt,bottom=3pt,
  boxsep=2pt, before skip=6pt, after skip=6pt
]
\scriptsize
\begin{tabular}{@{}l p{0.84\linewidth}@{}}
Node $i$ & Neural mass (parcel/column or nucleus) represented as a linear damped resonator.\\
\textbf{$N$} & Number of nodes (parcels).\\
\textbf{$z_i(t)=x_i+i\,y_i$} & Complex mesoscopic activity; $x,y$ are quadrature “push–pull” components (in‑phase / quadrature of a local E/I loop).\\
\textbf{$\alpha_i$} & Linear damping (leak/gain). $\alpha_i<0$ gives a stable focus with decay time $-\!1/\alpha_i$; reflects local E/I balance, membrane and synaptic dissipation.\\
\textbf{$\omega_i$} & Intrinsic angular frequency (preferred local timescale); set by effective synaptic/membrane constants. $f_i=\omega_i/(2\pi)$.\\
\textbf{$G$} & Global coupling gain controlling the overall strength of inter‑areal drive.\\
\textbf{$C_{ij}$} & Connectivity from region $j$ to $i$ (structural connectivity (SC) from tractography by default; functional/effective connectivity (FC/EC) alternatives or synthetic graphs if needed).\\
 \textbf{$\tau_{ij}$} & Propagation delay on pathway $j{\to}i$ (axonal conduction \& synaptic latency); \rev{typically estimated as $\tau_{ij}=d_{ij}/v$ from tract length $d_{ij}$ and an effective conduction velocity $v\in[5,10]\,\mathrm{m/s}$};\ induces frequency‑dependent phase offsets.\\
\textbf{$\hat F_{e;i}(t)$}& Exogenous drive  (external forcing from nodes or elements outside the network or an electric field, see Equation~\ref{eq:forcing}). Complex or real valued.
\end{tabular}
\end{tcolorbox}

\begin{tcolorbox}[
  colback=gray!5,
  colframe=slate,
  colbacktitle=softblue!25,
  coltitle=softblue!60!black,
  fonttitle=\bfseries\small,
  title={\small When to Use It},
  boxrule=0.4pt, enhanced, sharp corners,
  left=4pt,right=4pt,top=3pt,bottom=3pt,
  boxsep=2pt, before skip=6pt, after skip=6pt
]
\scriptsize
\textbf{Use this when} you want brain‑scale resonance with decay (stable focus): fitting FC/covariances, lag structures, and power spectra; probing linear entrainment and susceptibility.\\
\textbf{Assumptions} small fluctuations around equilibrium ($\alpha_i<0$), additive noise and/or weak drive; linear coupling over the connectome (optionally with delays).\\
\textbf{Best for} closed‑form second‑order statistics (covariances, cross‑spectra), rapid parameter sweeps, effective‑connectivity estimation, graph‑spectral analyses.\\
\textbf{Avoid} if self‑sustained oscillations, limit cycles, or multistability are essential—use full Stuart–Landau/Hopf bifurcation dynamics instead.
\end{tcolorbox}

\begin{tcolorbox}[
  colback=gray!5,
  colframe=slate,
  colbacktitle=softblue!25,
  coltitle=softblue!60!black,
  fonttitle=\bfseries\small,
  title={\small How to Use It},
  boxrule=0.4pt, enhanced, sharp corners,
  left=4pt,right=4pt,top=3pt,bottom=3pt,
  boxsep=2pt, before skip=6pt, after skip=6pt
]
\scriptsize
\textbf{Provide} $C$ (SC default; FC/EC or synthetic graphs if SC unavailable), $G$, $\alpha_i{<}0$, $\omega_i$ (centered on target band), optional $\tau_{ij}$, $F_i(t)$, and $\hat \eta_i(t)$.\\
\rev{\textbf{Defaults} normalize $C\!\leftarrow\!C/\lambda_{\max}(W)$; pick $\alpha_i=-1/\tau_i$ with $\tau_i$ in a plausible range; set $\tau_{ij}=0$ initially.\\ \rev{\textbf{Integration ($\Delta t$).} With non-zero delays $\tau_{ij}>0$ and stochastic forcing, Eq.~\eqref{eq:linear_damped_connectome} is a stochastic delay differential equation: prefer a method-of-steps DDE solver in the deterministic limit, or use fixed-step Euler--Maruyama on the delayed system as a controlled approximation with $\Delta t \le \min\{1/(100\,f_{\max}),\ \min_{ij}\tau_{ij}/10\}$ and convergence verified by step halving (cf.\ the Kuramoto-box discussion above).}}\\
\textbf{Readouts} stationary covariance/lagged covariance, PSD and cross‑spectra; impulse/transfer functions for linear response; reconstruct $x_i(t)=\Re\,z_i(t)$ if a real observable is needed.
\end{tcolorbox}

\subsubsection{Applications} 

Equation~\eqref{eq:damped_complex} is the small-amplitude (linearised) limit of the canonical Stuart–Landau (SL) oscillator, whose full version is introduced below. In this regime, each node behaves as a damped complex oscillator whose real and imaginary parts jointly encode a local E/I loop: the real part plays the role of in-phase activity, the imaginary part the quadrature component, and the linear coefficient $\alpha <0$ sets the decay time back to equilibrium while $\omega$ fixes the resonance frequency. 

\rev{Linearity means that second-order statistics --- instantaneous and lagged covariances, cross-spectra, and power spectral densities --- admit \emph{closed-form} expressions involving the Jacobian and the connectome (with optional delays), so functional connectivity and spectra follow without simulation.} This is worked out explicitly for the SL whole-brain model by Ponce-Alvarez and Deco, who derive analytic formulas and show their accuracy near the stable focus~\cite{PonceAlvarez2024}.

Because of that tractability, the linear damped-oscillator network has become a \emph{standard baseline} in whole-brain modeling pipelines: (i) as a linearized SL model with analytically tractable functional connectivity (FC) and power spectral density (PSD) for rapid parameter sweeps and state comparisons; (ii) as a multivariate Ornstein–Uhlenbeck (MOU) model on the connectome to fit time-shifted covariances and estimate \emph{effective connectivity}; and (iii) as graph-spectral neural-field approximations in which connectome eigenmodes diagonalize the dynamics. Representative examples include the SL-linearisation for closed-form statistics and fast grid searches~\cite{PonceAlvarez2024}, MOU-based estimation of directed effective connectivity from fMRI~\cite{Gilson2016}, and analytic graph-neural-field or spectral-graph models that link Laplacian eigenmodes to MEG/EEG spectra and spatial patterns~\cite{Aqil2021,Raj2020,Verma2022}. Together, these works show that much of the large-scale resting-state phenomenology of the brain can be captured by linear (or linearised) dynamics, a point underscored by systematic model-selection studies arguing that macroscopic resting-state activity is often best described by linear models~\cite{Nozari2024}. 

Biologically, this phase-linear description retains the features most relevant at mesoscopic scales. The damping $\alpha$ reflects local E/I balance and neuromodulatory tone; $\omega$ captures dominant local timescales; coupling rescales long-range synaptic, ephaptic or subcortical gain; and inputs $F(t)$ represent afferents or stimulation. In this light, \emph{stimulation} acts as designed forcing that reveals the network's linear frequency response (and hence entrainment bandwidths). In contrast, \emph{pharmacology} primarily shifts effective gain or damping and can therefore alter susceptibility and resonance. 

\rev{\paragraph{Stimulation and pharmacology as perturbations of the linearised focus.} Linearising any of the network models in this review around a stable focus $\mathbf{x}^{*}$ and writing the deviation $\mathbf{u}=\mathbf{x}-\mathbf{x}^{*}$
recovers the complex damped network of Eq.~\eqref{eq:linear_damped_connectome},
$\dot{\mathbf{u}}=J\,\mathbf{u}+\mathbf{f}(t)+\boldsymbol{\xi}(t)$,
with Jacobian $J=-A+i\Omega+GC$, $A=\mathrm{diag}(\alpha_i)$,
$\Omega=\mathrm{diag}(\omega_i)$, and $\boldsymbol{\xi}(t)$ Gaussian white noise of covariance $D$. For $\mathbf{f}\equiv 0$ the stationary covariance
$\Sigma$ satisfies the Lyapunov equation
$J\Sigma+\Sigma J^{*}=-D$, and the cross-spectral density factorises through
the resolvent $G(\omega)=(i\omega I-J)^{-1}$:
$S(\omega)=G(\omega)\,D\,G^{*}(\omega)$~\cite{PonceAlvarez2024HopfLinear,Gilson2016}.
A pharmacological intervention enters as a parameter shift $\delta J$ inherited
from changes in local gain/damping ($\delta\alpha_i$) and effective coupling
($\delta C_{ij}$). To first order the covariance and PSD respond as
\begin{align}
J\,\delta\Sigma+\delta\Sigma\,J^{*} &= -\bigl(\delta J\,\Sigma+\Sigma\,\delta J^{*}\bigr),
\label{eq:dSigma}\\
\delta S(\omega) &= G(\omega)\bigl[\delta J\,\Sigma+\Sigma\,\delta J^{*}\bigr]G^{*}(\omega).
\label{eq:dPSD}
\end{align}
A drug that nudges a region toward Hopf shrinks the real part of the
corresponding eigenvalue of $J$ and sharpens the spectral peak as the
inverse of the eigenvalue's distance to the imaginary axis; an intervention that rescales effective coupling redistributes power across modes through the eigenstructure of $C$. Stimulation reads the same equation with $\mathbf{f}(t)$ active. A periodic
drive $\mathbf{f}(t)=\mathbf{f}_{0}\,e^{i\omega_{s}t}$ (e.g., transcranial alternating current stimulation (tACS) at carrier frequency $\omega_{s}$) admits the steady-state response
$\langle\mathbf{u}\rangle(t)=G(\omega_{s})\,\mathbf{f}_{0}\,e^{i\omega_{s}t}$: the same resolvent that sets the spectrum also sets the entrainment bandwidth and its spatial selectivity, with regional response amplified along
eigenmodes of $J$ whose imaginary part lies near $\omega_{s}$. Pharmacology and stimulation are two readouts of the same linearised susceptibility, identifiable from the same fits~\cite{duchetHowDesignOptimal2024}.}

Finally, the linear network sits naturally at the base of a modeling “ladder”, where it has already provided evidence for the functional relevance of oscillations in the brain~\cite{effenbergerFunctionalRoleOscillatory2025}. When nonlinear terms are reintroduced (full SL), one recovers amplitude dynamics, multistability, and turbulence-like regimes exploited to explain state-dependent changes (e.g.,\ wake vs.\ sedation, psychedelic modulation) and nonequilibrium signatures. Yet even there, linear-response or linear-noise approximations remain invaluable for deriving analytic perturbation metrics (e.g.\ fluctuation–dissipation measures of nonequilibrium) and for mapping empirical changes onto interpretable parameter shifts~\cite{Deco2023,cabral2024remote,deco2020modeling,decoTurbulentlikeDynamicsHuman2020,ponce-alvarezHopfWholebrainModel2024}.


\definecolor{coral}{HTML}{F57558}
\definecolor{softblue}{HTML}{668EF2}
\definecolor{slate}{HTML}{76819D}


\section{Stuart–Landau Oscillator (SL)} \label{sec:Stuart-Landau}
\begin{tcolorbox}[colback=gray!6,colframe=gray!6,enlarge left by=0mm,
  boxrule=0pt,enhanced, sharp corners]

  \rev{The Stuart--Landau (SL) oscillator extends the linear damped oscillator of Section~\ref{sec:linear-phase} by adding a cubic amplitude saturation, producing a self-sustained limit cycle through a Hopf bifurcation. As the normal form of that bifurcation, SL is the minimal phase--amplitude model into which every nonlinear neural mass model considered in later sections (Wilson--Cowan, NMM1, NMM2) locally reduces near oscillation onset, so it serves as the universal local bridge for the cross-walk this review develops. Biologically, the cubic term abstracts a family of self-regulating mechanisms---synaptic depression, spike-frequency adaptation, inhibitory plasticity~\cite{vogels_2011}, and other homeostatic feedbacks~\cite{nicola_2018,aldarabsah_2024}---which together prevent the runaway dynamics of unregulated linear models.}
 
\end{tcolorbox}

In \textit{polar form} writing \(z = r\,e^{i\theta}\) yields
\begin{align}
  \dot{r}      &= \alpha\,r \;-\; \gamma\,r^{3}, 
  \label{eq:sl_polar_r}\\
  \dot{\theta} &= \omega \;-\; \beta\,r^{2}.
  \label{eq:sl_polar_theta}
\end{align}

\begin{figure}[t]
    \centering
    \includegraphics[width=0.65\linewidth]{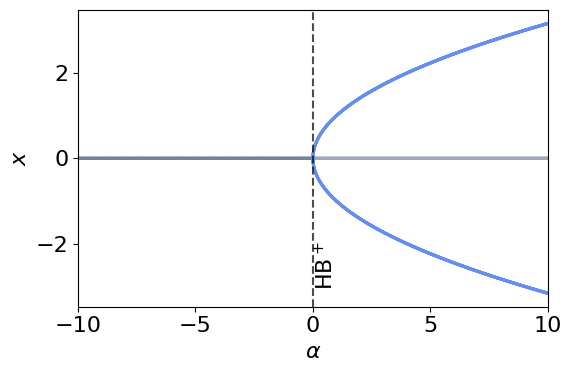}
    \caption{Bifurcation diagram of the Stuart-Landau oscillator with bifurcation parameter $\alpha$. Dark (light) gray lines indicate stable (unstable) fixed points. The amplitude of oscillations is indicated by the blue lines. The dashed line marks the supercritical Hopf bifurcation. See Table~\ref{tab:bif-params} for details.}
    \label{fig:SL_bif}
\end{figure}

Here, the linear terms \((\alpha,\,\omega)\) generate growth and rotation, while the amplitude‐dependent nonlinearities \((\gamma,\,\beta)\) self‐regulate both amplitude and frequency, stabilizing the limit cycle at \(r=r^{*}\). More specifically, \(\alpha\in\mathbb{R}\) is the linear growth (\(\alpha>0\)) or decay (\(\alpha<0\)) rate, \(\omega>0\) is the intrinsic oscillation frequency, \(\gamma>0\) governs nonlinear amplitude saturation, \(\beta\in\mathbb{R}\) introduces nonlinear frequency modulation, coupling amplitude and phase. Equations (\eqref{eq:sl_polar_r}, \eqref{eq:sl_polar_theta}) represent the normal form of a Hopf bifurcation\cite{kuznecovElementsAppliedBifurcation1995}, which gives rise to the sustained oscillations characteristic of the SL (or sometimes called Hopf) model (see Figure~\ref{fig:SL_bif} and Appendix~\ref{sec:bifurcations} for details).

The amplitude equation drives \(r\) toward
\begin{equation}
  r^{*} = \sqrt{\frac{\alpha}{\gamma}},
\end{equation}
while the phase evolves at an amplitude‐dependent rate \( \dot\theta = \omega - \beta\,r^{2}\).

In \textit{Cartesian form}
with \(x = r\cos\theta\) and \(y = r\sin\theta\), one obtains
\begin{align}
  \dot{x} &= \alpha\,x - \omega\,y
           - \gamma\,(x^{2}+y^{2})\,x
           + \beta\,(x^{2}+y^{2})\,y,
  \label{eq:sl_cart_x}\\
  \dot{y} &= \alpha\,y + \omega\,x
           - \gamma\,(x^{2}+y^{2})\,y
           - \beta\,(x^{2}+y^{2})\,x,
  \label{eq:sl_cart_y}
\end{align}
which makes explicit once again the dominating,  push-pull motif in the first-order terms.
In \textit{complex form}, in turn, we have:
\begin{equation}
\dot{z} \;=\; (\alpha + i\,\omega)\,z \;-\; (\gamma + i\,\beta)\,\lvert z\rvert^{2}z.
\end{equation}

\subsection {Effect of Forcing}
To model constant or time-varying synaptic inputs, we add a complex forcing term $F(t)$,
\begin{equation}
\dot{z} \;=\; (\alpha + i\,\omega)\,z \;-\; (\gamma + i\,\beta)\,\lvert z\rvert^{2}z \;+\; F(t).
\label{eq:sl_forcing}
\end{equation}
A constant \(F\) displaces the limit cycle to a new fixed point \(z^{*}\), implicitly given by
\[
  (\alpha + i\omega)\,z^{*} - (\gamma + i\beta)\,\lvert z^{*}\rvert^{2}z^{*} + F = 0,
\]
generally yielding \(\lvert z^{*}\rvert \neq \sqrt{\alpha/\gamma}\). Thus, \(F(t)\) provides a biologically plausible mechanism for input‐dependent modulation of both amplitude and frequency, although it may also alter the stability of the attractor (see Appendix~\ref{app:shiftedoscillators}).

\subsection {Coupling}
Extending to a network of \(N\) diffusively coupled SL oscillators, each node \(i\) obeys
\begin{equation}
\dot{z}_{i} = (\alpha + i\,\omega_{i})\,z_{i}
             - (\gamma + i\,\beta)\,\lvert z_{i}\rvert^{2}z_{i}
             + \frac{G}{N}\sum_{j=1}^{N}\bigl(z_{j}-z_{i}\bigr)
             + \hat F_{e;i}(t),
\label{eq:sl_coupled}
\end{equation}
where \(\omega_{i}\) is the intrinsic frequency of node \(i\), \rev{\(G\)} the coupling strength, and \rev{\(\hat F_{e;i}(t)\)} any external input. The diffusive term synchronizes the network by pulling each oscillator toward the ensemble mean, while nonlinear saturation ensures bounded amplitudes, giving rise to collective phenomena such as synchronization, amplitude death, and clustering.

\noindent
For whole‑brain simulations with nonlinear nodes, we couple Stuart–Landau oscillators over a weighted connectome with optional delays, additive drive, and noise:
\begin{tcolorbox}[
  colback=gray!5,
  colframe=gray!5, 
  boxrule=0pt,     
  enhanced, sharp corners,
  left=4pt, right=4pt, top=3pt, bottom=3pt,
  boxsep=2pt,
  before skip=6pt, after skip=6pt
]
\begin{equation}
\begin{split}
\dot{z}_i(t)
= {}& (\alpha + i\,\omega_i)\,z_i(t)
      - (\gamma + i\,\beta)\,|z_i(t)|^{2} z_i(t) \\
&\quad + G \sum_{j=1}^{N} C_{ij}\,\bigl[z_j\!\bigl(t-t_{ij}\bigr)-z_i(t)\bigr]
      + \hat F_{e;i}(t), \qquad i=1,\dots,N.
\end{split}
\label{eq:sl_connectome}
\end{equation}
\end{tcolorbox}

where $z_i=x_i+i\,y_i$ is the complex state; $\alpha,\gamma,\beta\in\mathbb{R}$ and $\omega_i\in\mathbb{R}$ are the local SL parameters; $G\!\ge\!0$ is a global coupling gain; $C=[C_{ij}]\in\mathbb{R}_{\ge0}^{N\times N}$ is the (possibly directed) connectivity matrix (set $C_{ij}\!=\!1/N$ for all‑to‑all if no connectome is used); $\tau_{ij}\!\ge\!0$ are propagation delays (set $\tau_{ij}\!\equiv\!0$ if ignored) and $ \hat F_{e;i}(t)$ is the external forcing. \rev{We can rewrite these equations in polar coordinates to reconnect with the Kuramoto model~\cite{millanSynchronizationCoupledStuartLandau2025}. 
For phase-oscillator networks with distance-dependent conduction delays---directly relevant to whole-brain SL models with heterogeneous tract lengths---Budzinski et al.~\cite{budzinski_2023} derive analytical predictions for the resulting spatiotemporal patterns by analysing the spectrum of a delay operator combined with the adjacency matrix, providing closed-form access to the locked, twisted, and wave-like states observed in delayed Kuramoto-type whole-brain simulations.}

\begin{tcolorbox}[
  colback=gray!5,
  colframe=slate,
  colbacktitle=softblue!25,
  coltitle=softblue!60!black,
  fonttitle=\bfseries\small,
  title={\small \textit{Stuart-Landau Network (Eq.~\ref{eq:sl_connectome})}\\ Whole-brain Simulations Parameters \& Physiological Meaning },
  boxrule=0.4pt, enhanced, sharp corners,
  left=4pt,right=4pt,top=3pt,bottom=3pt,
  boxsep=2pt, before skip=6pt, after skip=6pt
]
\scriptsize
\begin{tabular}{@{}l p{0.84\linewidth}@{}}
Node $i$ & Neural mass (parcel/column or nucleus) with nonlinear self–limiting dynamics.\\
\textbf{$N$} & Number of nodes (parcels).\\
\textbf{$z_i(t)=x_i+i\,y_i$} & Complex mesoscopic state; $x,y$ are quadrature “push–pull” components (E/I‑like in‑phase and quadrature).\\
\textbf{$r_i,\;\theta_i$} & Amplitude and phase: $z_i=r_i e^{i\theta_i}$. Isolated steady amplitude (if $\alpha_i>0$): $r_i^{*}=\sqrt{\alpha_i/\gamma_i}$; mean rotation $\Omega_i=\omega_i-\beta_i r_i^{*2}$.\\
\textbf{$\alpha_i$} & Linear growth/decay (Hopf bifurcation parameter). $\alpha_i>0$: self‑sustained rhythm; $\alpha_i<0$: decay to rest. Physiologically: net local gain/E–I balance, neuromodulatory tone.\\
\textbf{$\omega_i$} & Intrinsic angular frequency (preferred local timescale) set by effective membrane/synaptic constants.\\
\textbf{$\gamma_i>0$} & Nonlinear amplitude saturation (self‑limiting gain); prevents runaway activity, sets $r_i^{*}$.\\
\textbf{$\beta_i$} & Amplitude–phase coupling (“shear”): amplitude changes shift instantaneous frequency; captures nonlinear dispersion/adaptation effects.\\
\textbf{$G$} & Global coupling gain scaling inter‑areal drive over the connectome.\\
\textbf{$C_{ij}$} & Connectivity weight from region $j$ to $i$ (SC by default; FC/EC or synthetic graphs when needed).\\
\textbf{$\tau_{ij}$} & Propagation delay $j{\to}i$ (conduction + synaptic latency); introduces frequency‑dependent phase offsets.\\
\textbf{$\hat F_{e;i}(t)$}& Exogenous drive  (external forcing from nodes or elements outside the network or an electric field, see Equation~\ref{eq:forcing}). Complex or real valued.\\

\end{tabular}
\end{tcolorbox}

\begin{tcolorbox}[
  colback=gray!5,
  colframe=slate,
  colbacktitle=softblue!25,
  coltitle=softblue!60!black,
  fonttitle=\bfseries\small,
  title={\small When to Use It},
  boxrule=0.4pt, enhanced, sharp corners,
  left=4pt,right=4pt,top=3pt,bottom=3pt,
  boxsep=2pt, before skip=6pt, after skip=6pt
]
\scriptsize
\textbf{Use this when} you need \emph{self‑sustained rhythms with amplitude regulation} and phase–amplitude coupling; to place nodes near a Hopf bifurcation point and study metastability, envelopes, and input‑dependent modulation.\\
\textbf{Coupled form} Diffusively coupled SL nodes on a connectome; nonlinear saturation keeps amplitudes bounded and supports synchrony, clustering, chimeras, waves, and amplitude death (all‑to‑all in~\eqref{eq:sl_coupled}; connectome form in~\eqref{eq:sl_connectome}).\\
\textbf{How forcing enters} Add a complex input per node to shift the working point, entrain, or reshape amplitude and instantaneous frequency; constant biases displace the attractor, periodic drives yield nonlinear locking (single‑node in~\eqref{eq:sl_polar_r}–\eqref{eq:sl_polar_theta}; forcing in~\eqref{eq:sl_forcing}).\\
\textbf{Strengths} Minimal nonlinear phase–amplitude model; explicit control of oscillation onset via $\alpha$; captures band‑limited power, envelope FC/FCD, metastability/turbulence; reduces to linear analytics near the stable focus.\\
\textbf{Limitations} Abstract (few biophysical knobs); parameter identifiability harder far from the Hopf bifurcation; sensitive to delays/coupling very close to the bifurcation; multi‑timescale adaptation needs extensions.\\
\textbf{Typical analyses} Fit FC/FCD and spectra; map working point vs.\ coupling/delays; quantify metastability/turbulence; assess entrainment and phase–amplitude metrics; graph‑eigenmode/wave analyses.\\
\textbf{Good defaults} Set $\gamma>0$; place $\alpha\approx 0$ (slightly negative for subcritical “fluctuating” regime, slightly positive for limit cycles); modest $\omega$ heterogeneity; empirical $W$ (optional delays $\tau_{ij}$); small noise $\sigma$; tune $\beta$ to match frequency–amplitude shifts.\\
\textbf{Report} Working point ($\alpha$), coupling/delay model, frequency distribution, input/noise statistics, and fitted readouts (PSD, FC/FCD, envelopes, metastability).
\end{tcolorbox}

\begin{tcolorbox}[
  colback=gray!5,
  colframe=slate,
  colbacktitle=softblue!25,
  coltitle=softblue!60!black,
  fonttitle=\bfseries\small,
  title={\small How to Use It},
  boxrule=0.4pt, enhanced, sharp corners,
  left=4pt,right=4pt,top=3pt,bottom=3pt,
  boxsep=2pt, before skip=6pt, after skip=6pt
]
\scriptsize
\textbf{Model (drop‑in):}
\textbf{Provide} $C$ (SC default; FC/EC or synthetic graphs if SC unavailable), $G$, local $(\alpha_i,\omega_i,\gamma_i>0,\beta_i)$, optional $\tau_{ij}$, $F_i(t)$, and $\hat \eta_i(t)$.\\
\textbf{Defaults} normalize $
C\!\leftarrow\!C/\lambda_{\max}(W)$; choose small $\alpha_i$ (tune distance to Hopf bifurcation: $<0$ damped, $>0$ limit cycle), set $\gamma_i{=}1$ for scaling, $\beta_i\approx 0$ initially; initialize $z_i(0)$ with small random amplitude; Euler–Maruyama or RK methods with $\Delta t\!\le\!1/(200\,f_{\max})$. \rev{With non-zero delays $\tau_{ij}>0$ and stochastic forcing, Eq.~\eqref{eq:sl_connectome} is an SDDE; prefer method-of-steps DDE solvers in the deterministic limit, otherwise use Euler--Maruyama on the delayed system as a controlled approximation with the dual bound $\Delta t \le \min\{1/(200 f_{\max}),\,\min_{ij}\tau_{ij}/10\}$ and convergence verified by step halving.}\\
\textbf{Readouts} amplitudes $r_i(t)$, phases $\theta_i(t)$, PSD/cross‑spectra, spatial/temporal FC, order parameters for phase and amplitude; compare $r_i$ to $r_i^{*}=\sqrt{\alpha_i/\gamma_i}$ when $\alpha_i>0$.
\end{tcolorbox}

\subsection{Applications}

\rev{The Stuart--Landau equation arose as the weakly nonlinear amplitude equation for the laminar--turbulent transition in shear flows~\cite{Landau1944,Stuart1958,Stuart1960} and is now understood as the normal form of a Hopf bifurcation. Any sustained oscillation that emerges from an instability with weak nonlinearity admits a local SL description, which is what makes SL the natural meeting point for the rate, oscillator, and exact mean-field formalisms compared in this review.}

In computational neuroscience, these same properties make SL a natural generative model for whole‑brain dynamics. At the node level, the bifurcation parameter $a$ controls the local working point—negative $a$ gives noise‑driven, damped fluctuations; $a\approx 0$ yields large, susceptible excursions; and positive $a$ produces self‑sustained rhythms—while $\omega$ sets the intrinsic timescale. Network coupling on the structural connectome, together with finite conduction delays and stochastic drive, then determines how local oscillators form transient coalitions, travel as waves, or lock into metastable patterns. In resting‑state fMRI, SL networks tuned close to the edge of the Hopf bifurcation reproduce both static functional connectivity (FC) and its time variability (FCD), with the best fits occupying a narrow corridor of high \emph{metastability} that effectively defines a dynamical cortical core \cite{Deco2017Metastability}. Incorporating realistic delays in the same framework explains how fast local generators can coalesce into slow, spatially organized \emph{metastable oscillatory modes} (MOMs) that appear and dissolve at reduced collective frequencies, linking anatomy to itinerant large‑scale patterns observed across modalities \cite{Cabral2022MOMs}.

Electrophysiologically, SL models provide a bridge between structure and spectra. Allowing one or multiple resonant channels per region improves the correspondence to resting MEG: SL networks account for band‑limited envelope correlations, the location of spectral peaks, and the transient alignment of modes, with multi‑frequency instantiations offering the best cross‑band fits \cite{Deco2017MEG}. These same phase–amplitude dynamics, when embedded on the connectome with delays, rationalize why modest shifts in global coupling or delay reorganize band‑limited power and envelope FC without changing anatomy, providing a mechanistic map from structure to oscillatory phenomenology \cite{Cabral2022MOMs,siebenhuhnerSpectralPatternsMEG2025}. Multimodal comparisons (fMRI+MEG) using common structural priors further demonstrate that SL’s small, interpretable parameter set can jointly capture FC, FCD, and transient mode structure across measurements \cite{castaldo2023multi}.

Casting SL network behavior in the language of \emph{turbulence} sharpens the operational regime. When the model is fit to empirical amplitude turbulence and FC, the best‑performing working point is typically \emph{subcritical fluctuating}—just below the Hopf bifurcation—where susceptibility and information‑encoding capacity are maximal; strength‑dependent perturbations in this regime reveal richer responsiveness than in supercritical limit cycling \cite{decoTurbulentlikeDynamicsHuman2020,SanzPerl2022StrengthDependent}. 
\rev{A particularly clean neuroscience instantiation of these ideas is provided by Freyer et al.~\cite{freyer_2012}, who showed that an SL-type model with multiplicative noise operating near a subcritical Hopf bifurcation reproduces the scale-invariant bistability of the alpha rhythm observed in resting-state MEG. In their formulation, noise drives stochastic switching between a low-amplitude focus and a coexisting limit cycle, generating the heavy-tailed amplitude distributions and $1/f$-like spectral signatures characteristic of empirical alpha activity. This canonical example illustrates that noise near criticality is an organizing principle that supports physiologically realistic multistability in neural mass dynamics, and motivates the subcritical-fluctuating working point favored by SL whole-brain fits.}
Using closely related SL formulations, local‑coherence (“turbulence”) readouts distinguish wake, sleep, anesthesia, and pharmacologically altered states, positioning SL as a compact generative lens on state‑dependent information flow \cite{Escrichs2022Turbulence}. \rev{The same perturbative logic extends to psychedelics: after fitting LSD and placebo states, SL-based \emph{in silico} stimulations predict enhanced sensitivity to strong inputs and characteristic turbulence signatures under LSD, consistent with empirical observations of expanded dynamical repertoires~\cite{Jobst2021LSD,cruzatEffectsClassicPsychedelic2022}. A complementary perspective frames such state changes through the geometry and complexity of the underlying state space, with plasticity-induced reshaping of attractors as the mechanistic substrate of psychedelic-driven dynamics~\cite{ruffiniNeuralGeometrodynamicsComplexity2024}.}

Biologically, SL’s parameters expose the very levers experiments can turn. Interpreting the bifurcation parameter as net local E/I gain and neuromodulatory tone frames drugs as \emph{parameter shifts} that move regions toward or away from oscillatory onset; interpreting external input as \emph{forcing} turns stimulation into a probe of the network’s susceptibility, mode by mode. \rev{Personalizing SL fits thus yields subject-specific maps of working points and delays that generate testable predictions about which regions and frequencies will respond most --- and how those responses reorganize whole-brain dynamics. Building on this, dynamic sensitivity analysis quantifies how localized stimulation perturbs subject-specific SL/WC fits and identifies the regions whose targeted modulation most efficiently drives transitions between brain states~\cite{vohryzek_dynamic_2023}.} Finally, because SL admits a linear approximation near the stable focus, one can derive closed‑form spectra and covariances for rapid exploration and uncertainty quantification, then reintroduce nonlinearity as needed; this provides a transparent bridge between analytic tractability and the rich nonlinear phenomena that motivate the model in the first place \cite{PonceAlvarez2024HopfLinear}.



\definecolor{coral}{HTML}{F57558}
\definecolor{softblue}{HTML}{668EF2}
\definecolor{slate}{HTML}{76819D}

\section{Interlude: Synapses  and Transfer Functionals}
\label{sec:Interlude}

\begin{tcolorbox}[colback=gray!6,colframe=gray!6,enlarge left by=0mm,
  boxrule=0pt, enhanced, sharp corners]
The discussion below interchanges neurotransmitter release, ionic currents, and membrane potential perturbations. These quantities are related by linear operators, and composition of linear operators is itself a linear operator  (see Appendix~\ref{app:L-op}); 
The chain from presynaptic action potential through conductance, synaptic current, and membrane voltage is the receiving cell therefore collapses to a single first- ore second-order differential equation.
\end{tcolorbox}

\subsection{Synaptic Dynamics and operator formalism}\label{sec:synaptic-dynamics}
Neurons have two main forms of communication: chemical and electrical.
The first happens through neurotransmitters emitted by the presynaptic
neuron after a spike~\cite{hillis2020}.
The second is bound to the existence of gap-junctions between nearby cells~\cite{goodenough2009}.
It is estimated that chemical synapses drive the vast majority of
neural communication in the mammalian brain, and for this reason, electrical
coupling has often been considered as a secondary character in neural dynamics (see, however, ~\cite{pereda2014,alcami2019,vaughn2022}).
Thus, earlier work on NMMs focused on chemical coupling only. 

For a single synaptic connection, the dynamics of neurotransmitter binding
can be modelled as kinetic reactions~\cite{destexhe1994,Destexhe1998,dayan2001}.
Ultimately, both experimental and modeling results show that, typically, upon receiving a spike, the effect of neurotransmitter release to a postsynaptic neuron will first undergo an exponential increase of post-synaptic potential, followed by an exponential decay. 
Therefore, the postsynaptic-potential $s_j(t)$ (mV) of a single, isolated neuron
is usually modeled by the \textit{second order ordinary differential equation}~\cite{dayan2001,gerstner2014}
\begin{equation}\label{eq:synapses_single}
    \tau_r\tau_d\ddot s_j = \gamma r(t) - (\tau_r+\tau_d) \dot s_j -s_j
\end{equation}
where $\tau_r$ and $\tau_d$ are the rise and decay times (ms), $\gamma$ is the amplitude of the PSP (mV/kHz), and $r(t)$ is the input term, modeling the arrival of presynaptic spikes.

While Eq.~\eqref{eq:synapses_single} corresponds to that of a single neuron $j$, its linearity allows us to  quantify the mean synaptic activity of $N$ neurons receiving and reacting identically to the same inputs as
$
s(t)=\frac{1}{N}\sum_{j=1}^N s_j(t)
$.
Then, we obtain the same equation for the mean synaptic activity,
\begin{equation}\label{eq:synapses}
    \tau_r\tau_d\ddot s(t)= \gamma \, r(t) - (\tau_r+\tau_d) \dot s -s\;.
\end{equation}
Interestingly, Eq.~\eqref{eq:synapses} shares the same structure as the equation of a damped, driven harmonic oscillator, which, as we will show in the following sections, allows it to display oscillations and exhibit diverse dynamical responses.

For a single pulse received at time zero ($r(t)=\delta(t)$), the solution to \eqref{eq:synapses} is
\begin{equation}
s(t)=\frac{\gamma}{\tau_d-\tau_r}\left( e^{-t/\tau_d} - e^{-t/\tau_r}\right)\;.
\end{equation}

Usually, different types of neurotransmitters will result in different timescales.
Table~\ref{table:synapses} contains some putative quantities for these parameters.
Considering this variety of time scales, 
there are two limiting cases of Eq.~\eqref{eq:synapses} that are widely used in the literature.
In some cases, one can simplify this equation by assuming that the rise
and decay times are identical, $\tau=\tau_r=\tau_d$. 
Then Eq.~\ref{eq:synapses} reads
\begin{equation}\label{eq:synapses_2nd}
    \tau^2\ddot s(t)= \gamma \,  r(t) - 2\tau \, \dot s -s\;.
\end{equation}
The solution of this equation for a single pulse at time zero ($r(t)=\delta(t)$)
reads
\begin{equation}
s(t)=\frac{\gamma \, t}{\tau^2}\, e^{-t/\tau}
\end{equation}
which is sometimes referred to as \emph{alpha synapse}.

Time constants for the principal receptor types — $\alpha$-amino-3-hydroxy-5-methyl-4-isoxazolepropionic acid (AMPA), N-methyl-D-aspartate (NMDA), and $\gamma$-aminobutyric acid (GABA$_A$, GABA$_B$) — are summarised in Table~\ref{table:synapses}.

\definecolor{softblue}{HTML}{668EF2}
\definecolor{slate}{HTML}{76819D}

\begin{table}[t!]
\centering
\begingroup
\arrayrulecolor{slate} 
\begin{tabular}{ |c|c|c| }
  \hline
  \rowcolor{softblue!25}
  \textbf{\color{softblue!60!black} Neurotransmitter} &
  \textbf{\color{softblue!60!black} $\tau_r$ [ms]} &
  \textbf{\color{softblue!60!black} $\tau_d$ [ms]} \\
  \hline
  AMPA        & 0-2   & 2-5   \\
  NMDA        & 3-15  & 40-100 \\
  GABA$_\text{A}$ & 0-2   & 6-20  \\
  GABA$_\text{B}$ & 25-50 & 100-300 \\
  \hline
\end{tabular}
\caption{Generic ranges for the values of $\tau_r$ and $\tau_d$, extracted from~\cite{gerstner2014} and references therein.}
\label{table:synapses}
\endgroup
\end{table}


On the other hand, if one considers that the rise time is almost instantaneous, $\tau_r\to 0$,
then Eq.~\ref{eq:synapses} reads
\begin{equation}\label{eq:synapses_1st}
\tau_d \dot s = \gamma r(t) - s(t)
\end{equation}
which is just an equation for exponential decay, thus a single
pulse at $r(t)=\delta(t)$ gives the solution
\begin{equation}
s(t)=\frac{\gamma}{\tau_d}e^{-t/\tau_d}.
\label{eq:epsp}
\end{equation}

Other types of neurotransmitter kinetics might result in more complex
synaptic dynamics~\cite{destexhe1994,Destexhe1998}.
Of particular importance is the glutamate NMDA receptors, whose
dynamics also depend on the voltage of the postsynaptic neuron~\cite{dayan2001,gerstner2014}.

 \subsubsection*{The synapse as filter: operator form}
How should the synapse equations be read? To shed some light on this, we recast Eq.~\eqref{eq:synapses} into operator form by rewriting it first,
$$
    \tau_r\tau_d\ddot s(t) + (\tau_r+\tau_d) \dot s +s = \gamma r(t), 
$$
and defining the synapse linear operator
\begin{equation}
\hat L_s [s(t)]= \frac{1}{\gamma} \big[\tau_r\tau_d \frac{d^2}{dt} + (\tau_r+\tau_d) \frac{d}{dt}  +1\big]s(t).
\end{equation}
Then Eq.~\eqref{eq:synapses}  can be written succinctly as 
\begin{equation} \label{eq:L-operator-synapse-equation}
\hat L_s [s(t)] =  r(t) .
\end{equation}
The synapse operator $\hat{L}$ can be derived using the properties of linear time-invariant systems, and the impulse response of the neural mass has the form of Eq. \eqref{eq:epsp}, which is a model supported by experimental and theoretical studies \cite{rallDendriticLocationSynapses1967,burkeCompositeNatureMonosynaptic1967,nelsonAnomalousRectificationCat1967,rallDistinguishingTheoreticalSynaptic1967} (see Appendix~\ref{app:L-op} for details).

The synapse operation can be cast as a causal filter. Since $\hat L$ is a linear differential operator, using proper boundary conditions, we can define its inverse $\hat L ^{-1}= \hat K$ (another linear operator), 
\begin{equation} \label{eq:K-operator-synapse-equation}
s(t) = \hat K_s[ r(t)] .
\end{equation}
This shows that the synapse response (membrane voltage perturbation) is a filtered version of its firing rate input. 

The same logic applies to Eq.~\ref{eq:synapses_1st}, but this time the operator is first order,
\begin{equation}
\hat L_s [s(t)]= \frac{1}{\gamma}  [\tau_d\, d/dt  +1]s(t).
\end{equation}

\begin{figure}[t!]
    \centering
    \includegraphics[width=0.65\linewidth]{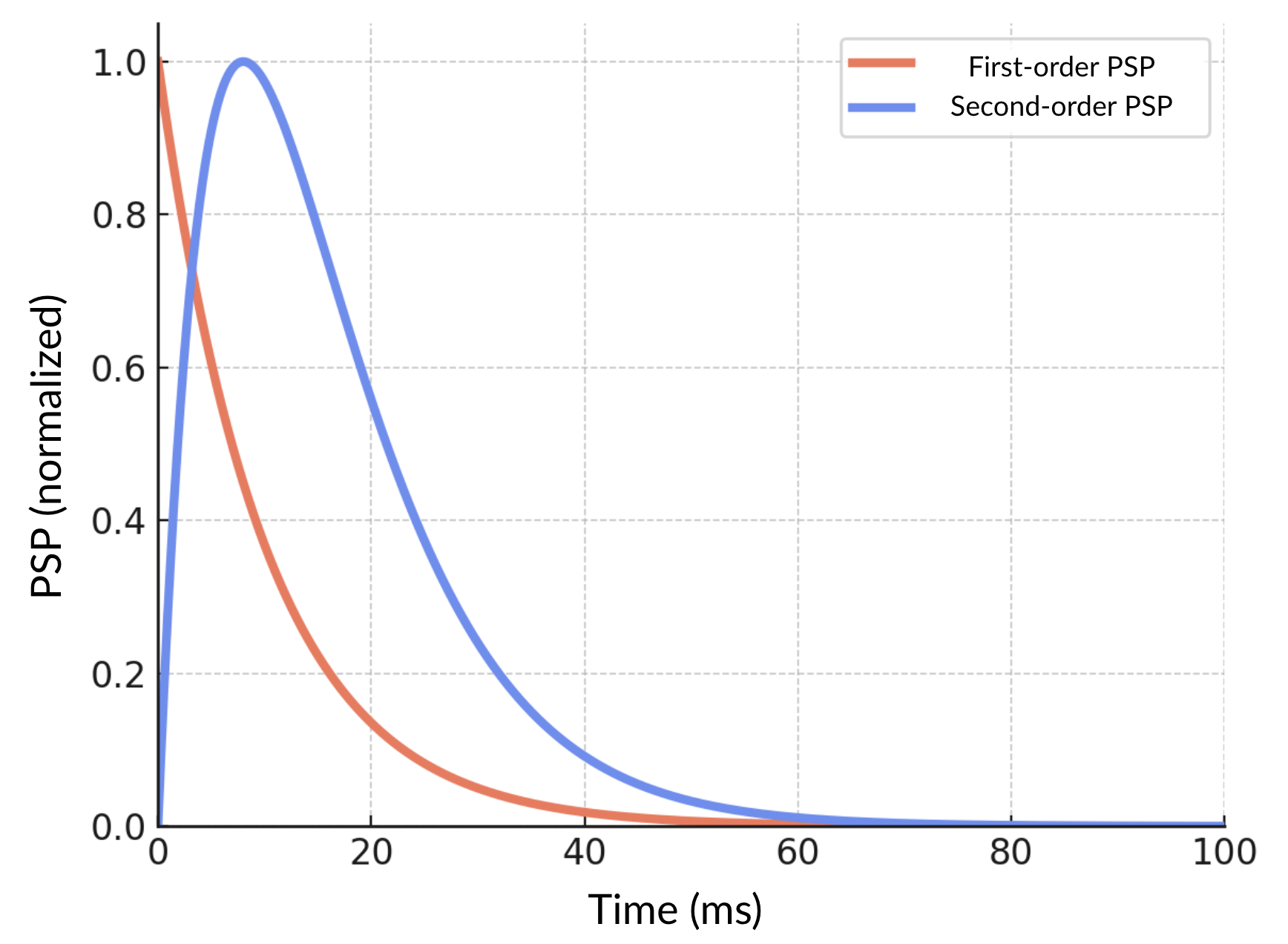}
    \caption{Post synaptic potentials in the first-order or the more realistic second-order synapse models. These plots are the response to an ``impulse" at time zero, i.e., the solutions to $\hat L[x] = \delta (t)$.
    \label{fig:psps}}
\end{figure}

Synapses are \textit{linear operators}:  filters that transduce incoming firing rate inputs into currents or PSPs.
 The effect of such a filter is to distort and delay the incoming signals.
Their action can be characterized by the impulse response, i.e.,   how it responds to a sharp (delta function) input, $\hat L[x] = \delta (t-t_0)$. In the case of Wilson-Cowan, where the operator is first order, the response is simply a decaying exponential (see Fig.~\ref{fig:psps}).

In more realistic models, such as Jansen-Rit (JR)~\cite{jansen1995}, Wendling~\cite{wendlingEpilepticFastActivity2002}, or LaNMM~\cite{Ruffini2020P118,SanchezTodo2023}, the linear operator is \textit{second order}.
The second-order operator is essential for realistically capturing synaptic conductance dynamics, as biological PSPs do not rise instantaneously. While a simple first-order exponential decay adequately describes passive membrane voltage relaxation, it neglects the finite time required for neurotransmitter binding, ion channel opening, and resulting conductance increase. Introducing separate rise ($\tau_{\text{rise}}$) and decay ($\tau_{\text{decay}}$) time constants (which may be the same numerically), the second-order operator accurately represents this two-phase process: a rapid initial conductance increase followed by slower conductance closure. This explicitly accounts for physiological delays and ensures PSP dynamics include a biologically realistic temporal scale, crucial for correctly modeling neuronal integration and timing-dependent neural computations.

Synaptic dynamics depend on the firing rate of the presynaptic population, which we now turn to. 

\rev{\subsection{Current-based versus conductance-based synapses}
\label{sec:current-vs-conductance}
Throughout this review we adopt a \emph{current-based} description of synaptic interaction, in which the synaptic input is treated as an additive perturbation to a postsynaptic state variable through the linear filter $\hat{L}_s$. This is a deliberate idealisation. Real chemical synapses are \emph{conductance-based}: the synaptic current depends explicitly on the postsynaptic membrane potential through the reversal potential $E_{\rm rev}$ of the relevant ion channel,
\begin{equation}
I_{\rm syn}(t) \;=\; g(t)\,\bigl(E_{\rm rev} - V(t)\bigr),
\label{eq:conductance-syn}
\end{equation} 
where $g(t)\ge 0$ is the time-dependent synaptic conductance, generated by the same kinetic mechanisms that produce the second-order PSP shape derived above~\cite{destexhe1994,Destexhe1998}. The dependence on $V(t)$ is the source of \emph{shunting}: as the membrane potential approaches $E_{\rm rev}$, the driving force $(E_{\rm rev}-V)$ vanishes and the synapse delivers no current despite the elevated conductance, and if $V$ overshoots $E_{\rm rev}$ the effective sign of the interaction reverses. An ``inhibitory push'' can therefore become a ``pull'' or vanish entirely depending on the operating point, a property absent from the current-based filter. For the cross-walk developed in this review, the current-based form is retained because the rate-based and exact-mean-field neural mass models compared in \S\S\ref{sec:WILCO}--\ref{sec:NMM2} are themselves formulated in current-based terms, and because near a stable operating point the conductance-based interaction linearises to a current-based effective coupling with state-independent effective gain. The conductance-based / shunting Liley-type mass listed in block~(B) of the roadmap Table retains the full $V$-dependence and is the natural starting point for readers requiring explicit shunting effects, including the voltage-dependence of NMDA-receptor dynamics noted at the end of \S\ref{sec:synaptic-dynamics}}.

\subsection{Transfer functional}

Experimental results characterize the firing rate 
of a population of neurons as a function of its total input current or membrane potential. This approach extends the concept of $f$-I curve, used to characterize 
the dynamics of single neurons~\cite{ermentroutMathematicalFoundationsNeuroscience2010,dayan_theoretical_2001}, to a neural population.
Accordingly, a population receiving an input current $I(t)$  produces a firing rate given by
\begin{equation}\label{eq:transfer-functional}
r(t)=\hat \Phi[I(t)],
\end{equation}
where $\hat \Phi$ is some operator (the \textit{transfer functional}) that describes the process, including delays in response and non-linearity. For simplicity, this is sometimes simplified to a static  and instantaneous nonlinear relation between the input and the 
output of a neural mass.  The functional becomes then  a function $\varphi$ with
\begin{equation}\label{eq:transfer-function}
r= \varphi(I),
\end{equation}
which receives the name of \emph{transfer function}.  


\begin{table*}\label{table1} \small
\centering
\begingroup
\arrayrulecolor{slate}              
\setlength{\arrayrulewidth}{0.4pt}  
\begin{tabular}{|c|c|c|} \hline
  \rowcolor{softblue!25}
  \textbf{\color{softblue!60!black} Model} &
  \textbf{\color{softblue!60!black} $\varphi[I]$} &
  \textbf{\color{softblue!60!black} Refs.} \\
  \hline
  & & \\  
  Sigmoid & $\displaystyle{\frac{2e_0}{1+e^{\rho(I_0-I)}}}$ & \cite{Wilson1972,jansen1995}\\[15pt]
  LIF (Gaussian noise) & $\left\{\displaystyle{\tau_m\sqrt{\pi} \int_\frac{V_r-I}{\sigma}^\frac{V_\text{th}-I}{\sigma} \exp(x^2)[1+\erf(x)]\dd x}\right\}^{-1}$ & \cite{siegert1951,amit1997,brunel1999} \\[15pt]
  EIF (Gaussian noise) & $\displaystyle{\left\{\frac{2\tau_m}{\sigma^2}\int_{-\infty}^{V_{th}}
  \int_{\max(V,V_r)}^{V_{th}}\mkern-32mu\mkern-18mu e^{\frac{2}{\sigma^2}(F(v)-F(u))}
  \dd u \dd v \right\}^{-1} }$ & \cite{fourcaud-trocme2003,fourcaud-trocm2005} \\[15pt]
  QIF (Gaussian noise) & $\displaystyle{\left\{\tau_m\sqrt{\pi} \int_{-\infty}^\infty \exp\left(-\frac{\sigma^4}{48}x^6 - Ix^2 \right) \dd x\right\}^{-1}}$ & \cite{brunel2003a} \\[15pt]
  QIF (Cauchy noise) & $\displaystyle{\frac{1}{\tau_m\pi\sqrt{2}}}\sqrt{I+\sqrt{I^2+\Delta^2}}$ & \cite{montbrio2015,clusella2022a} \\
  & & \\ 
  \hline
\end{tabular}
\endgroup
\caption{Analytical expressions for the transfer functions of some heuristic models and 
static approximations of integrate-and-fire models, including leaky integrate-and-fire (LIF), exponential integrate-and-fire (EIF) and quadratic-integrate-and-fire (QIF). In the case
of EIF, $F(v) = -\frac{v^2}{2} + Iv + \Delta_T^2\exp\left(\frac{v-V_T}{\Delta_T}\right)$.
} \label{table:transfers}
\end{table*}

There are several possible choices for the shape
of the transfer function $\varphi$ (see Table~\ref{table:transfers}). 
A common choice is the sigmoid function considered by Wilson and Cowan~\cite{Wilson1972},
\begin{equation}\label{eq:sigmoid}
\varphi(I) = \frac{2e_0}{1+e^{\rho(I_0-I)}} \equiv \sigma [I] 
\end{equation}
where $e_0$ is half of the maximum firing rate of each neuronal population, $I_0$ is the value of the current when the firing rate is $e_0$, and $\rho$ determines the slope of the sigmoid at the central symmetry point $(I_0, e_0)$.
Wilson and Cowan argued that such a sigmoidal shape accounts for heterogeneity in either neural connectivity
or excitability threshold.
Freeman proposed another expression that he fitted to experimental data,
obtaining a similar sigmoid shape~\cite{freeman1975,freeman1979}.

Further theoretical and experimental research has validated that, indeed, neurons subject to an increasing input current $I(t)$ produce a firing rate that, in many cases, can be captured by a nonlinear function.
However, these functions do not always follow a sigmoid function.
For instance, Amit and Brunel derived the transfer function that follows
from considering a network of leaky integrate-and-fire neurons subject to Gaussian noise~\cite{amit1997}
Similarly, transfer functions for exponential integrate-and-fire, quadratic integrate-and-fire
and others have been derived analytically~\cite{fourcaud-trocme2003,brunel2003a,fourcaud-trocm2005,montbrio2015}.
We provide the transfer functions of some of these cases in Table~\ref{table:transfers}.

Dynamical transfer functions, where the response of the population is not instantaneous, were addressed by Wilson and Cowan, who proposed a filtered, exponential decay dynamics towards the firing rate activity, 
\begin{equation}\label{eq:basic_firing}
\tau_m \dot r = -r + \varphi( I (t)),
\end{equation}
 where $\tau_m$ is a time scale controlling the response time of the neuron.
 As mentioned at the beginning of this section, this exponential decay response has often been used interchangeably with the exponential decay
dynamics presented in Eq.~\eqref{eq:synapses_1st}, although notice that the time
constants $\tau_m$ and $\tau_d$ account for different biophysical quantities. 

Using operator language with $ \hat L_m = \tau_m d/dt +1$ and inverse $\hat K_m = \hat L_m^{-1}$ (a linear filter or convolution), this can be expressed as a non-linear filter
\begin{equation}\label{eq:tranfer-functional-relation}
r = \hat \Phi_m[ I(t)]
\end{equation}
with  the \textit{transfer functional} is 
$
\hat \Phi_m [ \cdot]= \hat K_m \big[ \varphi[ \cdot]\big].
$
In this case, we talk of a transfer \textit{functional} and not function: a nonlinear operator mapping total input currents to firing rate time series with delay governed by the time scale $\tau_m$.

 \begin{figure}[t!]
    \centering
    \includegraphics[width=0.4\linewidth]{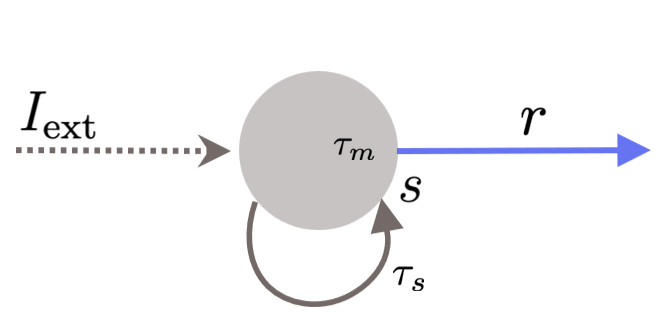}
    \caption{Simple recurrent population model. Here, the population receives an external current  $I_\text{ext}$ and self-input via a synapse $s$, and outputs its firing rate $r$. Two time scales (delays) enter: the membrane response time constant $\tau_m$ for updating the firing rate $r$, and the synaptic current time constant $\tau_s$. See Eq.~\eqref{eq:simpleNMM} in the text for details.}
    \label{fig:simplemodelrecurrent}
\end{figure}

\subsection{A simple self-coupled model and different limits} 
Let's consider now a population of neurons of the same type that receive an external input $I_\text{ext}(t)$
and has recurrent connectivity given with a strength $\kappa$ (see Figure~\ref{fig:simplemodelrecurrent}). 
Then, we can assume that the total input received by the population is
\begin{equation}\label{eq:input}
I(t) = \kappa s(t) + I_\text{ext}
\end{equation}
where $\kappa$ is an electrical admittance relating membrane potential perturbation and current.
Notice that to derive the expression of the total input~\eqref{eq:input} we have assumed
that the current flux generated by the recurrent coupling is proportional to the post-synaptic potential.
This is an additional hypothesis that we will revisit later on.

Therefore, the dynamics of the average PSP of this population is given by
\begin{equation}\label{eq:simpleNMM}
\begin{aligned}
 \tau_m \dot r &= -r + \varphi[\kappa s +  I_\text{ext}] \;,\\
 \hat L_s[s] &= r.
 \end{aligned}
\end{equation}
which we can rewrite in full operator form as the\textit{ master neural mass model equation}
\begin{equation}\label{eq:master_nmm}
 \boxed{\begin{aligned}
 r &=  \hat \Phi_m [\kappa s +  I_\text{ext}] ,\\
 s &= \hat K_s [r].
 \end{aligned}}
\end{equation}
with $\hat \Phi_m$ the transfer functional and  $\hat K_s$ the usual synaptic filter inverse operator characterized by some time scales $\tau_s$ --- see Eqs.~\eqref{eq:tranfer-functional-relation} and~\eqref{eq:K-operator-synapse-equation}.

Equation~\ref{eq:master_nmm} provides the minimal building blocks to derive a range
of NMMs from simple principles.  As we discuss next, taking
$\tau_m\to 0$ or $\tau_s\to 0$ with first or second order synaptic operators, it covers the Wilson-Cowan (see Section~\ref{sec:WILCO}), NMM1 (Section~\ref{sec:NMM1}) and NMM2 (Section~\ref{sec:NMM2}) formalisms.   
Finally, it also encompasses other logical variants — for example, a Wilson–Cowan style transfer function paired with second-order synapses. Although extensions of the Wilson–Cowan model do incorporate synaptic filtering or plasticity dynamics, to the best of our knowledge there exists no commonly adopted neural-mass model that preserves the original Wilson–Cowan architecture while explicitly modelling separate finite time constants for the membrane and the synaptic stages and, in particular, uses second-order synaptic operators — as proposed in our ‘master NMM’ formalism.

\paragraph{Limit of fast synapses (Wilson-Cowan model).} Taking the synaptic time constant to be much smaller than the membrane constant,  $\tau_s 
\rev{\ll} \tau_m$, leads to the \textit{limit of fast synapses, }
\begin{equation}\label{eq:wilco0}
\begin{aligned}
r &=  \hat \Phi_m [\kappa s +  I_\text{ext}] \;,\\
 s &= \gamma r,
 \end{aligned}
\end{equation}
with $\gamma$ the scaling constant associated with the $\hat L$ operator. This is the stance taken by Wilson-Cowan (see  Section~\ref{sec:WILCO}), which we can rewrite in firing rate form as 
\begin{equation}\label{eq:wilco0-op}
 r =  \hat \Phi_m [\omega r +  I_\text{ext}] ,
\end{equation}
with $\omega= \gamma \kappa$, or, equivalently in synapse potential (PSP) form as
\begin{equation}\label{eq:wilco0-op-s}
 s =  \gamma \hat \Phi_m [\kappa s +  I_\text{ext}].
\end{equation}
In this limit, we can think of firing rate and synaptic PSPs as the same up to scale (there is no delay, only a rescaling). Thus, the Wilson-Cowan equations can be read as firing rate or synaptic equations, although the dynamics are due to membrane capacitance rather than synaptic currents.

 \paragraph{Limit of fast membrane.} Conversely, if \rev{$\tau_m \ll \tau_s$} --- the \textit{limit of fast membrane} ---, the firing rate ODE becomes a static equation,
\begin{equation}
\begin{aligned}
 r  &=  \varphi[\kappa s +  I_\text{ext}] \;,\\
 \hat L_s[s] &=  r.
 \end{aligned}
\end{equation}
This is the approach taken in the Jansen-Rit formalism (NMM1) when the synaptic equations are second-order, for example.

\paragraph{Delays and refractory period.} Finally, Equation~\eqref{eq:simpleNMM} contains two time constants ($\tau_m$ and $\tau_s$).
However, there are two more time constants that play an important role in neural dynamics, and which we briefly mention here: 
 \textit{time delays} and \textit{refractory periods}.
Time delays introduce additional time scales that account for the latency of neurons or connecting circuit elements
for reaching the threshold voltage and releasing the neurotransmitters.
This effect might be due to different elements, for instance, the propagation of signals along the axon.
This might be simply modeled by adding a delay in the synaptic dynamics
\begin{equation}
\hat L[s]  =  r(t-\tau_l).
 \end{equation}
Notice that delay-differential equations have some special properties.
For instance, their dynamics are infinite-dimensional, and thus, phenomena such as
oscillatory activity can arise in instances where this was not previously considered.

On the other hand, refractory periods correspond to the inability of neurons to respond
to external stimuli for a few milliseconds after an action potential.
Whether refractory periods have a major role in the collective dynamics and function of neural populations is still a matter of debate.
In the simple framework explained so far, there are two main ways to include a refractory period.
On one hand, some researchers assume that transfer functions with a sigmoid function, i.e., a saturation for higher input, are already accounting for these refractory periods, hence the saturation.
However, many researchers consider transfer functions derived from models, which do not saturate for large inputs.
On the other hand, following Wilson and Cowan \cite{wilson_excitatory_1972}, the fraction of neurons susceptible to external inputs
at any time $t$ is given by
\begin{equation}
    1-\int_{t-\tau_f}^t r(t) \dd t
\end{equation}
where $\tau_f$ (ms) is the refractory period.
Therefore, Eq.~\eqref{eq:basic_firing} should instead read
\begin{equation}
    \tau_m \dot r = -r + \left(1-\int_{t-\tau_{f}}^t r(t) \dd t\right)\Phi[ I(t) ]\;.
\end{equation}
By considering that $\tau_{f}\ll \tau_m$, the integral can be approximated 
as $\int_{t-\tau_{f}}^t r(t) \dd t\approx \tau_{f} r(t)$ and the previous equation then reads
\begin{equation}
    \tau_m \dot r = -r + \left(1-\tau_{f}r\right)\Phi[ I(t) ]\;.
\end{equation}

For the sake of completeness, we now  reproduce Eq.~\eqref{eq:simpleNMM}
with both time delays and refractory period:
\begin{equation}\label{eq:simpleNMMwithdelays}
\begin{aligned}
\tau_m \dot r &= -r + (1-\tau_{f}r)\Phi[\kappa s(t) +  I_\text{ext}] \;,\\
 \hat L[s]&=  r(t-\tau_l) 
 \end{aligned}
\end{equation}
 
\section{The Wilson–Cowan model (WILCO)} \label{sec:WILCO}

\begin{tcolorbox}[colback=gray!6,colframe=gray!6,enlarge left by=0mm, boxrule=0pt, enhanced, sharp corners]
As a phenomenological normal form, the Stuart–Landau (SL) oscillator captures the universal signature of a Hopf bifurcation.  Yet real neural tissue is not a single abstract amplitude, but excitation and inhibition speaking in tandem.  Wilson–Cowan (WILCO) portrays that biological dialogue transparently: two real variables \(x,y\), representing firing rates or PSPs, interacting via synaptic weights, membrane time‑constants, sigmoid gains, and neural drives.  
\rev{ But compared with the simpler oscillator models, WILCO is not only biologically transparent but dynamically richer.
Near a Hopf bifurcation that it sits on the edge between a quiescent (damped) fixed point and a self-sustaining oscillation, WILCO reduces on its centre manifold to the SL normal form (see Appendix~\ref{app:WILCO2SL}). This equivalence is local: away from the bifurcation, and depending on parameter choices, WILCO supports saturation regimes, multistability between fixed points, and pattern-formation regimes that are not captured by the SL normal form.  } Finally, while the Wilson-Cowan model was originally conceived with $x$ and $y$ representing firing rates, its dynamics can also be used to represent first-order synaptic activity, with $x,y$ representing PSPs.  

\end{tcolorbox}

Following Eq.~\eqref{eq:wilco0} (fast synapse limit, where firing rate input and PSP are related linearly without delay), or more concretely the  form in Eq.~\eqref{eq:wilco0-op-s},  for a single population, we consider the  model for a pair of coupled populations introduced by Wilson and Cowan \cite{wilson_excitatory_1972},
\begin{equation}
    \begin{aligned}
  \tau_x \dot{x}+x &= \gamma_x\,\sigma_x\!\bigl(\kappa_{xx}x - \kappa_{xy}y + P_x\bigr),\\
  \tau_y \dot{y}+y &= \gamma_y\,\sigma_y\!\bigl(\kappa_{yx}x - \kappa_{yy}y\bigr),
  \label{eq:WILCOcore-psp}
\end{aligned}
\end{equation}
 with $x$ and $y$ representing synapse PSPs   and \(\tau_{\alpha}\) are membrane time‑constants, \(\kappa_{\alpha\beta}\) the effective synaptic weights (positive for excitation, negative for inhibition), \(P_x\) a tonic slowly varying drive that acts as the bifurcation (control) parameter, and \(\gamma_{\alpha}\) the synapse gain term 
\cite{Ermentrout2010}.  The sigmoid functions (see Eq.~\ref{eq:sigmoid}) are specific for each population.  

In the form  corresponding to Eq.~\eqref{eq:wilco0-op}, where $x,y$ represent firing rates, we have
\begin{equation}
\begin{aligned}
  \tau_x \dot{x}+x &=  \sigma_x\!\bigl(w_{xx}x - w_{xy}y + P_x\bigr),\\
  \tau_y \dot{y}+y &=  \sigma_y\!\bigl(w_{yx}x - w_{yy}y\bigr).
  \label{eq:WILCOcore}
\end{aligned}
\end{equation}

Regardless of its synaptic or firing rate form, the original WILCO model describes dynamics stemming from transfer function delays. 
Despite its origins, the model is sometimes interpreted as describing first-order synapse dynamics. The form remains the same, but the interpretation changes (dynamics are induced by synaptic, not membrane delays). 

A Taylor expansion shows that Eq.~\eqref{eq:WILCOcore} contains the  Hopf bifurcation machinery in the SL normal form plus additional bifurcations. Moreover,  each coefficient is anchored to a biophysical mechanism (e.g.\ self‑excitation \(w_{xx}\) or synaptic delay \(\tau_x\)).

\rev{
WILCO contains the Stuart--Landau oscillator as one of its parameter-tuned limits. When the cross-coupling is weak, the model supports a clean Hopf bubble (HB$^-\!\to$ LC $\to$ HB$^+$) with no fold or SNIC anywhere on the slice, and the center-manifold reduction at either end is the Stuart--Landau normal form. In this limit WILCO is SL. Once the cross-coupling product $\kappa_{xy}\kappa_{yx}$ crosses a threshold set by the diagonal terms, a Bogdanov--Takens point enters the oscillatory window. The saddle-node, SNIC, and Hopf curves emerge from it as three branches of a single codimension-2 unfolding: the cusp brings bistability, the SNIC brings Class~I excitability, and the system displays dynamics genuinely beyond SL's reach. In this regime WILCO is more than SL. The two regimes are 1-parameter slices through the same equations, separated by exactly one codimension-2 point in parameter space. This is what gives WILCO its place in the Rosetta Stone: it specializes to SL --- and to the PING/Jansen--Rit gamma rhythm --- at one limit, extends to bistability and Class~I excitability at the other, and the hinge between them is a single locatable point.}
\subsection{Effect of forcing}

Shifting the tonic drive \(P_x\) moves the mean input along the sigmoid with respect to the sigmoid threshold, i.e., the population \textit{operating point}; because the slope \(\sigma_x'(\cdot)\) changes with input, the effective linear gain and thus the distance to Hopf vary smoothly.\cite{breakspear_dynamic_2017}

Introducing an additional term \(\hat F_e(t)\) to the excitatory equation in firing rate form reads
\begin{align}
\tau_x\dot{x}+x = \sigma_x\!\bigl(w_{xx}x - w_{xy}y + P_x + \hat F_e(t)\bigr).
  \label{eq:WILCOforcing}
\end{align}
Near its bifurcation, the model becomes a forced SL oscillator\cite{fredrickson-hemsingModelockingDynamicsHair2012}.  
\(\hat F_e(t)\) may represent noise or a time-varying (e.g.\ sinusoidal, noisy, pulsed), or constant but transiently switched on and off forcing; in either case, it probes or entrains the dynamics without altering the underlying bifurcation point set by \(P_x\). Sinusoidal forcing carves out Arnol'd tongues of \(n{:}m\) phase locking whose geometry parallels that of the analytical SL normal form, yet the effective tongue width is filtered by the sigmoid slope \(\sigma_x'\), a feature absent from purely polynomial descriptions.

\subsection{Minimal ingredients for a Hopf bifurcation}

Linearizing \eqref{eq:WILCOcore} around its fixed point \((x_0,y_0)\) and applying the Routh–Hurwitz criteria, a Hopf bifurcation occurs when \(\text{tr}\,J=0\) with \(\det J>0\) \cite{guckenheimerNonlinearOscillationsDynamical2002}.
Physiologically, this translates into three rules:

\begin{enumerate}
  \item \emph{Reciprocal E–I coupling} (\(w_{xy}w_{yx}<0\)) is indispensable; a single self‑exciting pool cannot realize a  Hopf bifurcation.
  \item \emph{Sigmoid slope.} Either recurrent excitation \(w_{xx}>0\) \rev{or} an external drive \(P_x\) must position the operating point on the steep part of \(\sigma_x\) so that \(\sigma_x'(\cdot)\neq0\), recovering Ermentrout’s criterion for “sufficient self‑coupling or bias”.\cite{Ermentrout:2010aa}
  \item \emph{Cubic saturation.} The curvature of \(S_{\alpha}\) supplies the cubic term that tames the growth of infinitesimal oscillations; piece‑wise linear gains suppress this mechanism.
\end{enumerate}

Projecting the WILCO equations onto their two‑dimensional center manifold yields—after normal‑form transformation—the complex Stuart–Landau equation
\cite{pietras2019network}
\begin{equation}
\dot{z} \;=\; (\alpha + i\,\omega)\,z \;-\; (\gamma + i\,\beta)\,\lvert z\rvert^{2}z,
\label{eq:sl_complex}
\end{equation}
where the real coefficients \(\alpha,\gamma\) and their imaginary counterparts \(\omega,\beta\) are explicit functions of \(\tau_{\alpha},w_{\alpha\beta}\) and the first two derivatives of \(S_{\alpha}\); their signs decide whether the Hopf is super‑ or sub‑critical.
Conversely, the physiological variables follow $(x,y)^\top \approx \Re[z\,v]$, with $v$ the critical eigenvector. See Appendix~\ref{app:WILCO2SL} for more details.

The logistic transfer curve packages dendritic saturation, membrane noise, and threshold heterogeneity into one analytic function.\cite{deco_dynamic_2008}. 
Its first derivative shapes the linear gain, its second derivative injects the cubic non‑linearity that stabilizes the limit cycle—precisely the \(-(\gamma+i\beta)|z|^{2}z\) term in \eqref{eq:sl_complex}.  In other words, the WILCO sigmoid is a \emph{built‑in normal‑form generator}.

\begin{figure}[t]
  \centering
  \includegraphics[width=\textwidth]{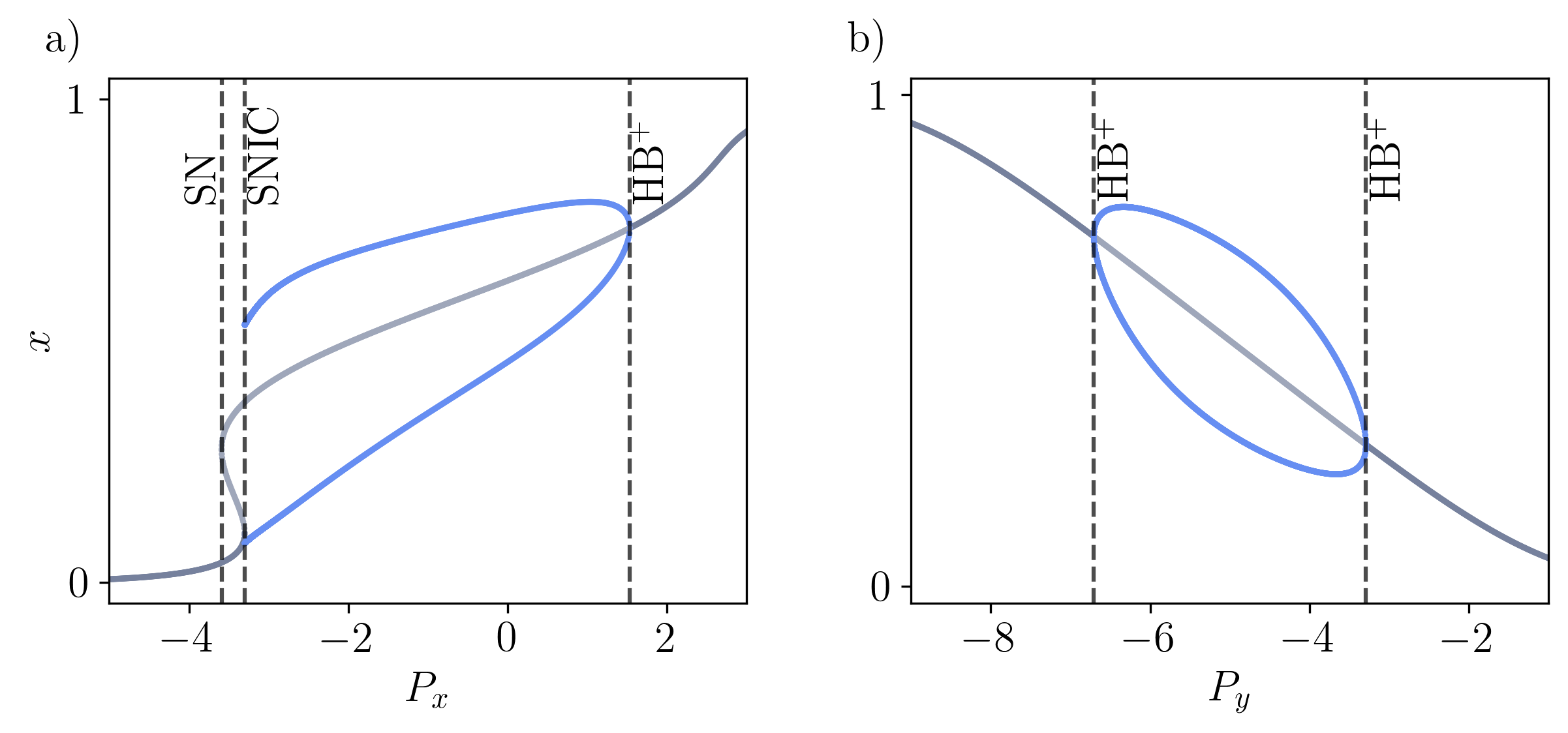}
 \caption{ \rev{The two oscillator personalities of WILCO. Both panels display the $x$-component of fixed points and limit cycles as a function of external inputs. Solid grey lines indicate stable/unstable fixed points, while blue curves denote limit-cycle minima and maxima. (a) Transitioning through $P_x$ from left to right, the system exhibits a progression from a saddle-node (SN) to a saddle-node-on-invariant-circle (SNIC) bifurcation, eventually terminating at a supercritical Hopf bifurcation ($\text{HB}^+$). This configuration allows the model to switch between Class I excitability (near the SNIC) and Class II excitability (near the $\text{HB}^+$). (b) The pure Hopf bubble: In this regime, the oscillatory activity is isolated within a "bubble" bounded by two supercritical Hopf bifurcations ($\text{HB}^+$), with no fold or SNIC bifurcations present. Here, the system functions as a pure Class II oscillator, locally equivalent to a Stuart–Landau oscillator.}}
  \label{fig:wilco_two_personalities}
\end{figure}

\rev{
Two examples of WILCO bifurcation diagrams are provided in
Fig.~\ref{fig:wilco_two_personalities}; they are 1-parameter slices
through the same model equations, differing only in which external input parameter is varied ($P_x$ or $P_y$) while the others remain fixed.

The model in panel~(a) --- $P_x$-driven regime (see Table~\ref{tab:bif-params} for parameters) --- exhibits three different
bifurcations as the external input to the excitatory population,
$P_x$, is increased. Initially, for low $P_x$, the system has a
single stable fixed point. As $P_x$ increases, the system undergoes
a saddle-node (SN) bifurcation, where two unstable fixed points
emerge. Further increasing $P_x$ leads to the collision and
annihilation of a stable and an unstable fixed point, and, beyond
this point, a limit cycle emerges. This transition corresponds to
a saddle-node on an invariant circle (SNIC) bifurcation. Finally,
the limit cycle vanishes in an HB$^+$ bifurcation, and from that
point the system only has a stable fixed point. Thus, near HB$^+$,
WILCO reduces on its center manifold to a Stuart--Landau oscillator
(Class~II onset, finite frequency, vanishing amplitude); near SNIC
it reduces to a theta-neuron / QIF normal form (Class~I onset,
finite amplitude, vanishing frequency); and near SN, to a
saddle-node normal form (bistability, no oscillation).
All three reductions are local pieces of the same
Bogdanov--Takens (BT) unfolding, of which panel~(a) is a
1-parameter slice.

The model in panel~(b) --- the $P_y$ driven regime-coupling (see Table~\ref{tab:bif-params} for parameters) --- exhibits a different bifurcation structure: a clean Hopf bubble bounded by HB$^-$ at one end and HB$^+$ at the other, with no fold and no SNIC anywhere on the slice. The cycle is born of small amplitude at HB$^-$, grows smoothly across the bubble, and dies of small amplitude at HB$^+$. The center-manifold reduction at \emph{both} ends of the bubble is the Stuart--Landau normal form, and the cycle is locally Stuart--Landau across the whole window (Class~II only, no Class~I, no bistability).

By navigating this parameter space, WILCO effectively contains the Stuart–Landau oscillator as a parameter-tuned limit while adding the BT-organized cusp and SNIC structures as a functional extension. This allows the same model equations to switch between acting as a simple harmonic-like oscillator and a complex neural integrator.
}


\subsection{Coupling}

To build a whole-brain network of equal Wilson–Cowan E-I  motifs,  the excitatory populations are usually coupled via long-range glutamatergic projections \cite{castaldo2023multi}.  Introducing both tonic/exogenous inputs ($P_{x,i}$) and dynamic drives ($\hat F_e(t)$) at the excitatory nodes, the coupled system reads (firing rate version)
\begin{tcolorbox}[
  colback=gray!5,
  colframe=gray!5, 
  boxrule=0pt,
  enhanced, sharp corners,
  left=4pt, right=4pt, top=3pt, bottom=3pt,
  boxsep=2pt,
  before skip=6pt, after skip=6pt
]
\begin{equation} 
\begin{aligned}
  \tau_x \dot{x}_i + x_i 
    &= \sigma_x\!\Bigl(
         w_{xx}\,x_i 
       - w_{xy}\,y_i 
        + P_{x,i} 
       + \hat F_{e;i}(t) + \sum_{j\neq i }^N C_{ij}\,x_j\Bigr)
       ,
  \label{eq:WILCO_net_x}\\[4pt]
  \tau_y \dot{y}_i + y_i 
    &= \sigma_y\!\bigl(
         w_{yx}\,x_i 
       - w_{yy}\,y_i
       \bigr).
\end{aligned}
\end{equation}
\end{tcolorbox}

Here  
\({C}_{ij}\) denotes the (row-normalised) structural connectivity weight from node~\(j\) to node~\(i\), typically estimated from diffusion-MRI tractography.  

Long–range excitation \(\sum_j C_{ij}x_j\) is integrated inside the sigmoid:  distant axonal currents enter the same dendritic current pool through instantaneous synapses; the non-linearity then converts the total current into a bounded firing rate, preserving the physiological ceiling set by \(\sigma_x\). 

\rev{The network form in Eq.~\eqref{eq:WILCO_net_x} has been the subject of extensive analysis in the dynamical-systems and computational-neuroscience literature, going well beyond the original two-population formulation of Wilson and Cowan. Foundational mathematical analyses of coupled WILCO units characterised the phase-plane structure, oscillation onset, and bifurcation skeleton of the canonical pair~\cite{borisyuk_kirillov_1992,hoppensteadt_izhikevich_1997}, and subsequent work has mapped the multistability, metastability, and pattern-forming regimes that arise when many WILCO nodes are coupled through delayed connectivity~\cite{daffertshofer_vanwijk_2011,coombes_byrne_2018}. The treatment in the present section is therefore best understood as a unified-notation summary of an established line of work, repositioned to serve the cross-walk to the Stuart--Landau, NMM1, and NMM2 formalisms developed in subsequent sections.}

\rev{Of particular relevance to the cross-walk this review develops, WILCO networks admit phase and phase--amplitude reductions back to generalised Kuramoto-type oscillator networks of the form discussed in \S\ref{sec:linear-phase}. Hlinka \& Coombes~\cite{hlinka_coombes_2012} demonstrated that phase reductions of WILCO networks reproduce empirically observed structural--functional connectivity relationships in resting-state fMRI, providing an explicit bridge between the rate-based formulation and the phase-oscillator analyses of \S\ref{sec:linear-phase}. Daffertshofer \& van Wijk~\cite{daffertshofer_vanwijk_2011} further analysed how the amplitude dynamics of WILCO modulate phase connectivity, clarifying the regime in which the phase-only description remains predictive. These reductions are the operational bridge that makes the WILCO\,$\to$\,Kuramoto leg of the cross-walk explicit.}

Equations \eqref{eq:WILCO_net_x}  implement \emph{additive} coupling: each excitatory node \(i\) receives the transduced firing rates \(x_j\) of its peers weighted by \(C_{ij}\). This choice reflects the physiology of long‐range excitatory fibers and avoids the homeostatic constraints of diffusive schemes, allowing the network to exhibit rich collective dynamics—from large‐scale synchrony and traveling waves to stimulus‐induced entrainment—while preserving the local Wilson–Cowan bifurcation structure.  

Equation~\ref{eq:WILCO_net_x} is one among many possible combinations of populations, couplings, and forcings.

\begin{tcolorbox}[
  colback=gray!5,
  colframe=slate,
  colbacktitle=softblue!25,
  coltitle=softblue!60!black,
  fonttitle=\bfseries\small,
  title={\small \textit{Wilson-Cowan Network (Eq.~\ref{eq:WILCO_net_x})} \\ Whole-brain Simulations Parameters \& Physiological Meaning},
  boxrule=0.4pt, enhanced, sharp corners,
  left=4pt,right=4pt,top=3pt,bottom=3pt,
  boxsep=2pt, before skip=6pt, after skip=6pt
]
\scriptsize
\begin{tabular}{@{}l p{0.82\linewidth}@{}}
\textbf{Node $i$} & Coarse‑grained cortical/subcortical region represented by two interacting subpopulations: excitatory ($x_i$) and inhibitory ($y_i$).\\
\textbf{$N$} & Number of regions (E–I motifs) in the network.\\
\textbf{$x_i(t),\,y_i(t)$} & Population activities (firing rates or PSP proxies, depending on the form used); the E/I push–pull loop produces oscillations and multistability (\emph{cf.} \eqref{eq:WILCOcore-psp}, \eqref{eq:WILCOcore}).\\
\textbf{$\tau_x,\;\tau_y$} & Membrane/synaptic time constants (ms) setting local response speeds and the E–I timescale separation.\\
\textbf{$\sigma_x(\cdot),\,\sigma_y(\cdot)$} & Static input–output (sigmoid) of each population; typical logistic with slope $\rho_{\alpha}$ and threshold $\theta_{\alpha}$ controlling gain and saturation.\\
\textbf{$w_{xx},w_{xy},w_{yx},w_{yy}$} & Local coupling weights (E$\to$E, I$\to$E, E$\to$I, I$\to$I). Signs follow \eqref{eq:WILCOcore}: $-w_{xy}y_i$ implements inhibition of E by I; $-w_{yy}y_i$ self‑inhibition of I.\\
\textbf{$P_{x,i}$ ($P_{y,i}$)} & Tonic drives (bias currents) shifting the operating point along the sigmoids and hence the effective linear gains.\\
\textbf{$G$} & Global coupling gain scaling long‑range excitation from other regions into $x_i$.\\
\textbf{$C_{ij}$} & Long‑range (row‑normalized) connectivity weight from region $j$ to $i$ (typically structural; functional/effective or synthetic graphs are alternatives). Enters additively inside $\sigma_x$ in \eqref{eq:WILCO_net_x}.\\
\textbf{$\tau_{ij}$} & Propagation delay on pathway $j{\to}i$ (conduction + synaptic latencies); optional.\\
\textbf{$\hat F_{e;i}(t)$}& Exogenous drive  (external forcing from nodes or elements outside the network or an electric field, see Equation~\ref{eq:forcing}). 
\end{tabular}
\end{tcolorbox}

\begin{tcolorbox}[
  colback=gray!5,
  colframe=slate,
  colbacktitle=softblue!25,
  coltitle=softblue!60!black,
  fonttitle=\bfseries\small,
  title={\small When to Use It},
  boxrule=0.4pt, enhanced, sharp corners,
  left=4pt,right=4pt,top=3pt,bottom=3pt,
  boxsep=2pt, before skip=6pt, after skip=6pt
]
\scriptsize
\textbf{Use this when} explicit excitation–inhibition, saturating gains, and operating‑point control are central (Hopf onset, multistability, stimulus responses, seizure‑like dynamics).\\
\textbf{Assumptions} mean‑field population description; first‑order E/I kinetics; static sigmoids capturing dendritic saturation and threshold dispersion; long‑range inputs summed into E.\\
\textbf{Best for} whole‑brain simulations with biophysical levers (E/I balance, gains, delays) and macroscopic readouts (FC/FCD, spectra, waves, metastability).\\
\textbf{Avoid} if only phase relations matter (prefer Kuramoto) or small‑fluctuation linear analytics suffice (prefer linear damped resonators).
\end{tcolorbox}

\begin{tcolorbox}[
  colback=gray!5,
  colframe=slate,
  colbacktitle=softblue!25,
  coltitle=softblue!60!black,
  fonttitle=\bfseries\small,
  title={\small How to Use It},
  boxrule=0.4pt, enhanced, sharp corners,
  left=4pt,right=4pt,top=3pt,bottom=3pt,
  boxsep=2pt, before skip=6pt, after skip=6pt
]
\scriptsize
\textbf{Provide} $C$ (SC by default; FC/EC or synthetic if SC absent), $G$, local $(\tau_x,\tau_y)$, $(w_{\cdot\cdot})$, $\sigma_{\alpha}$ parameters $(\rho_{\alpha},\theta_{\alpha})$, biases $P_{x,i}$ (and optionally $P_{y,i}$), plus optional $\tau_{ij}$, $F_i(t)$, noise.\\
\textbf{Defaults} normalize $C\!\leftarrow\!C/\lambda_{\max}(C)$; choose $\tau_x\!>\!\tau_y$ for gamma‑like loops or comparable for alpha/theta; start near Hopf by tuning $P_{x,i}$ and $w_{xx}$ to place the operating point on the steep part of $\sigma_x$; Euler/Heun or RK with $\Delta t\!\le\!1/(100\,f_{\max})$.\\
\textbf{Readouts} $x_i(t),y_i(t)$ time series; PSD/cross‑spectra; FC/FCD; phase–amplitude metrics; wave/cluster structure; operating‑point maps (via linearization around $(x_0,y_0)$).
\end{tcolorbox}

\subsection{Applications}

Historically, the Wilson--Cowan (WILCO) equations were introduced in two seminal papers in the early 1970s to capture the coarse‑grained dynamics of interacting excitatory and inhibitory neuronal populations \cite{WilsonCowan1972,WilsonCowan1973}. By replacing spikes with smooth population activity and embedding saturating input–output nonlinearities, the framework established a tractable mean‑field language for multistability, oscillations, and pattern formation. Later syntheses clarified when the mean‑field approximation is valid, how fluctuations and delays can be incorporated, and how WC relates to other neural‑mass and field formalisms \cite{DestexheSejnowski2009,Cowan2016,ChowKarimipanah2020}.

As a modeling workhorse, WILCO now spans scales—from local microcircuits to whole‑brain simulations coupled with connectomes derived from diffusion MRI—precisely because its parameters map cleanly to biology (excitatory/inhibitory gains, operating points, time constants, inputs) while retaining enough nonlinearity to express the canonical dynamical regimes. On the electrophysiology side, delay‑coupled WC networks fitted to human MEG reproduce band‑limited amplitude‑envelope correlations and phase‑locking across subjects when equipped with biologically plausible ingredients such as inhibitory synaptic plasticity and heterogeneous conduction delays; they naturally exhibit waxing–waning synchrony (metastability) and predict how changes in E/I balance or propagation speed reshape macroscopic spectra and coupling structure \cite{Abeysuriya2018}. Related firing‑rate analyses show how excitatory and inhibitory feedback loops jointly regulate gamma rhythms—useful for interpreting resonance and entrainment bandwidths observed in M/EEG \cite{Keeley2019}.

For fMRI, WILCO nodes embedded on the structural connectome have been used in multiscale pipelines that connect anatomy to resting‑state BOLD statistics and clinical phenotypes. \rev{Personalized WC-based models in major depressive disorder, for example, identified executive--limbic dysregulation consistent with empirical FC and symptom profiles~\cite{Li2021MDD}. 
} Within \textit{The Virtual Brain} ecosystem, WILCO is a standard regional model for synthesizing MEG/EEG/fMRI observables and probing how inter‑regional coupling, local gains, and delays generate subject‑specific variability in FC and its dynamics \cite{SanzLeon2013TVB,SanzLeon2015TVB}. In task and method‑development contexts, WILCO local dynamics frequently serve as the generative “ground truth” for benchmarking estimators of frequency‑specific or task‑modulated connectivity from MEG/fMRI \cite{Tewarie2023Nonrev,Masharipov2024}.

Because WILCO encodes excitation and inhibition explicitly, it offers a clean bridge to perturbation and inference. In Dynamic Causal Modeling (DCM) for fMRI, replacing the usual bilinear neuronal state equation with a WILCO‑type nonlinearity improves model evidence on multiple datasets while preserving physiological interpretability of effective connectivity and local transfer functions \cite{Sadeghi2020}. Clinically, introducing a non‑monotone (depolarization‑block) activation into a single WILCO microcircuit reproduces focal epileptiform activity and its spread, providing mechanistic handles for presurgical hypothesis testing \cite{Meijer2015}; conversely, disease‑specific applications at the whole‑brain scale use WILCO oscillators to explore how regional vulnerabilities and synaptic downscaling alter global connectivity and responsiveness \cite{SanchezRodriguez2024}. Stochastic variants poised near critical points reproduce avalanche statistics and scaling of spontaneous activity, offering a testbed for hypotheses about critical brain dynamics and their departures under pathology \cite{deCandia2021,Apicella2022,Alvankar2023}.

Comparative studies situate WILCO among other whole‑brain models. Multi‑modal head‑to‑head work reports that WILCO and SL networks achieve broadly comparable fits to MEG and fMRI benchmarks once conduction delays and local E/I homeostasis are respected; WC often affords advantages on spatiotemporal measures (e.g., functional‑connectivity dynamics, the size distribution of transient oscillatory modes) thanks to its explicit rate saturation and E/I partition, whereas SL’s normal‑form compactness facilitates analytic reductions and turbulence‑style analyses \cite{castaldo2023multi}. Systematic benchmarking across cohorts further underscores that no single model dominates all metrics: for some summary statistics and parcellations, simpler linear baselines can rival or exceed nonlinear formalisms, and reliability/subject‑specificity depend strongly on the targets of fit \cite{Domhof2022}. In practice, WILCO is most compelling when questions hinge on mechanistic E/I balance, pharmacology, stimulation, seizure dynamics or nonequilibrium signatures; when phase‑only timing, graph‑spectral tractability or normal‑form universality are paramount, Kuramoto/SL (and linear surrogates) may be the better lens. 
\rev{In all cases, WILCO's strengths and limitations are transparent. WILCO is a phenomenological rate model, originally formulated at the population level rather than derived from spiking dynamics: a compact E/I mean-field that is expressive enough to capture oscillations, multistability, and metastability while staying close to the biological levers experiments can manipulate.}


\section{NMM with second-order synapses (NMM1)} \label{sec:NMM1}
\begin{tcolorbox}[colback=gray!6,colframe=gray!6,enlarge left by=0mm, boxrule=0pt, enhanced, sharp corners]
In the Jansen-Rit or NMM1 formalism\footnote{(We use the term here NMM1 to avoid confusion with the specific model of three populations of Jansen and Rit).}, synaptic and somatic stages are separated explicitly.  Rather than collapsing synaptic dynamics into a single firing–rate equation, NMM1 treats each post-synaptic potential $x(t)$ and $y(t)$ as the output of a biologically grounded linear filter $\hat L_{\tau}$ (with separate rise and decay time constants), sums these to form the membrane perturbation $v$, and then applies a sigmoidal transfer $S(v)$ to yield the population firing rate.  This separation of synapse (impulse-response filters) and soma (sigmoid) provides a clear physiological connection and endows the model with proper delay time scales and intrinsic phase shifts that can sustain oscillations without the need for self-coupling.

 The biological ontology includes synapses, post-synaptic potentials (PSPs), the population membrane potential (the perturbation from its baseline), and the transfer function from membrane potential to firing rate output of the population (the sigmoid).

Accordingly, the variables in the equations include the postsynaptic potentials or PSPs ($x$ and $y$ in the E-I model we will discuss), the membrane potential $v$, the firing rates $r_x$ and $r_y$, and the sigmoid $S$.  As usual, all dynamical quantities refer to  ``mean" population averages. 
\end{tcolorbox}

The equations in NMM1  are similar to the Wilson-Cowan, but introduce second-order derivatives. Here we present them without self-coupling for simplicity (unlike in WILCO, it is not needed for a stable limit cycle)---see Figure~\ref{fig:fourmodels} (a),
\begin{equation}
    \begin{aligned}
            \tau_x^{2}\,\ddot{x} + 2\,\tau_x\,\dot{x} +x
              &= \gamma_x\,\sigma_x\bigl(  - w_{xy}\,y + \hat F_e \bigr),\\
            \tau_y^{2}\,\ddot{y} + 2\,\tau_y\,\dot{y} +y
              &= \gamma_y\,\sigma_y\bigl(w_{yx}\,x  \bigr) 
          \end{aligned}
\end{equation}
which can be expressed in operator formalism as
\begin{equation} \label{eq:nmm1-simple}
    \begin{aligned}
            L_x[x]
              &= \sigma_x\bigl(  - w_{xy}\,y + \hat F_e \bigr),\\
           L_y[y] &= \sigma_y\bigl(w_{yx}\,x  \bigr) 
          \end{aligned}
\end{equation}
with second order operator notation. For example, for the case of equal rising and decay times, 
\begin{equation}\label{eq:L-operator-same-a}
L_\alpha = \frac{1}{\gamma_\alpha} \big [ \tau_\alpha^2 \frac{d^2}{dt^2} + 2 \tau_\alpha \frac{d}{dt} + 1 \big]
\end{equation}
As in WILCO, the sigmoid function $\sigma(\cdot)$ represents the integration carried out by the ``soma" of the population. The argument of the sigmoid is therefore the total membrane perturbation caused by the PSPs from all synapses and other effects, such as electric fields.  The  $\gamma$s represent the coupling strength of the synapse --- the synaptic gain.

Because the synaptic dynamics are governed by second-order operators (with rise and decay impulse response), the neuronal circuit can sustain oscillations even without explicit self-coupling. Specifically, the second-order filter introduces a frequency-dependent phase shift in the system response. According to the Barkhausen stability criterion, sustained oscillations occur if the total loop gain equals unity and the total loop phase shift reaches an integer multiple of 360$^\circ$. In a push-pull arrangement—excitatory coupled to inhibitory populations and back—this intrinsic phase shift, arising purely from synaptic kinetics (distinct rise and decay constants), can fulfill the Barkhausen criterion. Thus, unlike the Wilson-Cowan case,  even in the absence of explicit self-feedback, second-order dynamics inherently provide the necessary conditions for sustained oscillations. We make this argument precise in the next paragraph.

\rev{\paragraph{Why second-order synapses sustain oscillations: the Barkhausen condition.}
The defining dynamical feature of NMM1 is its capacity to sustain oscillations \emph{without} explicit self-coupling, in contrast to WILCO. The mechanism is most cleanly understood through the classical Barkhausen criterion for closed-loop oscillation, which states that a feedback loop sustains a self-oscillation at angular frequency $\omega_0$ when (a) the loop gain $|G(\mathrm{i}\omega_0)|$ equals unity and (b) the total loop phase shift $\arg G(\mathrm{i}\omega_0)$ is an integer multiple of $2\pi$. In the linear regime around a fixed point, the NMM1 push--pull motif (E$\to$I$\to$E) realises a closed loop whose loop transfer function reads
\begin{equation}
G(\mathrm{i}\omega) \;=\; -\,w_{xy}\,w_{yx}\,\sigma_x'(v_x^*)\,\sigma_y'(v_y^*)\, L_x(\mathrm{i}\omega)\,L_y(\mathrm{i}\omega),
\label{eq:nmm1-loop-transfer}
\end{equation}
where $\sigma_\alpha'(v_\alpha^*)$ is the sigmoid slope at the operating point and $L_\alpha(\mathrm{i}\omega)$ is the frequency response of the second-order synaptic operator (cf.\ Eq.~\eqref{eq:L-operator-same-a}. Two features distinguish NMM1 from WILCO here. First, the second-order filter $L_\alpha$ contributes a frequency-dependent phase shift bounded by $\pi$, so the two synapses together can supply the full $2\pi$ required to close the loop, with the explicit minus sign of the inhibitory pathway providing the remaining $\pi$. Second, the rise and decay constants $\tau_{\alpha,r}, \tau_{\alpha,d}$ control \emph{both} the magnitude and the phase of $L_\alpha(\mathrm{i}\omega)$, so synaptic kinetics alone determine where the loop-phase condition is met --- and hence the resonance frequency. The loop-gain condition then selects the operating point at which oscillations onset through a Hopf bifurcation. The push--pull architecture is in this sense the \emph{minimal} architecture that satisfies the Barkhausen criterion in NMM1: a single population with second-order synapses and no self-coupling cannot supply both the gain inversion and the loop phase. The full algebraic derivation of \eqref{eq:nmm1-loop-transfer} and the explicit oscillation-onset condition are given in Appendix~\ref{app:barkhausen}.}

\subsection{Forcing and coupling}
Because of its realistic biological origins, accounting for the effects of coupling to other populations or of an electric field is straightforward: they produce additive voltage perturbation terms to the membrane potential, i.e., the argument of the corresponding sigmoid. So more generally, the equations with multiple synapses in a population reflect the additive combination of synaptic inputs. 

The generalization of Equation~\ref{eq:nmm1-simple} to \textit{multi-populations node}s and \textit{whole brain models }consists of one for each synapse $(m,n)$ and neuron $m$,
\begin{equation}
\begin{aligned}    \label{eq:multiNMM1}
&\hat L_{m\leftarrow n} \big[ u_{m\leftarrow n} \big] = C_{m\leftarrow n} \, r_n \\
&v_m = \Lambda[\hat E] + \sum_{ n:\,C_{m\leftarrow n} \neq 0 } u_{m\leftarrow n}  \\
&r_m = \sigma_m\big(v_m\big) 
\end{aligned}
\end{equation}
(forcing can be represented through synaptic coupling). As before, the first equation transduces connectivity-weighted firing rate inputs into PSPs using the $\hat L$ operator, the second sums all the PSPs and other perturbations affecting the neuron (e.g., an electric field in the equation), and the last one produces the firing rate output using the sigmoid.

The first equation links the input firing rate $r_n$ to its associated membrane perturbation (PSP). It can be read as the synapse equation for the input from neuron $n$ to neuron $m$. The input $r_n$  may also reflect an input to the model from some external neuron, in which case  $ r_n = f(t)$ for some function (constant, noise, etc.). 

The second equation evaluates the membrane potential $v_m$ of neuron $m$ as the sum of the synaptic voltage perturbations plus an electrical field $E$ perturbation (if present) \cite{ruffini_optimization_2014}. 

The last equation is a static transfer function: it evaluates the firing rate of a cell as a function of its total membrane perturbation.

The connectome in Equation~\ref{eq:multiNMM1} includes intra-parcel (defining, e.g., Jansen-Rit or LaNMM nodes) and inter-parcel connectivities in  whole-brain models. The latter are usually derived from diffusion MRI or from Ising modeling of fMRI \cite{mercadalBridgingLocalGlobal2025}.

 As a simple network example of inter-parcel coupling from excitatory to excitatory populations in the simple E-I motif model in Eq.~\ref{eq:nmm1-simple} (with all parcels equal), we have
 \begin{tcolorbox}[
  colback=gray!5,
  colframe=gray!5, 
  boxrule=0pt,
  enhanced, sharp corners,
  left=4pt, right=4pt, top=3pt, bottom=3pt,
  boxsep=2pt,
  before skip=6pt, after skip=6pt
]
\begin{equation} \label{eq:nmm1-coupled-example}
    \begin{aligned}
            L_x[x_i]
              &= \sigma_x\bigl(  - w_{xy}\,y_i + \hat F_{e;i}(t) + \sum_{j\neq i}  C_{ij} \, x_j(t-\tau_{ij}) \bigr),\\
           L_y[y_i] &= \sigma_y\bigl(w_{yx}\,x_i  \bigr) 
          \end{aligned}
\end{equation}
\end{tcolorbox}
with $\hat F_e(t)$ as before, including noise or deterministic forcing. 

As a further example, an electric field perturbation can be computed from the local electric field in the population. For example, in the case of weak electric fields at low frequencies (transcranial electrical stimulation), the perturbation is the dot product of the coupling constant $\vec \lambda$ and the electric field vector $\vec E$ \cite{ruffiniTranscranialCurrentBrain2013,Ruffini:2014aa,aberraBiophysicallyRealisticNeuron2018,galan-gadeaSphericalHarmonicsRepresentation2023}. Adding this perturbation to the simple example leads to (see Equation~\ref{eq:forcing}) 

\begin{equation} \label{eq:nmm1-coupled-example-withE}
    \begin{aligned}
            L_x[x_i]
              &= \sigma_x\bigl(  - w_{xy}\,y_i + f_i(t) +\vec \lambda_i \cdot \vec E_i(t)+ \hat \eta_i(t)+ \sum_{j \neq i}^N  C_{ij}\,  x_j(t-\tau_{ij})  \bigr),\\
           L_y[y_i] &= \sigma_y\bigl(w_{yx}\,x_i  \bigr)
    \end{aligned}
\end{equation}

\rev{As with the WILCO networks of \S\ref{sec:WILCO}, the rate-based formulation here admits explicit phase reductions back to generalised Kuramoto-type oscillator networks. Forrester et al.~\cite{forrester_2020} derive such a reduction for Jansen--Rit-type NMM1 networks and use it to show how local node dynamics shape the emergent functional connectivity patterns observed in whole-brain simulations, providing the explicit operational bridge between NMM1 and the phase-oscillator descriptions of \S\ref{sec:linear-phase}.}
\begin{tcolorbox}[
  colback=gray!5,
  colframe=slate,
  colbacktitle=softblue!25,
  coltitle=softblue!60!black,
  fonttitle=\bfseries\small,
  title={\small \textit{NMM1  Network (Eq.~\ref{eq:nmm1-coupled-example})} \\ Whole-brain Simulations Parameters \& Physiological Meaning},
  boxrule=0.4pt, enhanced, sharp corners,
  left=4pt,right=4pt,top=3pt,bottom=3pt,
  boxsep=2pt, before skip=6pt, after skip=6pt
]
\scriptsize
\begin{tabular}{@{}l p{0.8\linewidth}@{}}
\textbf{Node $i$} & Neural mass with explicit \emph{synapse} and \emph{soma}: PSP states drive a membrane perturbation $v_i$, which passes through a sigmoid to yield a firing rate $r_i$.\\
\textbf{$x_i(t),\,y_i(t)$} & Excitatory and inhibitory \emph{postsynaptic potentials} (PSPs). They are the outputs of second‑order synaptic filters (rise/decay), not directly the rates.\\
\textbf{$v_i(t)$} & Membrane potential perturbation (sum of PSPs and exogenous terms), i.e., the \emph{input} to the soma/nonlinearity.\\
\textbf{$r_{x,i}(t),\,r_{y,i}(t)$} & Excitatory/inhibitory \emph{firing rates} (soma outputs). These feed other synapses locally and across the network.\\
\textbf{$\sigma_x(\cdot),\,\sigma_y(\cdot)$} & Static sigmoids (e.g., logistic) mapping $v$ to firing rate; slope controls effective gain; saturation bounds activity.\\
\textbf{$L_x,\,L_y$} & Second‑order synaptic operators (cf.\ \eqref{eq:L-operator-same-a}): implement biophysical PSP kinetics with rise/decay; supply intrinsic phase lags that can sustain oscillations.\\
\textbf{$\tau_{\alpha}$, $(\tau_{\alpha,r},\tau_{\alpha,d})$} & Synaptic time constants (single or separate rise/decay); set resonance frequency and phase lag of each synapse.\\
\textbf{$\gamma_{\alpha}$} & Synaptic gains (PSP amplitude scale).\\
\textbf{$w_{yx},\,w_{xy}$} & Local E$\!\to$I and I$\!\to$E coupling (push–pull loop); self‑coupling often omitted in NMM1 because synaptic phase lags can close the Barkhausen loop.\\
\textbf{$\hat F_{e;i}(t)$}& Exogenous drive  (external forcing from nodes or elements outside the network or an electric field)--- see Equation~\ref{eq:forcing}).\\
\textbf{$C_{ij}$} & Long‑range connectivity from node $j$ to $i$ (SC default; FC/EC or synthetic graphs if needed); typically targets the excitatory pathway.\\
\textbf{$N$} & Number of nodes (parcels).\\
\end{tabular}
\end{tcolorbox}

\begin{tcolorbox}[
  colback=gray!5,
  colframe=slate,
  colbacktitle=softblue!25,
  coltitle=softblue!60!black,
  fonttitle=\bfseries\small,
  title={\small When to Use It},
  boxrule=0.4pt, enhanced, sharp corners,
  left=4pt,right=4pt,top=3pt,bottom=3pt,
  boxsep=2pt, before skip=6pt, after skip=6pt
]
\scriptsize
\textbf{Use this when} explicit synaptic kinetics and their phase lags matter (evoked responses, resonance, photic entrainment, band‑limited power and envelope dynamics) or when linking parameters to PSP amplitudes/time‑constants is essential.\\
\textbf{Assumptions} linear synaptic filters (second order) feeding a static nonlinearity; E/I pathways combined at the soma; long‑range excitation enters via excitatory synapses.\\
\textbf{Best for} EEG/MEG/fMRI generative modeling (spectra and event-related potentials (ERPs)), seizure phenomenology (fast/slow inhibition variants), and whole‑brain simulations where synaptic time constants set rhythms.\\
\textbf{Relations} near a stable focus it reduces to linear resonators; near Hopf it displays SL‑like amplitude–phase dynamics but with biophysical PSP knobs (gains/time‑constants).\\
\textbf{Avoid} if only relative phase is of interest (Kuramoto) or if closed‑form linear statistics suffice (damped linear network).
\end{tcolorbox}

\begin{tcolorbox}[
  colback=gray!5,
  colframe=slate,
  colbacktitle=softblue!25,
  coltitle=softblue!60!black,
  fonttitle=\bfseries\small,
  title={\small How to Use It},
  boxrule=0.4pt, enhanced, sharp corners,
  left=4pt,right=4pt,top=3pt,bottom=3pt,
  boxsep=2pt, before skip=6pt, after skip=6pt
]
\scriptsize
\textbf{Model (drop‑in)} use the operator form \eqref{eq:multiNMM1} or the E–I pair with coupling \eqref{eq:nmm1-coupled-example}–\eqref{eq:nmm1-coupled-example-withE}.\\
\textbf{Provide} synaptic operators $L_{\alpha}$ (choose $\tau_{\alpha}$ or $(\tau_{\alpha,r},\tau_{\alpha,d})$) and gains $\gamma_{\alpha}$; local couplings $w_{yx},w_{xy}$; soma sigmoids $\sigma_{\alpha}$ (gain, midpoint, max rate); drives $F_i(t)$ (and optionally $\vec\lambda\!\cdot\!\vec E_i$).\\
\textbf{Network} supply $C$ (SC default; FC/EC or synthetic if SC is absent), optional delays $\tau_{ij}$; long‑range input enters the excitatory pathway.\\
\textbf{Defaults} normalize $C\!\leftarrow\!C/\lambda_{\max}(C)$; start with standard JR‑style kinetics (faster E than I or vice‑versa depending on band); small noise; Euler–Maruyama/RK with $\Delta t\!\le\!1/(200\,f_{\max})$. \rev{With non-zero delays $\tau_{ij}>0$ and stochastic forcing, the system is an SDDE; prefer method-of-steps DDE solvers in the deterministic limit, otherwise use Euler--Maruyama on the delayed system as a controlled approximation with the dual bound $\Delta t \le \min\{1/(200 f_{\max}),\,\min_{ij}\tau_{ij}/10\}$.}\\
\textbf{Readouts} PSPs $(x_i,y_i)$, membrane $v_i$, rates $r_i$; PSD/cross‑spectra and ERPs; envelope/FC/FCD; assess resonance by scanning $\tau$’s and gains.
\end{tcolorbox}

\subsection{Applications} 
\begin{figure}[t!]
    \centering
    \includegraphics[width=1.0\linewidth]{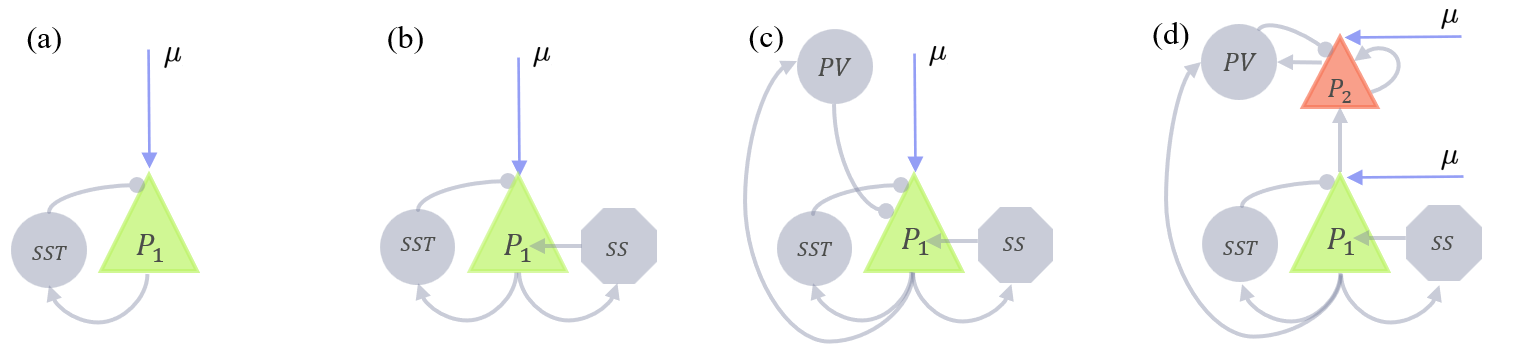}
 
        \caption{Four models using the second-order formalism: 
    a) PING-like push–pull motif, 
    b) Jansen–Rit model \cite{jansen_neurophysiologically-based_1993,grimbert_analysis_2006}, 
    c) Wendling model \cite{wendling_computational_2016}, 
    d) Laminar model \cite{ruffiniP118BiophysicallyRealistic2020,sanchez-todoPhysicalNeuralMass2023}. 
    Synapses are shown as arrowheads (excitatory) or buttons (inhibitory). 
    Noise or external inputs are indicated on pyramidal cells, although other targets ($\mu$) are also possible.}
    \label{fig:fourmodels}
\end{figure}

We review canonical models using the second-order synapse formalism
(see Figure~\ref{fig:fourmodels}) and summarize their characteristic features and uses.

\subsubsection*{Simple push–pull motif (PING-like) model}
The simplest second-order neural mass captures the push–pull loop between an excitatory and an inhibitory population, the canonical PING motif. Second-order synapses (distinct rise and decay) provide the phase lag needed to meet Barkhausen’s condition for sustained oscillations without explicit self-coupling. Minimal two-population models reproduce gamma-band rhythms and noise-sustained oscillations near Hopf; frequency depends mainly on inhibitory decay and E$\rightarrow$I gain and can be shifted by drive or kinetics. Applications include modeling high-frequency oscillations (HFOs) at seizure onset \cite{molaee-ardekani_computational_2010}; mechanistic context from spiking/mean-field work on PING/ING is reviewed in {Buzsáki \& Wang (2012)}\cite{BuzsakiWang2012} and  {Tiesinga \& Sejnowski (2009)}\cite{TiesingaSejnowski2009}, and synchronization analyses such as in {Börgers \& Kopell (2003)}\cite{BorgersKopell2003} and  {Whittington et~al. (2000)}\cite{Whittington2000}.

\subsubsection*{The Jansen–Rit model}
The Jansen–Rit (JR) model \cite{jansen1995} comprises three populations (pyramidal, excitatory interneurons, inhibitory interneurons) interconnected with second-order synapses and a static transfer function. It generates alpha-band activity and realistic evoked responses; its regimes (fixed point, alpha, spike-like) are organized by Hopf and other bifurcations \cite{GrimbertFaugeras2006,Spiegler2010}. JR serves as the neuronal model in Dynamic Causal Modeling (DCM) for M/EEG steady-state and evoked responses \cite{DavidFriston2003,Moran2009,Moran2013Review}, linking synaptic gains/time constants to observed spectra and ERPs.

\begin{figure}[h!]
    \centering
    \includegraphics[width=0.6\linewidth]{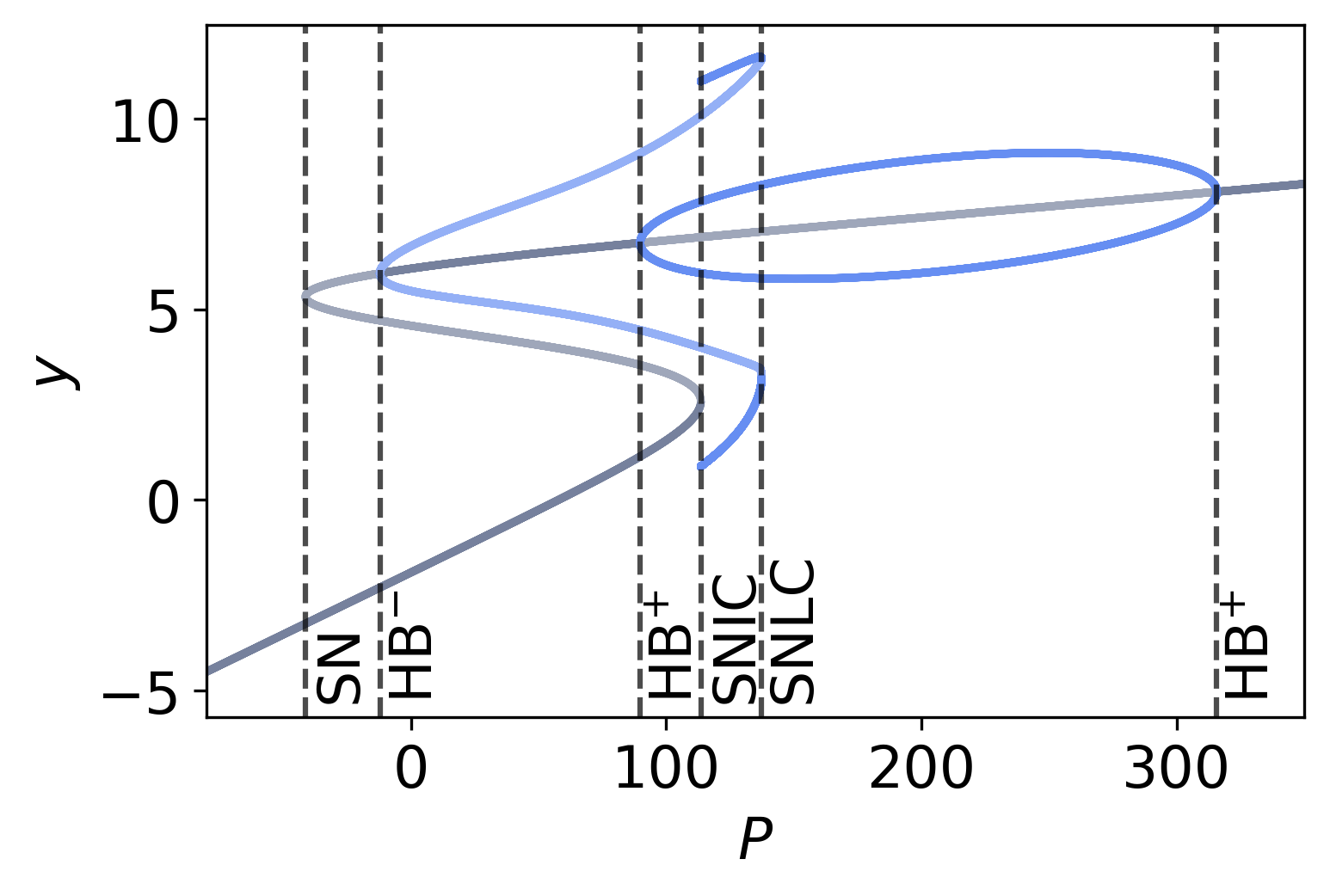}
    \caption{\rev{Bifurcation diagram of the Jansen-Rit model (Eq. \eqref{eq:nmm1-JR}) showing $y = y_1 - y_2$, with bifurcation parameter $P$. Dark (light) grey lines represent the stable (unstable) fixed points, while the dark (light) blue curves indicate the maximum and minimum amplitudes of the stable (unstable) limit cycles. See Table~\ref{tab:bif-params} for parameters.}}
    \label{fig:JR_bif}
\end{figure}

\rev{The model dynamics can be written as
\begin{equation}\label{eq:nmm1-JR}
\begin{aligned}
\hat{L}[y_0] &= A a\, \sigma(y_1 - y_2), \\
\hat{L}[y_1] &= A a\, \bigl(P + C_2\, \sigma(C_1 y_0)\bigr),\\
\hat{L}[y_2] &= B b\, C_4\, \sigma(C_3 y_0).
\end{aligned}
\end{equation}
The bifurcation diagram of the Jansen-Rit model is shown in Fig.~\ref{fig:JR_bif}. Going from left to right, an SN bifurcation creates two unstable fixed points. As the external input $P$ is increased, the system exhibits a subcritical Hopf bifurcation (HB$^-$); beyond this point, the stability of the fixed point shifts and an unstable limit cycle (depicted in light blue) is created, and the system exhibits two stable fixed points after the HB$^-$ bifurcation. This bistability persists as the system undergoes another Hopf bifurcation, this time a supercritical one (HB$^{+}$), where the upper fixed point loses stability and a stable limit cycle emerges. In this regime, the system can display either oscillatory or stationary behavior, depending on its initial conditions. Further increasing $P$ causes the two lower fixed points to collide in a saddle-node on invariant circle (SNIC) bifurcation. Beyond this point, the system exhibits two stable limit cycles, corresponding to oscillations with different frequencies. Eventually, a saddle-node of limit cycles (SNLC) bifurcation occurs, where an outer stable and an unstable limit cycle collide and annihilate, leaving a single remaining limit cycle. This final oscillatory state persists until a last supercritical Hopf bifurcation (HB$^{+}$), after which the system returns to a stationary regime.}

\subsubsection*{The Wendling model and its extensions}
Wendling’s CA1-inspired extension adds fast and slow inhibitory subpopulations (GABA$_A$, GABA$_B$) to the JR scaffold. By tuning inhibitory gains and kinetics, it reproduces background alpha, interictal spikes/spike–waves, and low-voltage fast activity; seizure onset emerges with impaired dendritic inhibition \cite{Wendling2002}. The model is widely used for interpreting stereo-electroencephalography (SEEG) / EEG patterns, exploring ictogenesis mechanisms, and assessing interventions \cite{Wendling2016Review}, making it a standard computational tool in epilepsy. Recent extensions include chloride dynamics and laminar integration \cite{lopez-solaPersonalizableAutonomousNeural2022}.
\textit{Whole-brain use:} Wendling nodes have been used for resting‑state whole‑brain network modeling that matches empirical FC \cite{Cui2024WendlingWBNM} and for patient‑specific whole‑brain simulations of interictal SEEG to aid clinical interpretation \cite{KoksalErsuz2024SEEGWB}.
They have been embedded in realistic forward models to synthesize SEEG and personalize local epileptogenic dynamics \cite{lopez-solaPersonalizableAutonomousNeural2022}; 
the same group reports personalised whole-brain seizure-propagation models that integrate SEEG, MRI, and dMRI to simulate patient-specific responses to surgical, stimulation, and pharmacological interventions within a unified physiological framework~\cite{LopezSola2025WB}.

\begin{figure}[h!]
    \centering
    \includegraphics[width=0.65\linewidth]{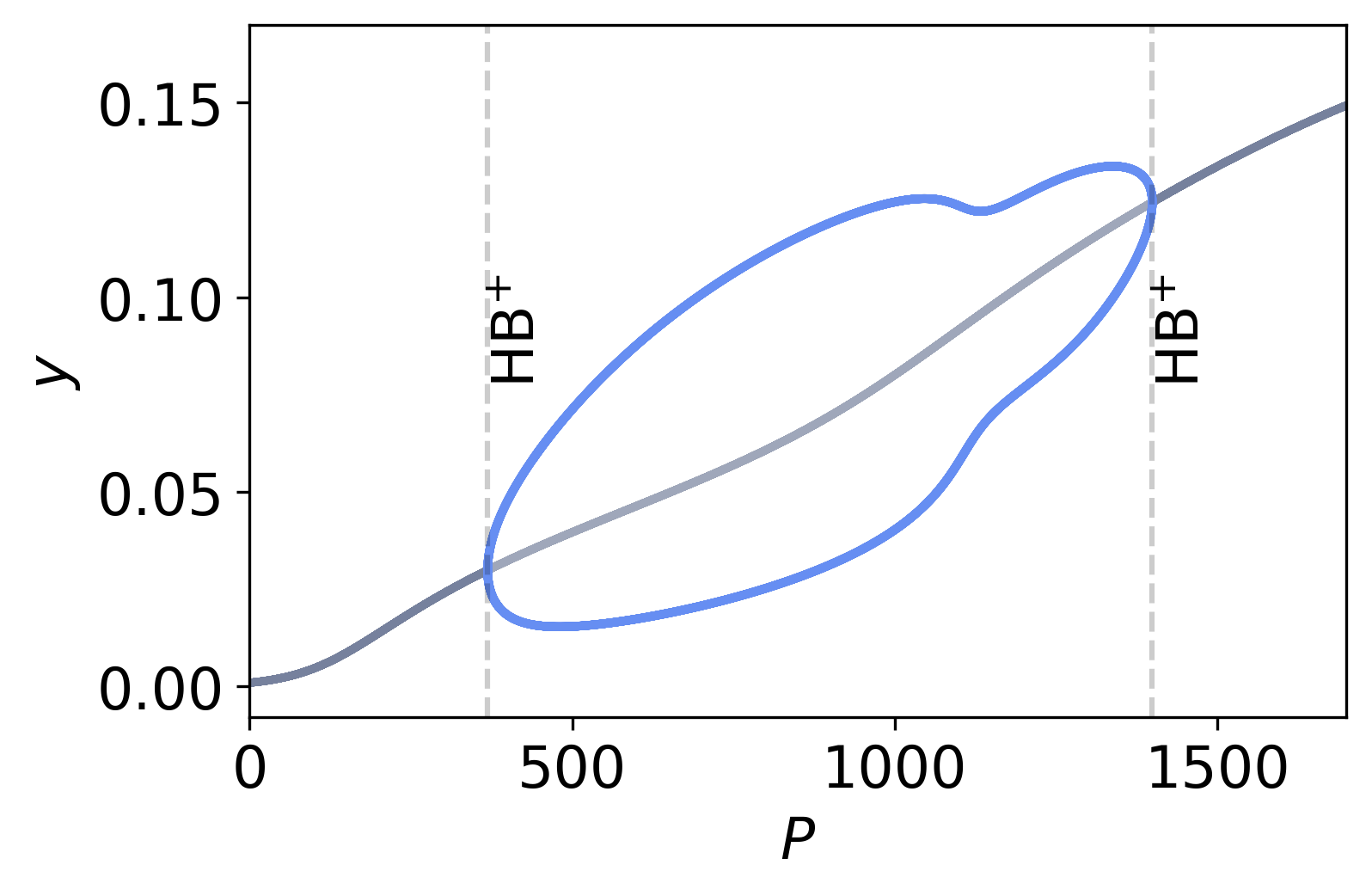}
    \caption{\rev{Bifurcation diagram of the Wendling model (Eq. \eqref{eq:nmm1-Wen}) showing $y = y_0$, with bifurcation parameter $P$. Dark (light) grey lines represent the stable (unstable) fixed points, while the dark (light) blue curves indicate the maximum and minimum amplitudes of the stable (unstable) limit cycles. See Table~\ref{tab:bif-params} for details.}}
    \label{fig:Wend_bif}
\end{figure}

\rev{As in the Jansen–Rit model, the core architecture is built around a pyramidal cell population interacting with both SS and SST neurons. The Wendling model extends this structure by introducing a parvalbumin-positive (PV) interneuron population, which both receives input from and projects back to the pyramidal cells. Incorporating this additional inhibitory loop results in the following system of equations:
\begin{equation}\label{eq:nmm1-Wen}
\begin{aligned}
\hat{L}[y_0] &= A b \, \sigma(y_1 - y_2 - y_3), \\
\hat{L}[y_1] &= A b \left(P - C_2 \, \sigma(C_1 y_0) \right),\\
\hat{L}[y_2] &= B b \, C_4 \, \sigma(C_3 y_0),  \\
\hat{L}[y_3] &= D d \, C_7 \, \sigma(C_5 y_0 - C_6 y_2).
\end{aligned}
\end{equation}
The bifurcation diagram in Fig.~\ref{fig:Wend_bif} illustrates the steady-state membrane potential of the pyramidal population as a function of its external input $P$. As $P$ increases, the system undergoes a supercritical Hopf bifurcation (HB$^{+}$) at approximately $P \approx 600$, giving rise to oscillations that persist until $P \approx 1750$.}

\subsubsection*{The laminar model (LaNMM)}
\rev{The laminar neural mass model (LaNMM)~\cite{Ruffini2020P118,SanchezTodo2023} extends the NMM1 framework to capture the layered organisation of cortical microcircuits, addressing observables that single-population JR or Wendling models cannot reach. LaNMM couples two NMM1 generators with distinct anatomical and dynamical roles. A \emph{deep} Jansen--Rit-like generator, situated in infragranular layers, produces alpha/theta-band rhythms through the standard three-population JR architecture. A \emph{superficial} PING-like generator, situated in supragranular layers, produces gamma-band rhythms through the second-order push--pull motif of \S\ref{sec:NMM1}. The two generators are coupled according to cortical laminar anatomy, with the deep generator modulating the superficial one through ascending projections that gate gamma amplitude on the slow phase, producing the empirically observed cross-frequency phase--amplitude coupling characteristic of laminar recordings.
Because the model is anchored to specific cortical layers, the synthetic local field potential (LFP) and current source density (CSD) signals it generates inherit realistic depth-dependent polarity and phase relations through volume-conduction physics, allowing direct comparison with depth-electrode recordings rather than only with surface EEG/MEG. 

\begin{figure}[t]
    \centering
    \includegraphics[width=0.65\linewidth]{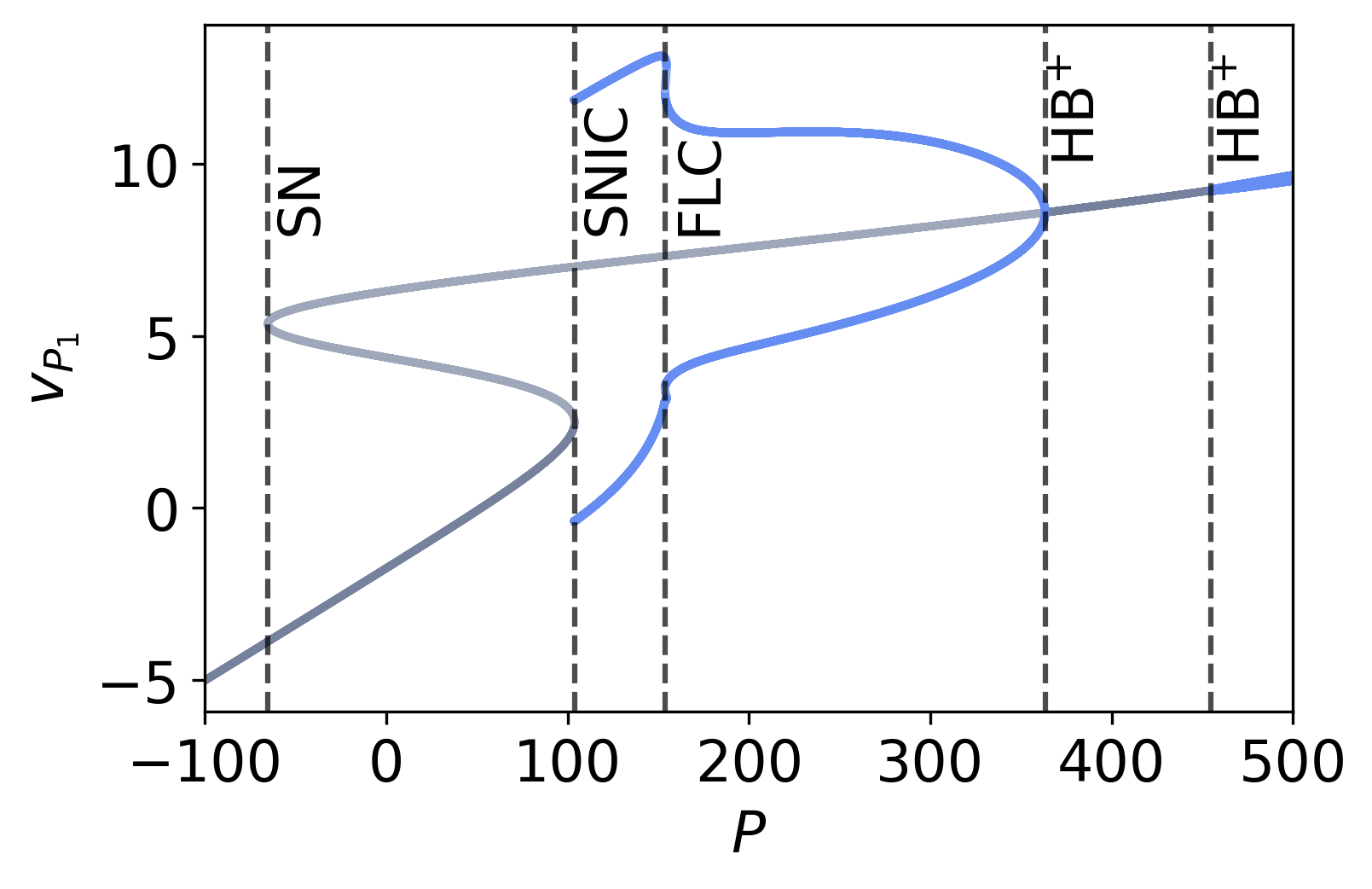}
    \caption{\rev{Bifurcation diagram of the LaNMM model (Eq. \eqref{eq:nmm1-lanmm}), showing $v_{P_1}$ with bifurcation parameter $P$. Dark (light) grey lines represent the stable (unstable) fixed points, while the dark (light) blue curves indicate the maximum and minimum amplitudes of the stable (unstable) limit cycles. See Table~\ref{tab:bif-params} for details.}}
    \label{fig:LaNMM_bif}
\end{figure}

Based on the connectivity scheme shown in Fig~\ref{fig:fourmodels} (d), the model is governed by the following system of equations}
\rev{\begin{align}
\hat{L} [y_1(t)] &= A_A a_A \sigma_{P_1}(C_1y_{2} + C_2y_{3} + C_{11}y_{4} + P)\\
\hat{L} [y_2(t)] &= A_A a_A \sigma_{SST}(C_4y_{1})\\
\hat{L} [y_3(t)] &= A_{G_s} a_{G_s} \sigma_{P_1}(C_5y_{1})\\
\hat{L} [y_4(t)] &= A_{A} a_A \sigma_{P_2}(C_6y_{4} +  C_7y_{5} +  C_{12}y_{1})\\
\hat{L} [y_5(t)] &= A_{G_f} a_{G_f} \sigma_{P_2}(C_9y_{4} +  C_{10}y_{5} +  C_{13}y_{1} )
\label{eq:nmm1-lanmm}
\end{align}}
\rev{
The bifurcation diagram of the LaNMM is shown in Fig.~\ref{fig:LaNMM_bif}. It depicts the steady-state membrane potential of the bottom pyramidal population ($v_{P1} = C_1y_{2} + C_2y_{3} + C_{11}y_{4} + P$) as a function of its external input ($P$). The overall structure of the diagram closely resembles that of the Jansen--Rit (JR) model, as expected, since the LaNMM combines features of both the JR and PING models. As the external input increases, the system undergoes a saddle-node (SN) bifurcation where two unstable fixed points emerge. However, unlike the JR model, the LaNMM does not exhibit bistability. Oscillations then arise through a saddle-node on invariant circle (SNIC) bifurcation. Further increases in $P$ lead to a pair of fold-of-limit-cycle (FLC) bifurcations, where the oscillation amplitude decreases while the frequency increases (see~\cite{depalmaaristidesEmergenceMultifrequencyActivity2025} for details). This behavior continues until the system passes through a Hopf bifurcation (HB$^{+}$). Beyond this point, and in contrast to the JR model, the system undergoes an additional HB$^{+}$ bifurcation, associated with the PING mechanism, giving rise to faster oscillatory activity. A systematic bifurcation analysis of the LaNMM identifies the parameter regimes that sustain coupled multifrequency oscillations when an external input is also introduced to the top pyramidal population, in addition to the previously considered input to the bottom population, and determines the conditions under which AD-like alterations disrupt these dynamics~\cite{depalmaaristidesEmergenceMultifrequencyActivity2025}.}

\rev{
Applications span (i) reproducing laminar spectral peaks and CSD sinks/sources observed in primate and rodent cortex ~\cite{mendoza-hallidayUbiquitousSpectrolaminarMotif2024,buffaloLaminarDifferencesGamma2011,spaakLayerspecificEntrainmentGammaband2012,sotero_laminar_2015}; (ii) explaining the cross-frequency coupling signatures observed in laminar recordings; (iii) capturing the alpha-slowing, gamma-loss, and disrupted theta--gamma coupling that characterise the oscillatory phenotype of Alzheimer's disease~\cite{sanchez-todoFastInterneuronDysfunction2025}, and the differential effects of psychedelic compounds in AD~\cite{gendraRestoringOscillatoryDynamics2024}; (iv) integration with stimulation physics for transcranial electrical stimulation (tES/tACS) studies through coupling to the local electric field as in Eq.~\eqref{eq:nmm1-coupled-example-withE}; and (v) deployment as a regional model in connectome-scale simulations to study gamma-band coordination and cooperative/competitive network rhythms~\cite{Mercadal2025PINGWB}. LaNMM has also been used as a generative model for predictive-coding interpretations of laminar dynamics~\cite{ruffiniCrossFrequencyCouplingNeural2025}. 

In the disease-modeling space, modulation of fast-interneuron coupling within LaNMM reproduces the biphasic oscillatory progression of Alzheimer's disease --- early hyperexcitability with elevated alpha and gamma power, followed by oscillatory slowing and reduced spectral power consistent with empirical biomarkers~\cite{sanchez-todoFastInterneuronDysfunction2025} --- and the corresponding modulation of layer-5 pyramidal excitability characteristic of 5-HT$_{2A}$-receptor activation reproduces the spectral signatures of serotonergic psychedelics, suggesting a candidate therapeutic axis for the prodromal and early phases of Alzheimer's disease~\cite{gendraRestoringOscillatoryDynamics2024}. 
}


\definecolor{coral}{HTML}{F57558}
\definecolor{softblue}{HTML}{668EF2}
\definecolor{slate}{HTML}{76819D}

\section{Next-generation models (NMM2)} \label{sec:NMM2}
\begin{tcolorbox}[colback=gray!6,colframe=gray!6,enlarge left by=0mm, boxrule=0pt, enhanced, sharp corners]
The neural mass models we have discussed so far, such as the Jansen-Rit, Wendling systems or LaNMM, provide a practical framework for representing and interpreting electrophysiological activity in both local and global brain models \cite{wilson_excitatory_1972,lopes_da_silva_model_1974,lopes_da_silva_model_1976,jansen_neurophysiologically-based_1993,jansen_electroencephalogram_1995,wendling_epileptic_2002,Ruffini:2018aa,sanchez-todo_personalization_2018,Ruffini2020P118,sanchez-todoTH230MechanisticUnderstanding2022,sanchez-todoFastInterneuronDysfunction2025,cabralMetastableOscillatoryModes2022,deco_awakening_2019}.  

 \rev{However, NMM1 mixes biophysically grounded and phenomenological elements: the synaptic operator $\hat L$ has direct biophysical content, since its impulse response is set by the kinetics of receptor binding and channel opening and can be matched to recorded post-synaptic potentials~\cite{Destexhe:1998aa,Pods:2013aa,Ermentrout:2010aa}, but the static wave-to-pulse sigmoid function~\cite{Freeman:1975aa,Kay:2018aa,Eeckman:1991aa} that maps membrane potential to firing rate rests on a weaker theoretical foundation, having been postulated phenomenologically by Freeman from population recordings rather than derived from spiking dynamics.}
 
Recently, Montbri\'o et al \cite{Montbrio:2015aa} derived an exact mean-field theory (MPR)  for a population of quadratic integrate-and-fire neurons under some simplifying assumptions, thereby connecting microscale neural mechanisms and meso/macroscopic phenomena. 
 The MPR model can be seen to replace Freeman's sigmoid function with a pair of differential equations for the mean membrane potential and firing rate variables---a dynamical relation between firing rate and membrane potential---, providing a more fundamental interpretation of the semi-empirical NMM sigmoid parameters. In doing so,  it sheds light on the mechanisms behind enhanced network response to weak but uniform perturbations. 
 
 In the exact mean-field theory,  intrinsic population connectivity modulates the steady-state firing rate sigmoid relation in a monotonic manner, with increasing excitatory self-connectivity leading to higher firing rates.  This provides a plausible mechanism for the enhanced response of densely connected networks to weak, uniform inputs such as the electric fields produced by non-invasive brain stimulation. This new, ``dynamic sigmoid"  also endows the neural mass model with a form of ``inertia'', an intrinsic delay to external inputs that depends on, e.g.,  self-coupling strength and state of the system.  
 
Models resulting from the MPR mean-field theory can be completed by adding the first or second-order equations for delayed post-synaptic currents and the coupling term with an external electric field \cite{Devalle:2017aa,ruffiniAnalysisExtensionExact2022,clusellaComparisonExactHeuristic2023}, bringing together the MPR and the usual NMM formalisms into a unified exact mean-field theory (NMM2, for short) displaying rich dynamical features.  In the single population model, we show that the resonant sensitivity to a weak alternating electric field is enhanced by increased self-connectivity and slow synapses.

\end{tcolorbox}

Classical NMMs are \emph{coarse‑grained} descriptions: they compress the high‑dimensional, spike‑resolved dynamics of local circuits into a few mesoscale order parameters (typically, population firing rate and a filtered postsynaptic potential). Conceptually, this follows Kadanoff–Wilson coarse‑graining: integrate out fast, microscopic degrees of freedom, retain slow collective variables, and allow parameters (gains, time constants, noise) to be \emph{renormalized} by the elimination of small scales \cite{kadanoffScalingLawsIsing1966,wilsonRenormalizationGroupCritical1975}. Empirically, data‑driven coarse‑graining of population activity can approach non‑Gaussian fixed forms with static and dynamic scaling—evidence that mesoscale statistics can be approximately scale‑invariant near special operating points \cite{meshulamCoarseGrainingFixed2019}.

\rev{Early NMMs (e.g., Wilson--Cowan, Jansen--Rit) thus combine a biophysically grounded synaptic filter with a phenomenological static nonlinearity --- the same split made explicit in the previous paragraph.}
Population‑density and mean‑field limits provide a more principled route from spiking to masses \cite{nykampPopulationDensityApproach2000,omurtagDynamicsNeuronalPopulations2000}, and field‑theoretic expansions make explicit how fluctuations and finite‑size effects correct mean‑field behavior—precisely the kinds of corrections induced by coarse‑graining \cite{buiceSystematicFluctuationExpansion2010}. In large‑scale modeling, these ideas motivate dynamic mean‑field reductions of biophysical networks into mesoscale nodes with a few state variables \cite{breakspearDynamicModelsLargescale2017}.

\rev{The mathematical foundation for these exact reductions was laid by the analysis of theta-neuron networks via the Ott--Antonsen ansatz: Luke, Barreto \& So~\cite{luke2013complete} provided the complete classification of macroscopic behaviour for heterogeneous theta-neuron networks, and So, Luke \& Barreto~\cite{so2014networks} extended this analysis to networks with non-trivial topologies, in both cases obtaining low-dimensional ODEs for collective phase coherence. Because the QIF and theta-neuron models are equivalent under the standard quadratic-to-theta change of variables $V = \tan(\theta/2)$, these reductions and the parallel reduction of QIF networks via the same Ott--Antonsen manifold capture the same macroscopic dynamics expressed in different coordinate systems. Montbri\'o, Paz\'o \& Roxin~\cite{montbrioMacroscopicDescriptionNetworks2015} (henceforth MPR) reformulated the reduction directly in physiological coordinates --- population firing rate $r(t)$ and mean membrane potential $v(t)$ --- yielding the two macroscopic ODEs that anchor the present section. The QIF coordinate system is convenient because it expresses the macroscopic dynamics directly in terms of measurable observables ($r$, $v$); the theta coordinate system is convenient for the underlying Ott--Antonsen analysis. NMM2 inherits results from both lines of work; see also subsequent reviews and extensions~\cite{coombesNextGenerationNeural2019,byrneNextgenerationNeuralMass2020,coombesNextGenerationNeural2023,ruffiniAnalysisExtensionExact2022,clusellaComparisonExactHeuristic2023,bickUnderstandingDynamicsBiological2020}.}
In the MPR framework, the steady‐state relationship between firing rate \(r\) and mean voltage \(v\) defines a \emph{sigmoid‐type} input–output curve (i.e., a static transfer function).  
However, when the full two‐dimensional dynamical system is considered (with \(r(t)\) and \(v(t)\) evolving in time), the effective gain becomes \emph{state‐ and history‐dependent}, rather than being a fixed static curve.

Positive self‑coupling ($J$) shifts fixed points to higher rates and can bring the system closer to resonant/oscillatory regimes, aligning with the intuition that denser local recurrence enhances responsiveness to weak, spatially uniform drives (e.g., uniform electric fields) \cite{montbrioMacroscopicDescriptionNetworks2015,Byrne2020}. Exact mean‑field extensions capture synaptic filtering, electrical coupling, and other biophysics \cite{Devalle:2017aa,ruffiniAnalysisExtensionExact2022,clusellaComparisonExactHeuristic2023}, and have been leveraged to model working‑memory circuits with short‑term plasticity \cite{taherExactNeuralMass2020}.

NMM2 can be read as a principled \emph{coarse‑grained} synthesis: it replaces Freeman’s static nonlinearity by the MPR \emph{dynamic} firing‑rate relation, and \emph{completes} it with biophysical synaptic filters (e.g., second‑order $\alpha$‑synapses) and exogenous field coupling. In renormalization group (RG), $r$ and $v$ are the relevant mesoscale variables; synaptic and coupling parameters are effective (\emph{renormalized}) couplings that depend on the level of coarse‑graining and circuit state. This yields (i) a physics‑based “dynamic sigmoid” with inertia and state‑dependent gain, (ii) a transparent link between microparameters $(\eta,\Delta,J)$ and mesoscale responsiveness, and (iii) a natural path to incorporate fluctuation corrections when needed (finite‑size, correlations) \cite{buiceSystematicFluctuationExpansion2010}.  

In their uniform,  mean-field derivation for a population of $N$  quadratic integrate and fire (QIF) neurons, Montbri\'o et al \cite{Montbrio:2015aa}  start from the equations  for the neuron membrane potential perturbation from baseline,
\begin{equation} \label{eq:qif}
\dot V_j = V_j^2 + \eta_j + J s(t) + I(t), \mbox{ if } V_j \ge V_p, \mbox{ then } V_r \leftarrow V_j
\end{equation}
In this equation, the total input current in neuron $j$ is  $I_j =  \eta_j + J s(t) + I(t)$ and includes a quenched noise constant component $\eta_j$ drawn from a Lorentzian (Cauchy) distribution, the input from other neurons $s(t)$ per connection received (the mean synaptic activation) with uniform coupling $J$, and a common input $I(t)$. The common input $I(t)$ can  represent both a common external input or the effect of an electric field, e.g.,
\begin{equation}
 I(t) =  p(t) + \vec \lambda \cdot \vec E(t)   
\end{equation}
for weak electric fields.
Here $p(t)$ is an external uniform current, and $\lambda$ is the dipole conductance term in the spherical harmonic expansion of the response of the neuron to an external, uniform electric field.   This is a good approximation if the neuron is in its subthreshold, linear regime and can be computed using realistic compartment models of the  (see, e.g., \cite{aberra_biophysically_2018} and \cite{Galan:2021aa}). 

The mean synaptic activation is given by
\begin{equation}
    s(t) = \frac{1}{N}\sum_{j=1}^N \sum_{k | t^k_j < t} \int_{-\infty}^t dt' \, a_\tau(t-t') \, \delta(t'-t^k_j)
\end{equation}
where  $t^k_j$ is  the arrival time of the $k$th spike from the $j$th neuron, and  $a(t)$ the synaptic activation function, e.g., $a(t) = e^{-t/\tau}/\tau$.  Note that we can write $J=j\cdot N$, where $j$ is the synapse coupling strength (charge delivered to the neuron per action potential at the synapse) of each synapse the cell receives from the network (there are $N$ of them in a fully connected architecture with $N$ neurons). 

We assume here, for simplicity, that all neurons are equally oriented with respect to the electric field. If the electric field is constant, variations in orientation can be absorbed by the quenched noise term.
The total input $p(t) + \vec \lambda \cdot \vec E(t) + J \,s(t)$ is thus homogeneous across the population (does not depend on the neuron).  

Starting from these, Montbri\'o et al 
 derive an effective   theory \textit{for a single population} in the large $N$ limit (Eq 12 in \cite{Montbrio:2015aa}),
\begin{align} \label{eq:montbrio}
\dot r &=\Delta /\pi + 2r v \\
\dot v &=v^2  - \pi^2 r^2 +  J \, s +\bar \eta + I(t) 
\end{align}
To this equation, we add the usual operator dynamics for the synapse activation,
\begin{equation}
    \hat L_s[s] = r, \text{ or, equivalently,  } s= \hat K_s [r]
\end{equation}
(see Figure~\ref{fig:fourmodels} (A)). 
Here $v$ and $r$ are the population mean membrane potential and firing rate, respectively. The new parameters $\bar \eta$ and $\Delta$ refer to the \rev{centre and half-width-at-half-maximum (HWHM)} of the Lorentzian \rev{(Cauchy) }distribution for the quenched noise input $\eta_j$. \rev{Note that the Lorentzian distribution does not admit moments of order $\geq 1$ in the conventional sense (the integral defining the mean is divergent), so $\bar\eta$ is a location parameter rather than an expectation; under the Cauchy principal value, $\bar\eta$ formally equals the median.}
The analysis in \cite{Montbrio:2015aa} hinges on the assumptions of all-to-all uniform connectivity (with synaptic weight $J$) and common input $I(t)$.

These equations can be read as a transfer functional mapping input currents into an output firing rate,  as in the master NMM Eq.~\ref{eq:master_nmm},
\begin{equation}
\label{eq:master_nmm2}
\begin{aligned}
r &=  \hat \Phi_m [J\, s +  I(t)] \;,\\
 s &= \hat K_s [r].
\end{aligned}
\end{equation}
The transfer functional is captured by Eq.~\ref{eq:montbrio}.  For a self-coupled inhibitory population (ING) with second-order synapses, the system of equations reads
\begin{equation}
\begin{aligned}
\tau_m \dot{v} &= \bar\eta - \left( \pi r \tau_m \right)^2 + v^2 - \tau_m \left( J_{ii} s \right) + I \\
\tau_m \dot{r} &= \Delta + 2 r v \\
\tau_s \dot{s} &= z \\
\tau_s \dot{z} &= r - 2z - s
\end{aligned}
\label{eq:nmm2-ING}
\end{equation}
where $\tau_s$ is the synaptic time constant, and where we explicitly introduce $\tau_m$, the membrane constant of the population~\cite{clusellaComparisonExactHeuristic2023}. 
  \begin{figure}[t!]
    \centering
    \includegraphics[width=1.0\linewidth]{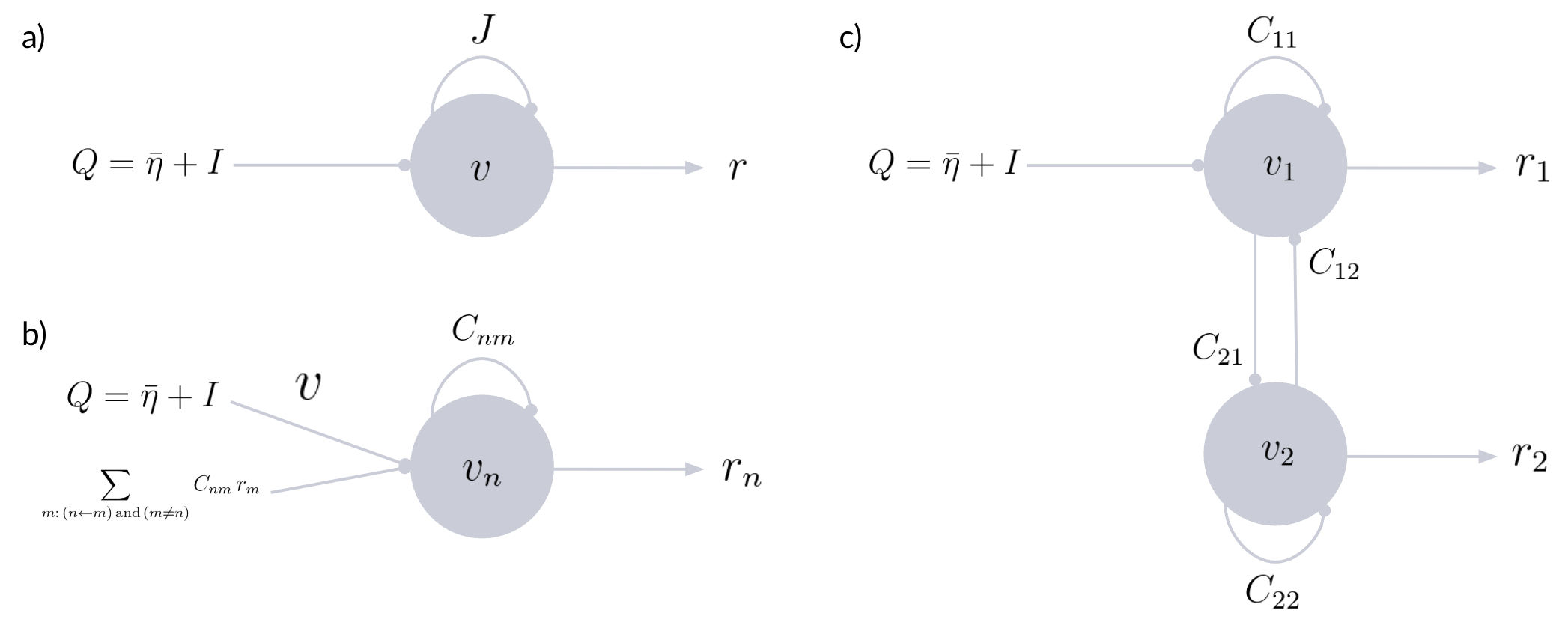}
    \caption{{\bf NMM2 diagrams}. (A): Diagram for self-coupled population with connectivity $J$ receiving and external input $Q$. (B): generalization for multiple populations. (C): A generic two-population model.}
    \label{fig:popdiagrams}
\end{figure}
 

\subsection{Forcing and coupling}
Forcing and coupling are natural in this model, inheriting naturally from the single neuron biophysics in Equation~\ref{eq:qif} through the term $I(t)$. We converge here on the notation used in Equation~\ref{eq:forcing}, where $\hat F_e$ represents the input external to the network.

\paragraph{E-I push-pull motif.}  For two coupled populations (see Figure~\ref{fig:popdiagrams}), we have the six equations, in dimensional (membrane-time) form
\rev{\begin{equation}
\begin{aligned}
\tau_x \dot r_x &= \frac{\Delta_x}{\pi\,\tau_x} + 2 r_x v_x, \\
\tau_x \dot v_x &= v_x^2 + \bar\eta_x - \left(\pi\, r_x\, \tau_x\right)^2
                   + \tau_x\!\left(J_x s_x - C_{xy} s_y\right) + I + \hat F_{e;x}, \\
s_x &= \hat K_x[r_x], \\
\tau_y \dot r_y &= \frac{\Delta_y}{\pi\,\tau_y} + 2 r_y v_y, \\
\tau_y \dot v_y &= v_y^2 + \bar\eta_y - \left(\pi\, r_y\, \tau_y\right)^2
                   + \tau_y\!\left(-J_y s_y + C_{yx} s_x\right), \\
s_y &= \hat K_y[r_y].
\end{aligned}
\label{eq:nmm2-EI-pushpull}
\end{equation}
Here $x\equiv E$, $y\equiv I$; $\tau_x,\tau_y$ are the membrane time constants of the
excitatory and inhibitory populations, and $\hat K_x,\hat K_y$ are the second-order
synaptic operators (with their own AMPA/GABA synaptic time constants). The membrane
time enters as in the exact dimensional MPR reduction~\cite{Montbrio:2015aa,clusellaComparisonExactHeuristic2023}:
the rate term becomes $\Delta/(\pi\tau)$, the quadratic curvature term $(\pi r\tau)^2$,
and the synaptic current is scaled by $\tau$; the external input $I$ and forcing
$\hat F_{e;x}$ are unscaled.
The sign conventions follow the rest of the manuscript: $J_x>0$ denotes excitatory self-coupling within the E population, $J_y>0$ denotes inhibitory self-coupling within the I population (entering with an explicit minus sign in the $\dot v_y$ equation), and $C_{yx}>0$, $C_{xy}>0$ denote the magnitudes of the cross-couplings (E$\to$I positive, I$\to$E negative as a consequence of the explicit minus in the $\dot v_x$ equation). The forcing term $\hat F_{e;x}$ enters only the excitatory population, consistent with the convention used throughout this review.}
In operator form, following Equation~\ref{eq:master_nmm2}, we express the complete set more compactly,
\begin{equation}
\begin{aligned}
     r_x &=  \hat \Phi_x\!\left[\tau_x\!\left(J_x\, s_x - C_{xy}\, s_y\right) + \hat F_{e;x}\right], &
 s_x &= \hat K_x [r_x], \\
  r_y &=  \hat \Phi_y\!\left[\tau_y\!\left(-J_y\, s_y + C_{yx}\, s_x\right)\right], &
 s_y &= \hat K_y[r_y],
\end{aligned}
\end{equation}
where $\hat\Phi_\alpha$ is the dimensional MPR transfer functional---the $(r_\alpha,v_\alpha)$ pair above with membrane time $\tau_\alpha$---and, as usual, we drive only the excitatory population. The push-pull motif is present through the nonlinearity.
\paragraph{General equations for multiple coupled populations.} The general equations for multiple interacting populations become 
\begin{equation}
\begin{aligned} \label{eq:nmm2-general}
\tau_n\,\dot r_n &= \frac{\Delta_n}{\pi\,\tau_n} + 2 r_n v_n, \\
\tau_n\,\dot v_n &= v_n^2 - \left(\pi\, r_n\, \tau_n\right)^2 + \bar\eta_n + \hat F_{e;n}
 + \tau_n\!\!\sum_{m:n\leftarrow m} C_{nm}\, s_{nm}, \\
s_{nm} &= \hat K_{nm}[r_m],
\end{aligned}
\end{equation}
where $\tau_n$ is the membrane time constant of population $n$ and the synaptic operators $\hat K_{nm}$ carry their own (synaptic) time constants, or
\begin{equation}
\begin{aligned}
 r_n &= \hat\Phi_n\!\Big[\tau_n\Big(J_n\, s_n + \!\!\sum_{m:n\leftarrow m} C_{nm}\, s_{nm}\Big) + \hat F_{e;n}\Big], \\
 s_{nm} &= \hat K_{nm}[r_m] .
\end{aligned}
\end{equation}

\paragraph{Equations for coupled E-I push-pull motif.} We can also write equations for multiple E-I motifs as in the previous sections, where each E-I motif is identical and we only connect excitatory to excitatory populations (with, e.g., $J_x = C_{xx}$ and only two types of synapses),
\begin{tcolorbox}[
  colback=gray!5,
  colframe=gray!5, 
  boxrule=0pt,
  enhanced, sharp corners,
  left=4pt, right=4pt, top=3pt, bottom=3pt,
  boxsep=2pt,
  before skip=6pt, after skip=6pt
]
\rev{\begin{equation} \label{eq:NMM2-EI}
\begin{aligned}
\tau_x\,\dot r_x^{(i)} &= \frac{\Delta_x}{\pi\,\tau_x} + 2\,r_x^{(i)} v_x^{(i)},\\
\tau_x\,\dot v_x^{(i)} &= \bigl(v_x^{(i)}\bigr)^2 - \bigl(\pi\, r_x^{(i)}\, \tau_x\bigr)^2 + \bar\eta_x
   + \tau_x\Bigl( C_{xx}\,s_x^{(i)} - C_{xy}\,s_y^{(i)} + \sum_{j\neq i} C_{ij}\,s_x^{(j)} \Bigr) + \hat F_{e}^{(i)},\\
s_x^{(i)} &= \hat K_x\bigl[r_x^{(i)}\bigr],\\[2pt]
\tau_y\,\dot r_y^{(i)} &= \frac{\Delta_y}{\pi\,\tau_y} + 2\,r_y^{(i)} v_y^{(i)},\\
\tau_y\,\dot v_y^{(i)} &= \bigl(v_y^{(i)}\bigr)^2 - \bigl(\pi\, r_y^{(i)}\, \tau_y\bigr)^2 + \bar\eta_y
   + \tau_y\Bigl( -C_{yy}\,s_y^{(i)} + C_{yx}\,s_x^{(i)} \Bigr),\\
s_y^{(i)} &= \hat K_y\bigl[r_y^{(i)}\bigr].
\end{aligned}
\end{equation}
All coupling magnitudes are taken non-negative, $C_{\alpha\beta}\geq 0$; the signs in the voltage equations encode the dynamical role of each connection: $+$ for excitatory drive ($C_{xx}$ self-recurrence, $C_{yx}$ E$\to$I, long-range $C_{ij}$ for E$\to$E) and $-$ for inhibitory drive ($C_{yy}$ self-recurrence, $C_{xy}$ I$\to$E). The swap-and-flip rule from the excitatory ($x$) to the inhibitory ($y$) population is then read directly off the equations: swap the population index and flip the sign of each cross-coupling term according to the role (excitatory vs.\ inhibitory) it plays in the receiving population.}
\end{tcolorbox}

\begin{figure}[t]
    \centering
    \includegraphics[width=0.65\linewidth]{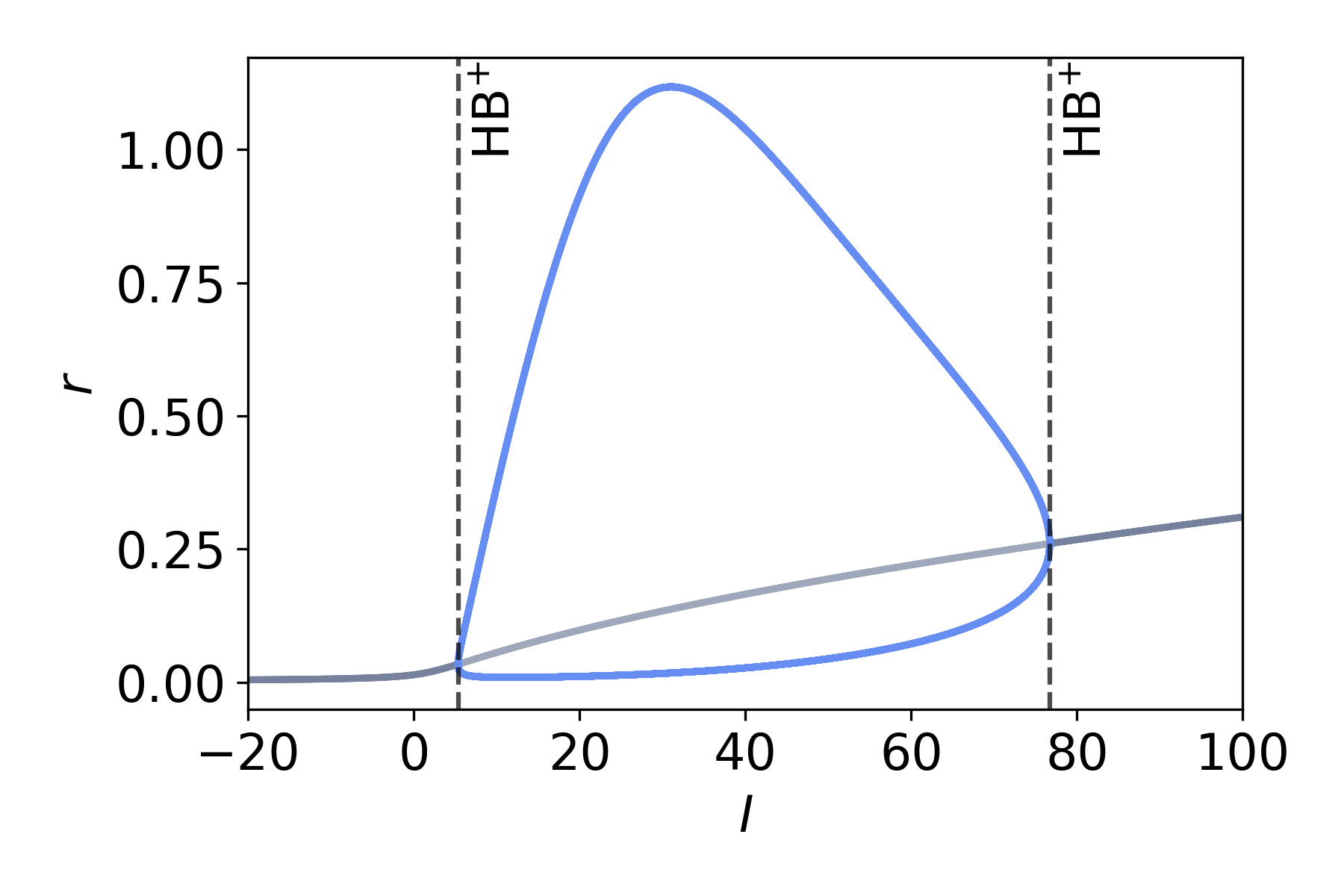}
    \caption{Bifurcation diagram of the interneuron-gamma (ING) NMM2 self-coupled model (Eq. \eqref{eq:nmm2-ING}) with bifurcation parameter $I$. Dark (light) grey lines represent the stable (unstable) fixed points, while the dark (light) blue curves indicate the maximum and minimum amplitudes of the stable (unstable) limit cycles. See Table~\ref{tab:bif-params} for details.}
    \label{fig:ING_bif}
\end{figure}

The bifurcation diagram of Eq.~\eqref{eq:nmm2-ING} is shown in Fig. \ref{fig:ING_bif}. It illustrates the steady-state firing rate $r$ as a function of the external input $I$. As in the previous models, increasing the external input induces a transition from
a stable fixed point to oscillatory activity via a supercritical Hopf bifurcation (HB$^{+}$). The resulting limit-cycle oscillations persist over a range of input values and are terminated at a second HB$^{+}$, with both bifurcation points indicated by the vertical dashed lines.

\begin{figure}[h!]
    \centering
    \includegraphics[width=0.65\linewidth]{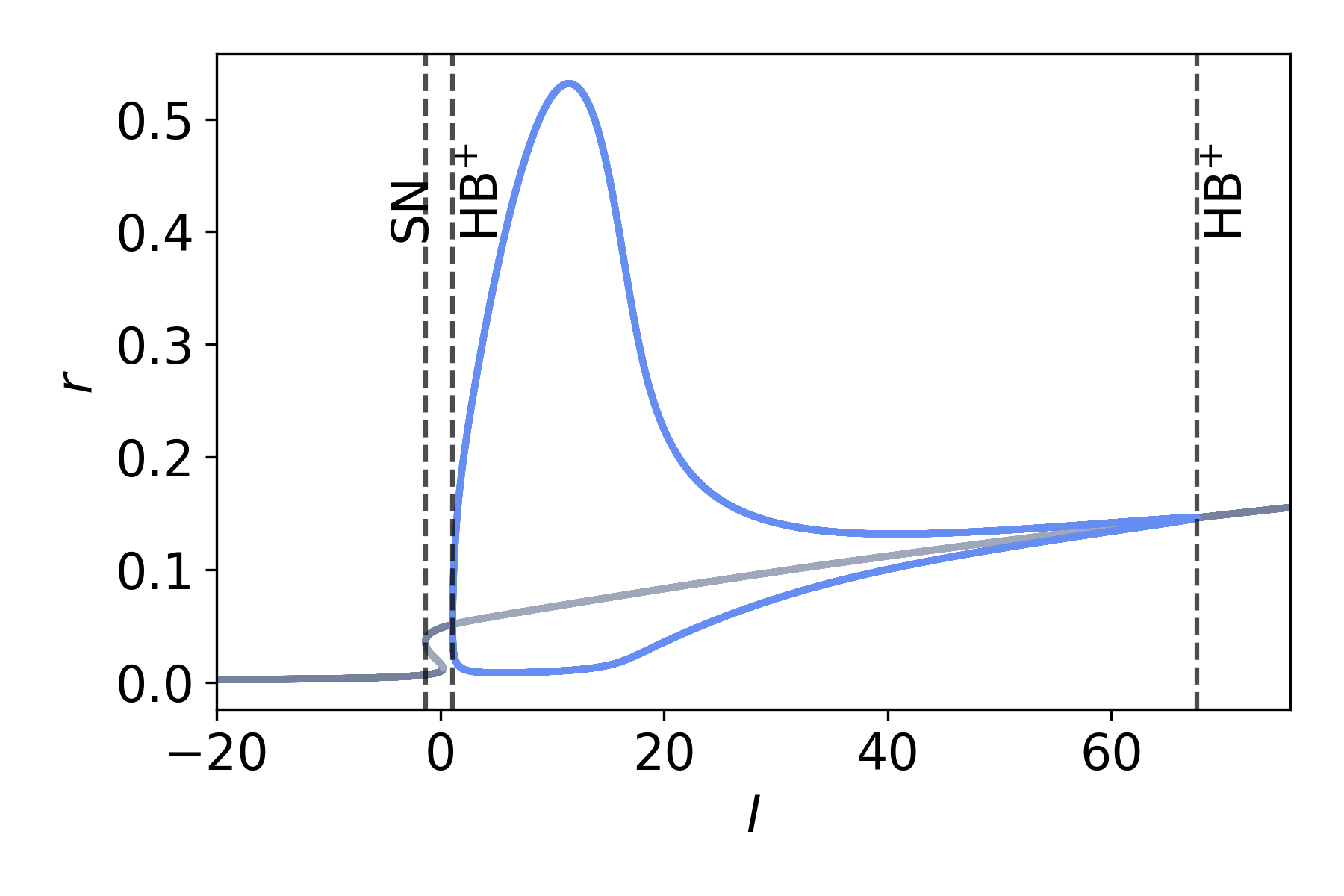}
    \caption{Bifurcation diagram of the pyramidal-interneuron (PING) NMM2 model with bifurcation parameter $I$ (Eq. \eqref{eq:nmm2-EI-pushpull}). Dark (light) grey lines represent the stable (unstable) fixed points, while the dark (light) blue curves indicate the maximum and minimum amplitudes of the stable (unstable) limit cycles. See Table~\ref{tab:bif-params} for details.}
    \label{fig:PING_bif}
\end{figure}

The bifurcation diagram of Eq.~\eqref{eq:nmm2-EI-pushpull}, which yields a  PING model, is shown in Fig. \ref{fig:PING_bif}. It depicts the steady-state firing rate as a function of the external input (I). From left to right, as the external input increases, the system first undergoes a pair of saddle-node (SN) bifurcations, indicating the creation and annihilation of fixed points. Following these, the system experiences
a supercritical Hopf bifurcation (HB$^+$), giving rise to stable oscillatory dynamics. The resulting limit-cycle oscillations persist over a range of input values until the system passes through another HB$^+$ bifurcation, at which the oscillations disappear and the system returns to a stable fixed point.

\begin{tcolorbox}[
  colback=gray!5,
  colframe=slate,
  colbacktitle=softblue!25,
  coltitle=softblue!60!black,
  fonttitle=\bfseries\small,
  title={
  \small \textit{NMM2  Network (Eq.~\ref{eq:nmm2-general})} \\ Whole-brain Simulations Parameters \& Physiological Meaning
  },
  boxrule=0.4pt, enhanced, sharp corners,
  left=4pt,right=4pt,top=3pt,bottom=3pt,
  boxsep=2pt, before skip=6pt, after skip=6pt
]
\scriptsize
\begin{tabular}{@{}l p{0.84\linewidth}@{}}
\textbf{Node $n$} & Neural mass with explicit \emph{synapse} and \emph{soma}: PSP states drive a membrane perturbation $s_{nm}$, which adds, with others, to $v_n$ and passes through a sigmoid to yield a firing rate $r_n$.\\
\textbf{$r_i(t)$}  & Population firing rate (Hz or normalized).  \\
\textbf{$v_n(t)$} & Mean membrane potential. \\
{$\tau_{m,x,y},\,  \tau_s$} &  Membrane and synaptic time constants. \\
\textbf{$\bar\eta_n,\;\Delta_n$} & \rev{Centre (location parameter) and HWHM} of the Lorentzian excitability distribution (bias and heterogeneity of population elements). Larger $\Delta_n$ broadens dispersion and damps coherence.\\
\textbf{$J_n$} & Local recurrent self‑coupling of population $n$ (excitatory $J_n\!>\!0$; inhibitory $J_n\!<\!0$) scaling the node’s own synaptic self-current $s_{nn}$. Tunes resonance and distance to oscillatory regimes.\\
\textbf{$s_{nm}(t)$} & Synaptic activation from presynaptic population $m$ to $n$; obtained by filtering $r_m$: $s_{nm}=\hat K_{nm}[r_m]$.\\
\textbf{$\hat K_{nm}$} & Synaptic filter (first‑ or second‑order). E.g., for equal rise/decay: $\hat K[r]$ solves $\gamma^{-1}(\tau^2\ddot s+2\tau\dot s+s)=r$; with distinct rise/decay use $(\tau_r,\tau_d)$.\\
\textbf{$\gamma_{nm}$,\,$\tau_{nm}$} & Synaptic gain and time constants (set per synapse type: AMPA/NMDA/GABA\textsubscript{A}/GABA\textsubscript{B}); determine amplitude, lag and band‑selectivity.\\
\textbf{$C_{nm}$} & Long‑range coupling weight from $m$ to $n$ (SC by default; FC/EC or synthetic graphs if needed).\\
\textbf{$t_{nm}$ (opt.)} & Propagation delay on pathway $m{\to}n$; introduces frequency‑dependent phase lags. Use by replacing $s_{nm}(t)$ by $s_{nm}(t-t_{nm})$ in Eq.~\ref{eq:nmm2-general}.\\
\textbf{$N$} & Number of nodes/populations.\\
\textbf{$\hat F_{e}^{(i)}(t)$}& Exogenous drive  (external forcing from nodes or elements outside the network or an electric field)---see Equation~\ref{eq:forcing}).\\
\end{tabular}
\end{tcolorbox}

\begin{tcolorbox}[
  colback=gray!5,
  colframe=slate,
  colbacktitle=softblue!25,
  coltitle=softblue!60!black,
  fonttitle=\bfseries\small,
  title={\small When to Use It — Next‑Generation Neural Mass (NMM2)},
  boxrule=0.4pt, enhanced, sharp corners,
  left=4pt,right=4pt,top=3pt,bottom=3pt,
  boxsep=2pt, before skip=6pt, after skip=6pt
]
\scriptsize
\textbf{Use this when} you need a \emph{first‑principles} link from spiking microdynamics to mesoscale variables: a dynamic “sigmoid” ($r$–$v$ equations) with state‑dependent gain/inertia, principled effects of self‑coupling $J$, and clean integration of synaptic filtering and uniform field inputs.\\
\textbf{Assumptions} all‑to‑all recurrence within each population, QIF spike mechanism, Lorentzian heterogeneity (MPR exactness), chemical synapses via $s=\hat K[r]$; extensions cover electrical synapses and plasticity.\\
\textbf{Best for} analytic studies of resonance/entrainment to weak uniform drives (tACS‑like), comparisons to heuristic NMMs.\\
\textbf{Avoid} if a static transfer is sufficient and parameters must map directly onto classic NMM1 fits; or if detailed channel/compartment biophysics is required (prefer conductance‑based masses).
\end{tcolorbox}

\begin{tcolorbox}[
  colback=gray!5,
  colframe=slate,
  colbacktitle=softblue!25,
  coltitle=softblue!60!black,
  fonttitle=\bfseries\small,
  title={\small How to Use It — Minimal Guide (NMM2 Network)},
  boxrule=0.4pt, enhanced, sharp corners,
  left=4pt,right=4pt,top=3pt,bottom=3pt,
  boxsep=2pt, before skip=6pt, after skip=6pt
]
\scriptsize

\textbf{Provide} $(\bar\eta_n,\Delta_n)$, $J_n$; synaptic filters $\hat K_{nm}$ with $(\tau,\gamma)$ (or $(\tau_r,\tau_d,\gamma)$); network $C_{nm}$ (SC/FC/EC or synthetic), optional delays $t_{nm}$; external $p_n(t)$ or field term $\vec\lambda_n\!\cdot\!\vec E_n(t)$.\\
\textbf{Defaults} start from JR‑style kinetics (use Table~\ref{table:synapses} for AMPA/NMDA/GABA ranges); row‑normalize $C$ and (optionally) scale by a global gain; initialize $(r_n,v_n)$ near the desired fixed point; \rev{With non-zero delays $\tau_{ij}>0$ in the network coupling and stochastic forcing in $\hat F_{e;n}$, Eqs.~\eqref{eq:nmm2-general}--\eqref{eq:NMM2-EI} are SDDEs and inherit the caveats discussed in \S\ref{sec:linear-phase}: prefer method-of-steps DDE solvers in the deterministic limit; if a fixed-step Euler--Maruyama scheme is used on the delayed system as a controlled approximation, satisfy both the oscillation-resolution bound $\Delta t \leq 1/(200\,f_{\max})$ and the delay-resolution bound $\Delta t \leq \min_{ij}\tau_{ij}/10$, verify convergence by step halving, and report the discretisation scheme, interpolation on the delay buffer, and random number generator (RNG) seed alongside other simulation parameters.}\\
\textbf{Readouts} $r_n(t)$, $v_n(t)$, and $s_{nm}(t)$; PSD/cross‑spectra; envelope/FC/FCD; resonance curves vs.\ $J,\tau$ and field amplitude; operating‑point maps from the $r$–$v$ nullclines.
\end{tcolorbox}

\subsection{Applications} 
\label{subsec:applications-nmm2}

The NMM2 formalism has begun to power a range of concrete applications:
\begin{enumerate}
 \item \textbf{Emergence of brain rhythms}: Several works have analyzed models based on the MPR theory
 to explore motifs allowing for the emergence of fast collective oscillations in one or two neural populations. The simplest instances include a single population of inhibitory neurons with synaptic delays\cite{Devalle:2017aa,Dumont:2019aa,Bi2020,Clusella2023}, 
 and excitatory-inhibitory population pairs \cite{Dumont:2019aa,Byrne2020,segneri2020,ReynerParra2022,MayoraCebollero2025}.
\item  \textbf{Whole‑brain modeling}: Next‑generation neural masses embedded on human connectomes offer an improved framework to analyze and reproduce brain dynamics captured by fMRI. 
    Some works have started to explore this venue, focusing on capturing pathology and healthy states \cite{Gerster2021,Rabuffo2021,perl_whole-brain_2023,Forrester2024} and performing detailed mathematical analyses for the emergence of complex spatiotemporal behavior \cite{clusellaComplexSpatiotemporalOscillations2023,delicado-moll2026}.
 \item \textbf{Non‑invasive stimulation theory}: Because NMM2 replaces a static sigmoid with dynamic firing‑rate equations, it predicts state‑dependent field sensitivity. With second‑order synapses, it explains enhanced resonant responses to weak uniform AC electric fields (tACS‑like) and how self‑coupling and synaptic time constants tune this sensitivity \cite{Clusella2023}. Relatedly, NMM2 has been used to study population responses to brief exogenous pulses (TMS‑like) and ERD/ERS phenomena \cite{Byrne2020,Byrne2022}.
\item \textbf{Cognition}: Exact mean‑field models reproduce working‑memory operations and associated oscillatory signatures, providing a coarse‑grained yet mechanistic account \cite{Schmidt2018,Taher2020}.
\item \textbf{Neuromodulation and DBS}: Extensions that include adaptation/neuromodulatory variables have been used to explore mechanism‑level effects of deep brain stimulation and dopamine on network dynamics \cite{Depannemaecker2024,Chen2022}.
\end{enumerate}

Additionally,
several   extensions to the MPR theory have extended its range
of applicability by challenging some of the simplifying assumptions of the QIF formulation \eqref{eq:qif}.
Paired with synaptic dynamics, these extensions provide even further refined versions of NMM2:

\begin{enumerate}
    \item \textbf{Dynamic noise}: Originally, the only source of microscopic disorder in the QIF formulation \eqref{eq:qif} was through the quenched variables $\eta_j$.
    Nonetheless, the exact mean-field theory also applies if $\eta_j(t)$ are considered to be
    dynamic Cauchy white noise \cite{clusella2024}. Additionally, approximation theories exist for the case of Gaussian white noise \cite{Goldobin2021,Goldobin2021_2}.
    \item \textbf{Quenched noise distribution}: The MPR theory requires $\eta_j$ to be Lorentzian-distributed to obtain a closed low-dimensional model for $r$ and $v$.
    Recent work generalized the theory to include $q$-Gaussian distributions, at the expense of
    increasing the number of independent variables in the resulting neural mass\cite{PP22,PP24}. 
    \item \textbf{Gap-junctions}: The MPR theory also applies for neurons communicating via electrical synapses mediated by gap-junctions and other diffusion-based coupling \cite{Pietras2019,Montbrio:2020aa}. This allows for deriving NMM2 including electrical coupling, a major breakthrough considering that heuristic formulations cannot account for this microscopic effect. 
    \item \textbf{Adaptation variables}: In spite of its significance, the QIF model is a simplified model for neuron dynamics. Further biophysical mechanisms in the single unit dynamics, such as spike-frequency adaptation or ion channel dynamics, require the inclusion of additional dynamical variables for which the exact mean-field theory might not apply.
    To address this problem, some works propose including a mean-field adaptation as an approximation to networks with single unit adaptation\cite{chenExactMeanfieldModels2022,ferraraPopulationSpikingBursting2023}. 
    A recent approach, instead, proposes including a suitable form of spike-frequency adaptation for which the MPR still holds exactly\cite{Pietras2025}. \rev{Beyond the QIF/theta line, Nicola \& Campbell~\cite{NicolaCampbell2013} derived an approximate mean-field reduction for heterogeneous networks of two-dimensional integrate-and-fire neurons.}
    \item \textbf{Full low-dimensional theory}: The MPR theory, which is strongly related to the Ott-Antonsen ansatz for the Kuramoto model\cite{Ott2008}, poses some challenging theoretical questions. The most relevant of them asks whether the low-dimensional manifold is attracting in the microscopic representation. A rigorous theoretical framework has been proposed recently, demonstrating that this is indeed the case provided the single units are not all identical ($\Delta>0$)\cite{Pietras2023}. 
    \rev{\item \textbf{Finite and asymmetric spikes}: The original theory requires QIF neurons with a 
    spike apex and reset at $V_\text{peak}=-V_\text{reset} = \infty$. This poses a limitation resulting neural mass model,
    especially when gap-junctions are included. Montbrió and Pazó \cite{montbrio2020} introduced a model accounting for spike asymmetry
     in order to properly capture these effects in networks with gap junctions. Recently, Cestnik \cite{cestnik2026} extended the exact low-dimensional reduction to a two-phase model capturing the dynamics for neurons with finite spikes ($|V_\text{peak}|,|V_\text{reset}| < \infty$).}
\end{enumerate}



\definecolor{coral}{HTML}{F57558}
\definecolor{softblue}{HTML}{668EF2}
\definecolor{slate}{HTML}{76819D}

\clearpage \section{Summary and Outlook}

\label{sec:Closing}

\begin{figure}[t!]
    \centering
    \includegraphics[width=0.95\linewidth]{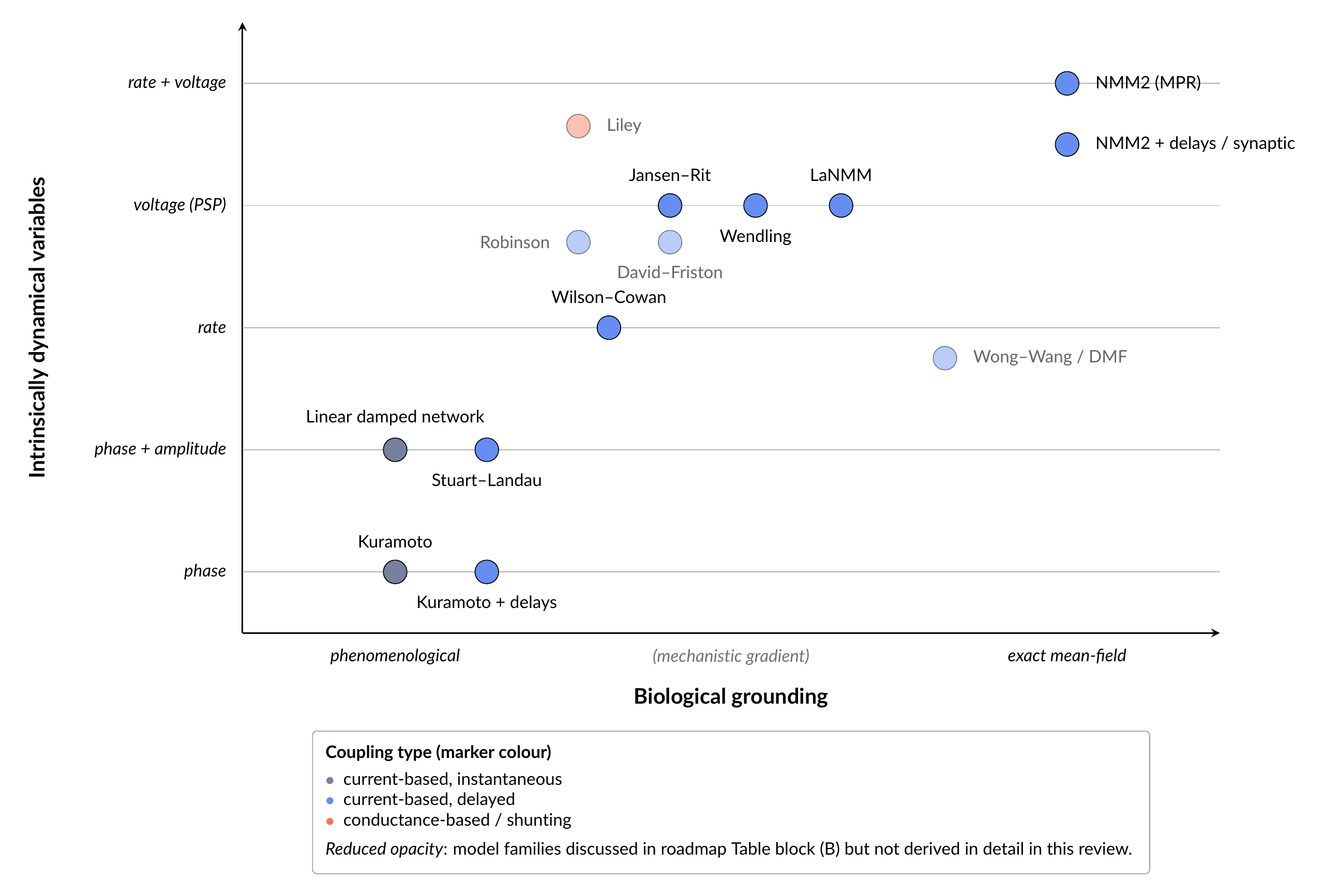}
    \caption{\rev{\textbf{Rosetta map of neural-mass and mean-field model families discussed in this review.} The horizontal axis orders models along a mechanistic gradient, from purely phenomenological constructions (chosen to capture observed dynamics with minimal commitment to mechanism) to exact mean-field reductions (linked to underlying spiking dynamics via formal derivation, e.g., the Lorentzian ansatz of MPR). The vertical axis tracks the \emph{intrinsically dynamical} variables, i.e., those that carry their own ODE, not the full set of state variables that appear in the equations. Wilson--Cowan sits at ``rate'' because only the rate variable has its own derivative; Jansen--Rit, Wendling, LaNMM, David--Friston, Liley and Robinson sit at ``voltage (PSP)'' because the post-synaptic potential filter supplies the dynamical equation while rate enters instantaneously through the sigmoid; NMM2 (MPR) and its delayed/synaptic extensions occupy ``rate + voltage'' because both variables have explicit ODEs. Wong--Wang/DMF is placed at ``rate'' because its gating variable evolves on synaptic timescale with first-order rate-like dynamics. Marker colour encodes coupling type: current-based instantaneous, current-based with delays, and conductance-based / shunting. Models referenced in roadmap Table block~(B) but not derived in detail in this review are shown at reduced opacity. Horizontal positions between the two endpoints are illustrative and reflect relative mechanistic specificity rather than a quantitative scale. The linear-versus-nonlinear coupling distinction is absorbed into model identity rather than encoded separately, and is discussed in Section~\ref{sec:current-vs-conductance}.}}
    \label{fig:rosetta-map}
\end{figure}

\rev{We have treated neural mass models as points on a continuous ladder connecting microscopic spiking descriptions to macroscopic whole-brain dynamics. Griffiths, Bastiaens \& Kaboodvand~\cite{griffiths_bastiaens_kaboodvand_2021}organise the same space along an orthogonal axis --- spatial scale (cells $\to$ circuits $\to$ networks), whereas the present review uses the derivation--representation axes \rev{summarised in Fig.~\ref{fig:rosetta-map}}. The two organisations are complementary.}

At the lowest rung, the harmonic and Stuart–Landau (SL) oscillators provide the normal-form language for local rhythms and their phase–amplitude structure. Wilson–Cowan (WILCO) models make the underlying excitatory–inhibitory push–pull architecture explicit and introduce sigmoidal transfer functionals as coarse-grained summaries of neuronal input–output relations. Second-order synaptic neural mass models (NMM1) add realistic synaptic filters with distinct rise and decay constants, thereby separating synapse from soma dynamics and producing realistic postsynaptic potentials, delays and phase shifts. Finally, next-generation models (NMM2) derived exactly from quadratic integrate-and-fire (QIF) networks close the loop from spikes to masses by replacing static sigmoids with analytically derived dynamic $(r,v)$ transfer equations, showing how population firing rates and mean membrane potentials evolve jointly under recurrent and external drive.\cite{montbrioMacroscopicDescriptionNetworks2015,coombesNextGenerationNeural2019,byrneNextgenerationNeuralMass2020,clusellaComparisonExactHeuristic2023}

A main message is that all these formalisms share a simple core: a push–pull motif between two effective degrees of freedom, coupled through linear filters and nonlinear transfer functions, and embedded in a network via structured coupling and delays. In the linear limit, this core reduces to damped resonators or complex Ornstein–Uhlenbeck processes, for which covariances and spectra can be obtained in closed form and mapped onto empirical functional connectivity.\cite{ponce-alvarezHopfWholebrainModel2024,Nozari2024} Close to Hopf, SL-like normal forms capture the onset of self-sustained oscillations, metastability, and turbulence-like dynamics under connectome coupling.\cite{decoTurbulentlikeDynamicsHuman2020,cabralMetastableOscillatoryModes2022,Escrichs2022Turbulence} WILCO and NMM1 add biophysical levers (synaptic gains, time constants, E/I balance, laminar circuits) without losing this dynamical structure, and NMM2 shows that, at least for QIF networks, these macroscopic descriptions can be derived exactly rather than postulated.

Looking ahead, one natural direction is to extend exact mean-field reductions beyond QIF neurons. Current next-generation neural mass models exploit specific analytic properties of the QIF nonlinearity and Lorentzian excitability distributions.\cite{montbrioMacroscopicDescriptionNetworks2015,coombesNextGenerationNeural2019} A major open problem is to obtain similarly low-dimensional closures for more realistic single-cell models, such as exponential integrate-and-fire or multi-variable Hodgkin–Huxley-type neurons, possibly via systematic approximations (e.g. moment closures, population-density expansions, or renormalization-group inspired coarse graining).\cite{nykampPopulationDensityApproach2000,buiceSystematicFluctuationExpansion2010,Goldobin2021} Such derivations would clarify which aspects of spiking physiology (e.g. spike-frequency adaptation, active dendrites, NMDA currents) survive coarse graining and how they renormalize effective gains, time constants, and non-Gaussian noise at the neural-mass level.

A second axis concerns more realistic local circuits. Laminar neural mass models already capture distinct deep and superficial generator loops and their contribution to laminar LFP, CSD, and cross-frequency coupling.\cite{sanchez-todoPhysicalNeuralMass2023,Ruffini2020P118} Next steps include: (i) explicitly modeling multiple inhibitory subtypes (PV, somatostatin (SOM), vasoactive intestinal peptide (VIP)) with layer-specific projections; (ii) incorporating both chemical and electrical synapses in a unified mean-field description; and (iii) embedding short-term synaptic plasticity and intrinsic adaptation into next-generation masses in an exact or controlled approximate way~\cite{Devalle:2017aa,taherExactNeuralMass2020,chenExactMeanfieldModels2022}. These refinements would allow laminar models to speak more directly to cell-type and layer-resolved data, calcium imaging, and perturbation experiments, and to test hypotheses about how microcircuit motifs implement canonical computations (e.g., gain control, winner–take–all, and gating).

At the whole-brain scale, three modeling ingredients become increasingly important. First, heterogeneity across regions: empirical work shows that cortical areas differ systematically in local time scales, recurrent excitation, and laminar architecture, which in turn shape their role in large-scale hierarchies.\cite{mejiasFeedforwardFeedbackFrequencydependent2016,rautOrganizationPropagatedIntrinsic2020} Future models should assign region-specific parameters (e.g., SL working points, WILCO gains, NMM1/NMM2 synaptic kinetics) informed by cytoarchitectonics, gene expression, and receptor densities, rather than using identical nodes everywhere. Second, directed and laminar-specific coupling: predictive-processing accounts suggest that feedforward and feedback pathways are implemented via distinct layers, with frequency-specific channels (gamma/beta) carrying prediction errors and predictions.\cite{bastosCanonicalMicrocircuitsPredictive2012,fristonFreeenergyPrincipleUnified2010} Embedding laminar neural masses in directed structural connectomes with frequency-dependent effective connectivity is a natural route to test such proposals against MEG/EEG and laminar recordings. Third, more refined parcellations and subcortical circuits: incorporating high-resolution cortical atlases (e.g. Glasser multimodal 360-area parcellation and its extensions)\cite{glasserMultimodalParcellationHuman2016} and explicit thalamic, basal ganglia, and cerebellar masses\cite{kringelbach_dynamic_2020,cofreWholeBrainModelsExplore2020,rautOrganizationPropagatedIntrinsic2020} will be essential to capture cortico–subcortical loops that shape rhythms, state transitions, and neuromodulatory control.

Neuromodulatory systems provide a complementary, low‑dimensional control axis over the same large‑scale circuits. Gradients of receptor densities and gene expression co‑localize with the structural and timescale hierarchies described above, suggesting that neuromodulatory tone shapes regional gain, effective time constants, and long‑range coupling in a spatially specific way \cite{demirtasHierarchicalHeterogeneityHuman2019,burtHierarchyTranscriptomicSpecialization2018}. From a modeling standpoint, even when neuromodulatory nuclei are not explicitly represented, their effects can be approximated by treating external drives and gain parameters as proxies for neuromodulatory axes—for example, using global or projection‑specific changes in background input, SL working points, or WILCO gains to implement neuromodulation‑like shifts in E/I balance, integration time and noise statistics, in line with classical circuit‑level neuromodulation work \cite{marderNeuromodulationNeuronalCircuits2012}. Recent whole‑brain modeling incorporates neurotransmission‑weighted connectivity to account for a wide repertoire of task‑evoked states, showing that neuromodulator‑dependent changes in effective coupling can reproduce diverse cognitive configurations within a single structural scaffold  \cite{decoEvolutionsBoldestTrick2025}. These findings motivate configuring “external inputs” in large‑scale models using receptor and gene‑expression maps, or fitting low‑dimensional neuromodulatory control variables to match state transitions induced by pharmacological, arousal, or task manipulations.

\rev{We see two further connections, with statistical inference and with information-theoretic analyses of state.} Linear and SL-based whole-brain models already support perturbation-based measures of nonequilibrium, susceptibility, and information routing across states (wake, sleep, anesthesia, psychedelics).\cite{decoTurbulentlikeDynamicsHuman2020,Escrichs2022Turbulence,cruzatEffectsClassicPsychedelic2022} NMM1 and NMM2 provide richer, mechanistically grounded arenas in which to define and compute such quantities, including algorithmic-information or compression-based characterizations of oscillatory dynamics. Coupling these models to modern inference frameworks (variational data assimilation, Bayesian model comparison, active inference) can turn them into forward models for multimodal data (fMRI, M/EEG, SEEG, laminar probes) and for in silico perturbation experiments (TMS, DBS, tES).\cite{breakspearDynamicModelsLargescale2017,Sadeghi2020,duchetHowDesignOptimal2024,ruffiniAlgorithmicAgentPerspective2024}


 \subsection*{Acknowledgments}
GR and FC have received funding from the European Research Council (ERC) under the European Union’s Horizon 2020 research and innovation programme (grant agreement No 855109, GALVANI) and from FET under the European Union’s Horizon 2020 research and innovation programme (grant agreement No 101017716, NEUROTWIN). 
PC has received financial support from the grant PID2024-155942NB-I00 funded by MCIN/AEI/10.13039/501100011033.
Raul de Palma Aristides and Jordi Garcia-Ojalvo are supported by the European Commission under European Union’s Horizon 2020 research and innovation programme Grant Number 101017716 (NEUROTWIN). Jordi Garcia-Ojalvo was also financially supported by the European Research Council (ERC) under the Synergy grant 101167121 (CeLEARN), by the Spanish Ministry of Science and Innovation and FEDER under project PID2024-160263NB-I00), and by the ICREA Academia program.

\textbf{Declaration of generative AI and AI-assisted technologies in the manuscript preparation process.}
During the preparation of this work the author(s) used ChatGPT in order to streamline the narrative. After using this tool/service, the author(s) reviewed and edited the content as needed and take(s) full responsibility for the content of the published article.

\bibliographystyle{unsrt}
\bibliography{references,giulio_references,pau_references}
\clearpage


\appendix

\clearpage 

\clearpage
\section{Linear Stability and Bifurcation Analysis}
\label{sec:bifurcations}

\subsection{Bifurcation diagrams}

Bifurcation diagrams provide a powerful way to visualize how the qualitative behavior of a dynamical system changes as key parameters vary, and they are widely used to study the emergence of oscillatory dynamics. In this section, we discuss the common bifurcations observed in neural mass models by examining their bifurcation diagrams. We emphasize that these diagrams depend strongly on the chosen parameter values, and this section is not intended to serve as an exhaustive catalogue of all possible dynamical regimes. For comprehensive analyses of each model, we will refer the reader to the dedicated literature.

We focus on one-parameter bifurcation diagrams, which we compute using the AUTO-07p software package. This tool allows us to identify fixed points and limit cycles, along with their stability properties. The scripts used to generate all bifurcation diagrams presented here, along with a complete list of parameters, are available at \cite{doedel2007auto07p}.

\subsubsection*{Stuart-Landau}
We begin with the Stuart-Landau (SL) model, which is given by
\begin{align}
  \dot{x} &= \alpha\,x - \omega\,y
           - \gamma\,(x^{2}+y^{2})\,x
           + \beta\,(x^{2}+y^{2})\,y,
  \label{eq:sl_cart_x2}\\
  \dot{y} &= \alpha\,y + \omega\,x
           - \gamma\,(x^{2}+y^{2})\,y
           - \beta\,(x^{2}+y^{2})\,x,
\end{align}
Setting $\gamma = 1$ and $\beta = 0$, we get
\begin{align}
  \dot{x} &= \alpha\,x - \omega\,y
           - (x^{2}+y^{2})\,x
           ,
  \\
  \dot{y} &= \alpha\,y + \omega\,x
           - (x^{2}+y^{2})\,y
\end{align}
The bifurcation diagram of the later, using $\alpha$ as the bifurcation parameter, is shown in Fig. \ref{fig:SL_bif}. The dark (light) gray line represents the stable (unstable) fixed points. For $\alpha < 0$, the system has complex eigenvalues with negative real parts, meaning that trajectories spiral towards the fixed point (damped oscillations). For $\alpha > 0$, the real parts of the eigenvalues are positive, so trajectories are repelled from the fixed point and attracted to the stable limit cycle, with amplitude represented in blue. This transition, in which a fixed point changes its stability while a stable (or unstable) limit cycle emerges, is known as a supercritical (or subcritical) Hopf bifurcation, here referred to as HB$^{+(-)}$.

\rev{The bifurcation diagrams for the Wilson--Cowan and NMM1 (Jansen--Rit, Wendling, LaNMM) models, previously included here, have been moved into the main body of the paper---see~\S\ref{sec:WILCO} and~\S\ref{sec:NMM1}. The diagrams below cover the remaining models in the cross-walk, namely the next-generation NMM2 ING and PING motifs.}

\subsubsection*{ING}
Our analysis now shifts to NMM2 models. We start with the ING model with second-order synapses, whose dynamics are governed by the following system of equations: 
\begin{equation}
\begin{aligned}
\tau_i \dot{v} &= \eta - \left( \pi r \tau_m \right)^2 + v^2 - \tau_i \left( J_{ii} s \right) + I \\
\tau_i \dot{r} &= \Delta + 2 r v \\
\tau_g \dot{s} &= z \\
\tau_g \dot{z} &= r - 2z - s
\end{aligned}
\end{equation}
The bifurcation diagram for the NMM2 model of interneuron-gamma (ING) oscillations is shown in Fig.~\ref{fig:ING_bif}. It illustrates the steady-state firing rate $r$ as a function of the external input $I$. As in the previous models, increasing the external input induces a transition from a stable fixed point to oscillatory activity via a supercritical Hopf bifurcation (HB$^{+}$). The resulting limit-cycle oscillations persist over a range of input values and are terminated at a second HB$^{+}$, with both bifurcation points indicated by the vertical dashed lines.

\subsubsection*{PING}
We continue our analysis of the NMM2 family by shifting our attention to the PING model, which can be viewed as a natural extension of the ING framework. The governing equations of the PING system are given by
\begin{equation}
\begin{aligned}
\tau_e \dot{v}_e &= \eta_e - \left( \pi r_e \tau_e \right)^2 + v_e^2 + \tau_e \left( J_{ee} s_e - J_{ei} s_i \right) + I \\
\tau_e \dot{r}_e &= \Delta_e + 2 r_e v_e \\
\tau_a \dot{s}_e &= z_e \\
\tau_a \dot{z}_e &= r_e - 2z_e - s_e \\
\tau_i \dot{v}_i &= \eta_i - \left( \pi r_i \tau_i \right)^2 + v_i^2 + \tau_i \left( J_{ie} s_e - J_{ii} s_i \right) + I \\
\tau_i \dot{r}_i &= \Delta_i + 2 r_i v_i \\
\tau_g \dot{s}_i &= z_i \\
\tau_g \dot{z}_i &= r_i - 2z_i - s_i
\end{aligned}
\end{equation}
The bifurcation diagram of the PING model is shown in Fig.~\ref{fig:PING_bif}. It depicts the steady-state firing rate as a function of the external input ($I$). 
From left to right, as the external input increases, the system first undergoes a pair of saddle-node (SN) bifurcations, indicating the creation and annihilation of fixed points. Following these, the system experiences a supercritical Hopf bifurcation (HB$^{+}$), giving rise to stable oscillatory dynamics. The resulting limit-cycle oscillations persist over a range of input values until the system passes through another HB$^{+}$ bifurcation, at which the oscillations disappear and the system returns to a stable fixed point.

\newpage

To summarize, bifurcation diagrams are particularly valuable for neural mass models because they:

\begin{itemize}
    \item \textbf{Reveal the onset of oscillations and other dynamical regimes.}
    They identify where fixed points lose stability and give rise to limit cycles, as well as where bistability or other qualitative transitions occur, providing a clear picture of the model’s possible behaviors.

    \item \textbf{Guide parameter selection and sensitivity analysis.}
    By showing how solutions depend on parameters such as coupling strengths or synaptic gains, bifurcation diagrams help identify which parameters critically shape the dynamics and which regimes are physiologically plausible.

    \item \textbf{Inform control and intervention strategies.}
    In applications ranging from neuromodulation to pharmacology, knowing how small parameter changes can shift the system between regimes is essential. Bifurcation diagrams make these transitions explicit by highlighting where stability changes occur.
\end{itemize}

\clearpage

\begin{landscape}
\setlength{\LTcapwidth}{\textwidth}
{\small
\begin{longtable}{@{}llp{4.8cm}p{8cm}}
  \caption{Parameters used for the bifurcation diagram in
  this paper. The code to reproduce the diagrams using \texttt{AUTO-07P} is available at \url{https://github.com/pclus/auto-tutorial}.}
  \label{tab:bif-params}\\
  \toprule
  \textbf{Fig.} & \textbf{Panel} & \textbf{Swept parameter} &
  \textbf{Fixed parameters} \\
  \midrule
  \endfirsthead
  \multicolumn{4}{@{}l}{\textit{Table~\ref{tab:bif-params} continued}}\\
  \toprule
  \textbf{Fig.} & \textbf{Panel} & \textbf{Swept parameter} &
  \textbf{Fixed parameters} \\
  \midrule
  \endhead
  \midrule
  \multicolumn{4}{r}{\textit{continued on next page\ldots}}\\
  \endfoot
  \bottomrule
  \endlastfoot

  \multicolumn{4}{@{}l}{\textit{Stuart--Landau}}\\
  Fig.~\ref{fig:SL_bif} & --- & $\alpha\in[-10,10]$ & $\omega=1$ \\ \midrule
  
  \multicolumn{4}{@{}l}{\textit{Wilson--Cowan}}\\
  \addlinespace[2pt]
  Fig.~\ref{fig:wilco_two_personalities} & (a) & $P_x\in[-5,5]$  &
   $\kappa_{xx}$ = $\kappa_{xy}=\kappa_{yx}=12$, $\kappa_{yy}=2$, $\tau_x = \tau_y = 1$, $\rho=1$, $\theta_x = \theta_y = 0$, $e_0 = 1$, $P_y=-6$ \\
    & (b) & $P_y\in[-10,0]$  \& same as (a) except for $I_x=0$ & \\
  \midrule
  \multicolumn{4}{@{}l}{\textit{Jansen--Rit}}\\
  Fig.~\ref{fig:JR_bif} & --- & $P\in[-80,350]$ & Same parameters as in \cite{JansenRit1995} and \cite{GrimbertFaugeras2006} \\ \midrule
  \multicolumn{4}{@{}l}{\textit{Wendling}}\\
  Fig.~\ref{fig:Wend_bif} & --- & $P\in[0,1700]$ & Same parameters as in \cite{wendlingEpilepticFastActivity2002} \\ \midrule
  \multicolumn{4}{@{}l}{\textit{LaNMM} }\\
  Fig.~\ref{fig:LaNMM_bif} & --- & $P\in[-100,500]$ & Same parameters as in \cite{sanchez-todoPhysicalNeuralMass2023} and \cite{depalmaaristidesEmergenceMultifrequencyActivity2025}\\ \midrule
  \multicolumn{4}{@{}l}{\textit{NMM2-ING} }\\
  Fig.~\ref{fig:ING_bif} & --- & $I\in[-20,100]$ & $\eta=0$, $J_{ii}=20$, $\tau_i=7.5$, $\tau_g = 2$, $\Delta=1$ \\ \midrule
  \multicolumn{4}{@{}l}{\textit{NMM2-PING}}\\
  Fig.~\ref{fig:PING_bif} & --- & $I\in[-20,75]$ & $\eta_e=\eta_i=0$, $J_{ee}=J_{ei}=J_{ie}=1$, $J_{ii}=2$,  $\tau_e = 15, \tau_a=10$, $\tau_i = 7.5, \tau_g = 2.5$, $\Delta_e=\Delta_i=1$\\
\end{longtable}
}
\end{landscape}
\section{From Wilson-Cowan to the Damped Harmonic Oscillator}

How do WILCO parameters map to harmonic oscillator parameters in the linearized regime (damping and oscillatory frequency)? 
Let \((x_0,y_0)\) be a fixed point under constant inputs \(I_{x_0},I_{y_0}\). Define
\[
u_x = x - x_0,\quad
u_y = y - y_0,\quad
\delta I_x = I_x - I_{x_0},\quad
\delta I_y = I_y - I_{y_0}.
\]
Also set
\[
h_{x_0} = w_{xx}\,x_0 + w_{xy}\,y_0 + I_{x_0}, 
\qquad
h_{y_0} = w_{yx}\,x_0 + w_{yy}\,y_0 + I_{y_0},
\]
so that 
\[
x_0 = S(h_{x_0}), 
\qquad
y_0 = S(h_{y_0}).
\]
The Wilson–Cowan equations read
\[
\begin{cases}
\tau_x\,\dot x = -\,x + S\bigl(w_{xx}x + w_{xy}y + I_x\bigr),\\
\tau_y\,\dot y = -\,y + S\bigl(w_{yx}x + w_{yy}y + I_y\bigr).
\end{cases}
\]
We expand each sigmoid around its equilibrium argument.  For the \(x\)–population:
\[
\begin{aligned}
S\bigl(w_{xx}x + w_{xy}y + I_x\bigr)
&= S\bigl(h_{x_0} + \Delta h_x\bigr), 
\quad
\Delta h_x = w_{xx}\,u_x + w_{xy}\,u_y + \delta I_x,\\
&= S(h_{x_0})
+ S'(h_{x_0})\,\Delta h_x
+ \tfrac{1}{2}\,S''(h_{x_0})\,(\Delta h_x)^2
+ \mathcal{O}\bigl(\lVert\Delta h_x\rVert^3\bigr).
\end{aligned}
\]

\rev{To make the truncation explicit, we adopt the asymptotic ordering $u_x, u_y = \mathcal{O}(\varepsilon)$ and $\delta I_x, \delta I_y = \mathcal{O}(\varepsilon^{1/2})$, the natural ordering when $\delta I$ is an externally controlled drive and $u$ is the system's response. Under this scaling the second-order terms in the expansion of $(\Delta h_x)^2$ scale as $u_x^2,\,u_x u_y,\,u_y^2 = \mathcal{O}(\varepsilon^2)$; $u_x \delta I_x,\,u_y \delta I_x = \mathcal{O}(\varepsilon^{3/2})$; and $\delta I_x^2 = \mathcal{O}(\varepsilon)$. The mixed cross-terms therefore dominate the pure-state quadratic terms by a factor $\varepsilon^{1/2}$, justifying their retention. The pure-input term $\delta I_x^2$ contributes a slowly varying input-only shift that we absorb into the redefinition of the operating point and do not display below. This ordering grounds the FM-modulation interpretation developed in the latter half of this appendix.}

Since \(\Delta h_x\) is small, keep only terms that are (i) linear in \(\{u_x,u_y,\delta I_x\}\) and (ii) the mixed cross‐terms \(u_x\,\delta I_x\) or \(u_y\,\delta I_x\).  Expand
\[
(\Delta h_x)^2 
= \bigl(w_{xx}u_x + w_{xy}u_y + \delta I_x\bigr)^2
= (w_{xx}u_x + w_{xy}u_y)^2 
\;+\; 2\,(w_{xx}u_x + w_{xy}u_y)\,\delta I_x 
\;+\; (\delta I_x)^2.
\]
Discard \(\mathcal{O}(u_x^2,u_y^2,u_xu_y,\,\delta I_x^2)\), but keep the cross‐term \(2\,(w_{xx}u_x + w_{xy}u_y)\,\delta I_x\).  Hence
\[
\begin{aligned}
S\bigl(w_{xx}x + w_{xy}y + I_x\bigr)
&\approx
S(h_{x_0})
+ S'(h_{x_0})\bigl[w_{xx}u_x + w_{xy}u_y + \delta I_x\bigr]\\
&\quad
+ \tfrac12\,S''(h_{x_0})\cdot 2\,(w_{xx}u_x + w_{xy}u_y)\,\delta I_x\\
&=
S(h_{x_0})
+ S'(h_{x_0})\bigl[w_{xx}u_x + w_{xy}u_y + \delta I_x\bigr]
+ S''(h_{x_0})\,(w_{xx}u_x + w_{xy}u_y)\,\delta I_x.
\end{aligned}
\]
Since \(S(h_{x_0}) = x_0\), substitute into 
\(\tau_x\,\dot x = -\,x + S(\cdots)\) and subtract the equilibrium relation \(\;0 = -\,x_0 + S(h_{x_0})\;\).  We get
\[
\begin{aligned}
\tau_x\,\dot u_x
&= \tau_x\,\dot x 
- 0 
= \bigl[-\,x + S(\cdots)\bigr] 
- \bigl[-\,x_0 + S(h_{x_0})\bigr]
= -\bigl(x - x_0\bigr) \\
+ \Bigl\{S(h_{x_0})
 &+ S'(h_{x_0})\bigl[w_{xx}u_x + w_{xy}u_y + \delta I_x\bigr]
+ S''(h_{x_0})\,(w_{xx}u_x + w_{xy}u_y)\,\delta I_x\Bigr\} \\
- S(h_{x_0}) \\
&= -\,u_x 
+ S'(h_{x_0})\bigl[w_{xx}u_x + w_{xy}u_y + \delta I_x\bigr]
+ S''(h_{x_0})\,(w_{xx}u_x + w_{xy}u_y)\,\delta I_x.
\end{aligned}
\]
Analogously for the \(y\)–population:
\[
\begin{aligned}
S\bigl(w_{yx}x + w_{yy}y + I_y\bigr)
&= S\bigl(h_{y_0} + \Delta h_y\bigr),
\quad
\Delta h_y = w_{yx}u_x + w_{yy}u_y + \delta I_y,\\
&\approx
S(h_{y_0})
+ S'(h_{y_0})\bigl[w_{yx}u_x + w_{yy}u_y + \delta I_y\bigr]
+ \tfrac12\,S''(h_{y_0})\cdot 2\,(w_{yx}u_x + w_{yy}u_y)\,\delta I_y,\\
&= S(h_{y_0})
+ S'(h_{y_0})\bigl[w_{yx}u_x + w_{yy}u_y + \delta I_y\bigr]
+ S''(h_{y_0})\,(w_{yx}u_x + w_{yy}u_y)\,\delta I_y.
\end{aligned}
\]
Subtracting the equilibrium yields
\[
\tau_y\,\dot u_y 
= -\,u_y
+ S'(h_{y_0})\bigl[w_{yx}u_x + w_{yy}u_y + \delta I_y\bigr]
+ S''(h_{y_0})\,(w_{yx}u_x + w_{yy}u_y)\,\delta I_y.
\]
Hence the first‐order system—including the cross‐terms where input multiplies state deviations—is
\[
\boxed{
\begin{aligned}
\tau_x\,\dot u_x 
&= -\,u_x 
+ S'(h_{x_0})\bigl[w_{xx}u_x + w_{xy}u_y + \delta I_x\bigr]
+ S''(h_{x_0})\,(w_{xx}u_x + w_{xy}u_y)\,\delta I_x,\\[0.7em]
\tau_y\,\dot u_y 
&= -\,u_y 
+ S'(h_{y_0})\bigl[w_{yx}u_x + w_{yy}u_y + \delta I_y\bigr]
+ S''(h_{y_0})\,(w_{yx}u_x + w_{yy}u_y)\,\delta I_y.
\end{aligned}
}
\]

Here we see that the self-coupling in the sigmoid affects, to first order, the frequency and also the time constant of the L-operator. To make this more explicit, we move the self-terms to the LHS, to provide a modified L-operator:
\[
\boxed{
\begin{aligned}
\tau_x\,\dot u_x + \bigl[1 -S'(h_{x_0}) w_{xx} - S''(h_{x_0})\,w_{xx} \,\delta I_x\bigr]u_x
&= 
\bigl[S'(h_{x_0}) + S''(h_{x_0})\,\delta I_x  \bigr]w_{xy}\, u_y + S'(h_{x_0})\delta I_x, \\[0.7em]
\tau_y\,\dot u_y 
+  \bigl[\,1 - S'(h_{y_0})\,w_{yy} - S''(h_{y_0})\,w_{yy}\,\delta I_y \bigr]\,u_y
&=
\bigl[S'(h_{y_0}) + S''(h_{y_0})\,\delta I_y\bigr]\,w_{yx}\,u_x 
+ S'(h_{y_0})\,\delta I_y.
\end{aligned}
}
\]

We now define two nonlinear differential operators \(L_x\) and \(L_y\) by grouping all ``self‐terms'' on the left.  Specifically:
\[
\begin{aligned}
L_x\bigl[u_x\bigr]
&:= \tau_x\,\dot u_x 
\;+\; \underbrace{\bigl[\,1 - S'(h_{x_0})\,w_{xx} \;-\; S''(h_{x_0})\,w_{xx}\,\delta I_x \bigr]}_{a_x},\\[0.6em]
L_y\bigl[u_y\bigr]
&:= \tau_y\,\dot u_y 
\;+\; \underbrace{\bigl[\,1 - S'(h_{y_0})\,w_{yy} \;-\; S''(h_{y_0})\,w_{yy}\,\delta I_y \bigr]}_{a_y}\,u_y.
\end{aligned}
\]
Next, define the ``instantaneous coupling‐(frequency)’’ coefficients and forcing terms:
\[
\begin{aligned}
\Omega_{xy}(t) 
&:= -\,\bigl[S'(h_{x_0}) + S''(h_{x_0})\,\delta I_x(t)\bigr]\,w_{xy}, 
&\quad 
F_x(t) &:= S'(h_{x_0})\,\delta I_x(t),\\[0.5em]
\Omega_{yx}(t)
&:= \phantom{-}\bigl[S'(h_{y_0}) + S''(h_{y_0})\,\delta I_y(t)\bigr]\,w_{yx},
&\quad 
F_y(t) &:= S'(h_{y_0})\,\delta I_y(t).
\end{aligned}
\]
With these definitions, each equation can be written in the familiar ``\(\,L[\cdot] = \pm\,\Omega\,(\text{other}) + F\)'' form:
\[
\boxed{
\begin{aligned}
L_x\bigl[u_x\bigr] 
&= -\,\Omega_{xy}(t)\,u_y \;+\; F_x(t),\\[0.7em]
L_y\bigl[u_y\bigr] 
&= +\,\Omega_{yx}(t)\,u_x \;+\; F_y(t).
\end{aligned}
}
\]
Here:
\[
\begin{aligned}
L_x[u_x] 
&= \tau_x\,\dot u_x 
\;+\; \bigl[\,1 - S'(h_{x_0})\,w_{xx} - S''(h_{x_0})\,w_{xx}\,\delta I_x \bigr]\,u_x,\\
L_y[u_y] 
&= \tau_y\,\dot u_y 
\;+\; \bigl[\,1 - S'(h_{y_0})\,w_{yy} - S''(h_{y_0})\,w_{yy}\,\delta I_y \bigr]\,u_y,\\
\Omega_{xy}(t) 
&= -\,\bigl[S'(h_{x_0}) + S''(h_{x_0})\,\delta I_x(t)\bigr]\,w_{xy},
\quad F_x(t) = S'(h_{x_0})\,\delta I_x(t),\\
\Omega_{yx}(t)
&= \bigl[S'(h_{y_0}) + S''(h_{y_0})\,\delta I_y(t)\bigr]\,w_{yx},
\quad F_y(t) = S'(h_{y_0})\,\delta I_y(t).
\end{aligned}
\]
In this packed form, \(L_x\) and \(L_y\) absorb all self‐interactions (including the \(\delta I\)-dependent shifts in effective gain), while the right‐hand side \(\pm\,\Omega(t)\cdot(\text{other variable}) + F(t)\) isolates the cross‐coupling—i.e.\ a time‐varying ``frequency'' \(\Omega_{xy},\Omega_{yx}\)—plus direct input forcing.

In summary, the sigmoid nonlinearity is not merely a saturating link; its first and second derivatives convert small input changes into dynamic adjustments of both damping and oscillatory frequency --- because $S'$ rescales all coupling strengths and $S''\,\delta I$ provides a direct input-dependent correction.

\paragraph{Functional Implications: Sigmoidal Frequency Modulation and Information Broadcast.}

The linearized equations
reveal that each population’s deviation \(u_i\) is not merely driven additively by inputs; the second derivative of the sigmoid, \(S''(h_{i0})\), multiplies the product of state deviations and input perturbations.  In other words, the term
\[
S''(h_{x0})\,\omega_{xy}\,u_y\,\delta I_x
\quad\Longleftrightarrow\quad
\text{(input to \(x\))} \times \text{(activity of \(y\))},
\]
acts as a form of \emph{frequency modulation} (FM): fluctuations in one population’s input \(\delta I_x\) directly tune the effective coupling through \(\omega_{xy}\,u_y\).  Since coupling strengths determine the natural frequency of oscillatory interactions, a small change in \(\delta I_x\) shifts the instantaneous frequency at which \(x\) and \(y\) exchange activity.

Physiologically, this FM-like mechanism allows one ``mass'' (population \(x\)) to broadcast information by modulating the oscillatory frequency of another ``mass'' (\(y\)).  A receiving population that is most sensitive at a particular frequency will selectively respond when the sender drives the shared sigmoid nonlinearity into a regime where \(S''(h_{x0})\,\delta I_x\) shifts the downstream frequency into that band.  In effect:
\begin{itemize}
  \item \textbf{Sender (population \(x\))}\enspace selects a desired frequency by adjusting \(\delta I_x\).  Because \(S''(h_{x0})\,\delta I_x\) scales the coupling coefficient \(\omega_{xy}\,u_y\), the instantaneous ``spring constant'' of the \(y\)–oscillator is tuned.
  \item \textbf{Receiver (population \(y\))}\enspace is predisposed to respond when its own input \(\delta I_y\) or baseline drive \(h_{y0}\) places it near that modulated frequency.  Thus, only those populations whose resonance matches the sender’s modulation will effectively ``hear'' the broadcast.
\end{itemize}

In summary, the presence of \(S''(\cdot)\) in the linearized dynamics gives rise to rich waveform‐shaping capabilities: a small change in one population’s input causes a proportional shift in the effective coupling to the other population, which in turn alters its oscillation frequency.  This FM‐style interaction can be exploited in networks to parcel information into distinct frequency channels, ensuring that only suitably tuned downstream circuits decode the message.

\clearpage

\section{From Wilson-Cowan to Stuart-Landau}
\label{app:WILCO2SL}
The Wilson-Cowan equations describe the interaction between excitatory ($E$) and inhibitory ($I$) populations,
\begin{align}
\tau_E \dot{E} &= -E + S_E(w_{EE}E - w_{EI}I + P_E),\\
\tau_I \dot{I} &= -I + S_I(w_{IE}E - w_{II}I + P_I),
\end{align}
where $S_{E,I}$ are sigmoidal firing-rate functions and $P_{E,I}$ represent external inputs. 
At equilibrium $(E^*, I^*)$, small deviations $u = E - E^*$ and $v = I - I^*$ evolve according to
\begin{equation}
\begin{pmatrix}\dot{u}\\[2pt]\dot{v}\end{pmatrix}
= J \begin{pmatrix}u\\[2pt]v\end{pmatrix},
\qquad
J =
\begin{pmatrix}
\dfrac{-1 + S_E' w_{EE}}{\tau_E} & \dfrac{- S_E' w_{EI}}{\tau_E}\\[6pt]
\dfrac{S_I' w_{IE}}{\tau_I} & \dfrac{-1 - S_I' w_{II}}{\tau_I}
\end{pmatrix}.
\end{equation}
The trace $T=\mathrm{Tr}(J)$ and determinant $D=\det(J)$ determine stability.  
A Hopf bifurcation occurs when
\begin{equation}
T = 0, \qquad D > 0,
\end{equation}
so that the eigenvalues are purely imaginary, $\lambda_{\pm} = \pm i\omega_0$, with
\begin{equation}
\omega_0 =
\sqrt{\frac{
(-1 + S_E' w_{EE})(-1 - S_I' w_{II}) + S_E' S_I' w_{EI} w_{IE}
}{\tau_E \tau_I}}.
\end{equation}
where $\omega_0$ marks the natural oscillation rate of the coupled E-I system. The linear analysis captures only the onset of oscillations. To understand how their amplitude stabilizes, one must include the curvature of the sigmoids. Because $S_{E,I}$ flatten at high input, their local expansion around the fixed point,
\begin{equation}
S_E(x) \simeq S_E^* + S_E'(x-x^*) + \tfrac{1}{2}S_E''(x-x^*)^2 + \tfrac{1}{6}S_E'''(x-x^*)^3,
\end{equation}
reveals that $S''' < 0$ produces a negative cubic nonlinearity.   As activity grows, the effective gain drops, providing nonlinear damping that prevents runaway oscillations.   This saturation is the key mechanism that limits amplitude after the Hopf bifurcation.

Close to the bifurcation point, the two-dimensional dynamics can be expressed more simply by moving to a complex coordinate
\begin{equation}
z = x + i y = M\begin{pmatrix}E - E^* \\ I - I^*\end{pmatrix},
\end{equation}
where $M$ is formed from the eigenvectors of $J$.  
In these coordinates, the system reduces to the canonical Stuart-Landau form,
\begin{equation}
\dot{z} = (\mu + i\omega_0)z - (a + ib)|z|^2z,
\end{equation}
which captures the slow modulation of amplitude and phase near the Hopf point.  
The parameter $\mu$ measures the distance from the bifurcation,
\begin{equation}
\mu = \tfrac{1}{2}\left.\frac{dT}{dp}\right|_{p_c}(p - p_c),
\end{equation}
for some control parameter $p$, such as $P_E$.  
Here $\omega_0$ is the linear oscillation frequency, $a>0$ arises from the negative curvature of the sigmoids ($S'''<0$), and $b$ accounts for a weak amplitude-dependent frequency shift.

Writing $z = r e^{i\phi}$ gives
\begin{align}
\dot{r} &= \mu r - a r^3,\\
\dot{\phi} &= \omega_0 - b r^2.
\end{align}
For small amplitudes, the linear term $\mu r$ drives growth; for large amplitudes, the cubic term $-a r^3$ dominates, stabilizing oscillations at
\begin{equation}
r_{\mathrm{ss}} = \sqrt{\mu/a}.
\end{equation}
Thus, the cubic nonlinearity captures the biological self-limiting effect of population saturation: as excitation and inhibition rise, firing rates approach their ceiling, reducing effective gain and fixing the oscillation amplitude.

In the real $(x,y)$ plane, the same dynamics can be written as
\begin{align}
\dot{x} &= \mu x - \omega_0 y - a r^2 x + \eta_x,\\
\dot{y} &= \omega_0 x + \mu y - a r^2 y + \eta_y,
\end{align}
where $(\eta_x, \eta_y)$ represent small fluctuations in the excitatory and inhibitory drives.
Each term can be interpreted in the underlying E-I dynamics:

\begin{center}
\renewcommand{\arraystretch}{1.2}
\begin{tabularx}{0.95\linewidth}{lX}
\toprule
\textbf{Term} & \textbf{Interpretation in E--I dynamics}\\
\midrule
$(\mu x, \mu y)$ & Bifurcation control: distance from Hopf; depends on gains, weights, or external drive.\\
$(-r^2x, -r^2y)$ & Nonlinear saturation: reflects sigmoid flattening; prevents unbounded growth of activity.\\
$(-\omega_0 y, +\omega_0 x)$ & Rotational coupling: captures the $90^\circ$ lag between excitation and inhibition (E drives I, I suppresses E).\\
$(\eta_x, \eta_y)$ & Noise inputs: fluctuations in excitatory and inhibitory drives.\\
\bottomrule
\end{tabularx}
\end{center}

In this reduced picture, the coordinates $x$ and $y$ correspond to rotated versions of the excitatory and inhibitory deviations. They form the two quadrature phases of the E-I oscillation: $x$ represents the excitatory component, while $y$ lags by roughly $90^\circ$ as the inhibitory counterpart.  
The terms $-r^2x,r^2y$ summarize the saturation of the firing-rate nonlinearity, whereas the rotational term $(-\omega_0 y, \omega_0 x)$ embodies the mutual E-I feedback that sustains rhythmic exchange between excitation and inhibition.


\clearpage \section{Stuart–Landau Oscillator: Parameters and Geometry}

Consider the Stuart–Landau normal form for a supercritical Hopf bifurcation:
\[
\dot z \;=\; (\mu + i\,\omega)\,z \;-\; (g + i\,\beta)\,\lvert z\rvert^2\,z,
\qquad z(t)\in\mathbb{C}.
\]
Below we summarize the role of each parameter and then clarify the geometric effect (or lack thereof) of \(\beta\). The parameters are:

 \begin{itemize}
  \item \(\mu\) (linear growth rate / distance from bifurcation).  
    \begin{itemize}
      \item If \(\mu<0\), the fixed point \(z=0\) is stable (all small perturbations decay).  
      \item At \(\mu=0\), a Hopf bifurcation occurs.  
      \item For \(\mu>0\), the origin is unstable and a limit cycle of amplitude \(r_{\infty}=\sqrt{\mu/g}\) emerges.  
      \item Thus, \(\mu\) measures how far one is past the Hopf threshold: \(\lvert\mu\rvert\) is the distance in parameter space.
    \end{itemize}

  \item \(\omega\) (linear frequency).  
    \begin{itemize}
      \item The term \(i\,\omega\,z\) induces a rotation of small perturbations at angular speed \(\omega\).  
      \item Even if \(\mu<0\), any decaying oscillation spins at frequency \(\omega\).
    \end{itemize}

  \item \(g\) (nonlinear damping / saturation).  
    \begin{itemize}
      \item Without saturation, \(\mu\,z\), $\mu>0$, would cause unbounded amplitude growth.  
      The term \(-\,g\,\lvert z\rvert^2\,z\) provides cubic damping in amplitude:
      \[
      \dot r = \mu\,r - g\,r^3,\quad r = \lvert z\rvert.
      \]
      For \(g>0\), a stable limit cycle of radius 
      \(\displaystyle r_{\infty} = \sqrt{\frac{\mu}{\,g\,}}\) 
      appears when \(\mu>0\).
    \end{itemize}

  \item \(\beta\) (nonlinear frequency shift / phase–amplitude coupling).  
    \begin{itemize}
      \item The term \(-\,i\,\beta\,\lvert z\rvert^2\,z\) means that as \(r=\lvert z\rvert\) grows, the instantaneous angular velocity becomes
      \[
      \dot\theta = \omega \;-\; \beta\,r^2.
      \]
        On the limit cycle \(r = r_{\infty} = \sqrt{\mu/g}\), the asymptotic frequency is
      \(\displaystyle \omega_{\rm LC} = \omega - \beta\,{\mu}/{\,g\,}.\)
     Thus, \(\beta\) governs how amplitude modulates the phase velocity (``shear'' or ``twist''), but does not directly affect the amplitude equation.
    \end{itemize}
\end{itemize}

\subsection{Geometry }

The limit‐cycle solution for \(\mu>0,\;g>0\) is
\[
z_{\infty}(t)
= r_{\infty}\,\exp\!\Bigl[i\bigl(\omega\,t + \phi_{0}\bigr)\Bigr]
\;\exp\!\Bigl[-\,i\,\beta\,r_{\infty}^2\,t\Bigr]
= r_{\infty}\,\exp\!\bigl[i\bigl(\omega - \beta\,r_{\infty}^2\bigr)\,t + i\phi_{0}\bigr],
\]
with constant amplitude \(r_{\infty}=\sqrt{\mu/g}\).  In Cartesian form,
\[
x(t) = \Re\bigl[z_{\infty}(t)\bigr] = r_{\infty}\cos\bigl((\omega - \beta\,r_{\infty}^2)\,t + \phi_{0}\bigr),
\quad
y(t) = \Im\bigl[z_{\infty}(t)\bigr] = r_{\infty}\sin\bigl((\omega - \beta\,r_{\infty}^2)\,t + \phi_{0}\bigr).
\]
Hence
\[
x(t)^2 + y(t)^2 \;=\; r_{\infty}^2 \quad\text{for all }t,
\]
so the trajectory in the \((x,y)\)‐plane is a \emph{perfect circle} of radius \(r_{\infty}\).  The parameter \(\beta\) appears only in the phase:
\[
\dot r = \mu\,r - g\,r^3, 
\qquad
\dot\theta = \omega - \beta\,r^2.
\]
Since \(\beta\) does not enter \(\dot r\), the equilibrium amplitude \(r_{\infty}\) is independent of \(\beta\).  Therefore, \(\beta\) does \emph{not} distort the circular shape into an ellipse.  Instead, \(\beta\) shifts the frequency of rotation once the amplitude has saturated.  Concretely, a nonzero \(\beta\) produces an \emph{amplitude‐dependent} frequency: as \(r\) grows, \(\dot\theta\) decreases by \(\beta\,r^2\).  
 On the limit cycle \(r=r_{\infty}\), the constant frequency is \(\omega - \beta\,(\mu/g)\). Finally, the \((x,y)\)‐trajectory remains a uniform circular orbit at this shifted frequency; there is no ellipticity.

  In conclusion, while \(\mu\) sets the radius of the circle and \(\omega\) (together with \(\beta\)) fixes its angular speed, only the pair \((\mu,g)\) determines the geometric shape (the radius) of the limit cycle.  The parameter \(\beta\) influences {\em when} around the circle the oscillator moves (phase), but not {\em how} it traces out space (shape).

\subsection{Parameter Redundancy and Scaling in the Stuart–Landau Normal Form}

 Must all four parameters appear, or can some be scaled away in the normal‐form reduction to simplify analysis of the dynamics?  

\subsubsection{Linear part: \(\mu\) and \(\omega\)}

Near the Hopf bifurcation of a real dynamical system, one obtains a conjugate‐pair of eigenvalues 
\(\lambda_{1,2}(\nu) = \alpha(\nu) \pm i\,\Omega(\nu)\), 
where \(\nu\) is the original system’s bifurcation parameter.  By definition, at the bifurcation point \(\nu=\nu_c\), \(\alpha(\nu_c)=0\) and \(\Omega(\nu_c)\neq0\).  In the SL reduction  \(\mu\) is chosen so that \(\mu(\nu_c)=0\) and \(\mu\approx \alpha'(\nu_c)\,(\nu-\nu_c)\) measures the distance from the Hopf point, and 
 \(\omega\) is (to leading order) the Hopf frequency \(\Omega(\nu_c)\).  

Because the \emph{fixed‐time‐units} normal form must preserve the linear spectral center (i.e.\ the imaginary part at the bifurcation), both \(\mu\) and \(\omega\) are \emph{in general} essential parameters.  However, one can make a \emph{rotating‐frame} transformation 
\[
z(t) \;=\; u(t)\,e^{\,i\,\omega\,t}
\]
to eliminate \(\omega\) entirely if one is only interested in autonomous amplitude dynamics.  Under that change, 
\(\dot z = e^{i\omega t}(\dot u + i\,\omega\,u)\) 
yields an equation for \(u\) with linear part \((\mu + i\,\omega)u - i\,\omega\,u = \mu\,u\).  In other words, 
\[
\dot u 
= \mu\,u \;-\; (g + i\,\beta)\,\lvert u\rvert^2\,u,
\]
so \(\omega\) \emph{drops out}.  Of course, if one cares about the absolute phase or wants to study phase interactions with external forcing, it may be convenient to keep \(\omega\).  

\subsubsection{Nonlinear part: \(\,g\) and \(\beta\)}

The cubic coefficient in the SL normal form appears as the complex constant \((g + i\,\beta)\).  In principle, one can also nondimensionalize time and rescale \(z\) to remove one additional parameter.  For example, define new units
\[
t' = g\,t, 
\qquad 
u(t') = \sqrt{\frac{g}{\mu_0}}\,z(t), 
\]
where \(\mu_0\) is some reference scale (e.g.\ \(\mu_0=1\)).  In these units, the equation becomes
\[
\frac{d u}{d t'} 
= \frac{\dot z}{g} 
= \frac{1}{g}\Bigl[(\mu + i\,\omega)\,z - (g + i\,\beta)\,\lvert z\rvert^2\,z\Bigr]. 
\]
Writing \(\mu = \tilde\mu\,g\) and \(\omega = \tilde\omega\,g\), and using \(z = \sqrt{\mu_0/g}\,u\), one finds
\[
\frac{d u}{d t'} 
= \bigl(\tilde\mu + i\,\tilde\omega\bigr)\,u 
\;-\; \Bigl(1 + i\,\tilde\beta\Bigr)\,\lvert u\rvert^2\,u,
\]
where \(\tilde\beta = \beta/g\).  In this scaled form:
\begin{itemize}
    \item The linear growth parameter becomes \(\tilde\mu=\mu/g\)
    \item The frequency becomes \(\tilde\omega=\omega/g\)
    \item The nonlinear amplitude coefficient is now unity.
    \item The phase‐amplitude coupling remains as the single ratio \(\tilde\beta = \beta/g\). 
\end{itemize}
Thus, by an appropriate choice of time and amplitude scaling, one \emph{can} reduce the SL normal form to
\[
\dot u 
= (\tilde\mu + i\,\tilde\omega)\,u 
\;-\; (1 + i\,\tilde\beta)\,\lvert u\rvert^2\,u,
\]
with only three \emph{essential} real parameters \(\tilde\mu,\;\tilde\omega,\;\tilde\beta\).  In many analyses, using the transformation described above, one chooses the rotating frame to set \(\tilde\omega=0\), leaving 
\[
\dot u 
= \tilde\mu\,u - (1 + i\,\tilde\beta)\,\lvert u\rvert^2\,u,
\]
with just two parameters \(\tilde\mu\) and \(\tilde\beta\).  In that minimal form,  \(\tilde\mu\) controls the distance from bifurcation (radial growth), while   
 \(\tilde\beta\) controls the amount of nonlinear frequency correction (phase–amplitude coupling).  

\subsubsection{Summary: Which Parameters Are ``Necessary''?}

\begin{itemize}
  \item At the \emph{level of the full, original SL equation}, one typically lists four real parameters \(\mu,\;\omega,\;g,\;\beta\).  
  \item \emph{By scaling amplitude} (i.e.\ set \(g=1\) in new units), the nonlinear damping coefficient is removed, leaving only the ratio \(\beta/g\) as the relevant nonlinear parameter.  
  \item \emph{By moving to a rotating frame}, the linear frequency \(\omega\) can be subtracted off, so the form no longer explicitly contains \(\omega\).  
  \item Consequently, the \emph{minimal} normal form that still captures amplitude growth and phase–amplitude coupling has only \(\tilde\mu\) and \(\tilde\beta\).  

  \item If one wishes to retain \emph{physical units} (time in seconds, amplitude in volts, etc.), then \(\omega\) may remain as the observable oscillation frequency, and \(g\) remains the precise cubic coefficient.  However, any analysis of scaling laws or bifurcation structure can be conducted in the reduced form with fewer parameters.
\end{itemize}

Therefore, while the \emph{generic derivation} of the SL normal form yields four parameters, \emph{two of them can be eliminated} by choosing convenient time and amplitude scales (and, if desired, a rotating frame).  The remaining parameters for the unfolded Hopf–SL dynamics are the real bifurcation parameter (distance \(\tilde\mu\)) and the dimensionless nonlinear frequency‐shift ratio (\(\tilde\beta\)).   

\subsection{Alternative Form Emphasizing  the Limit‐Cycle Radius}

Here we cast the Stuart-Landau equation as a damped harmonic oscillator with dynamical damping and frequency.
The  equation
\[
\dot z \;=\; (\mu + i\,\omega)\,z \;-\; (g + i\,\beta)\,\lvert z\rvert^2\,z
\]
can be rewritten to make explicit how \(\lvert z\rvert^2\) is driven toward its steady‐state value \(\mu/g\).  Group the real and imaginary parts of the nonlinear coefficient:
\[
\dot z 
\;=\; \bigl[\mu - g\,\lvert z\rvert^2 \;\bigr]\,z 
\;+\; i\,\bigl[\omega - \beta\,\lvert z\rvert^2 \;\bigr]\,z.
\]
Equivalently,
\[
\boxed{
\dot z 
= \bigl(\underbrace{\mu - g\,\lvert z\rvert^2}_{\text{real radial growth/damping}}\bigr)\,z 
\;+\; i\,\bigl(\underbrace{\omega - \beta\,\lvert z\rvert^2}_{\text{instantaneous frequency}}\bigr)\,z
}
\]
or
\[
\boxed{
\dot z 
= \bigl(\underbrace{\mu - g\,r^2}_{\text{real radial growth/damping}}\bigr)\,z 
\;+\; i\,\bigl(\underbrace{\omega - \beta\,r^2}_{\text{instantaneous frequency}}\bigr)\,z
}
\]
or
$$
\boxed{
\dot z = \big( \alpha(r)  + i \omega(r) \big)\, z}
$$
reminiscent of the \textit{damped harmonic oscillator,} (Equation~\ref{eq:damped_complex}) but with dynamical damping and frequency.

The first term acts as a \textbf{feedback controller} of the amplitude, a dynamical damping term keeping it close to the limit cycle radius, where $a(r_\infty)=\mu - g\,r_\infty^2=0$. It can be seen as an \textbf{oscillatory homeostatic mechanism.}

In this form:
\begin{itemize}
  \item The \emph{real part} \(\,\mu - g\,\lvert z\rvert^2\) multiplies \(z\).  Writing \(z = r\,e^{i\theta}\) yields the radial equation
  \[
  \dot r = \bigl(\mu - g\,r^2\bigr)\,r,
  \]
  so that \(\lvert z\rvert = r\) is driven toward 
  \[
  r_{\infty} = \sqrt{\frac{\mu}{\,g\,}}\quad(\mu>0).
  \]
  Thus one sees directly that \(\lvert z\rvert^2\to \mu/g\) as \(t\to\infty\).

  \item The \emph{imaginary part} \(\,\omega - \beta\,\lvert z\rvert^2\) multiplies \(i\,z\).  In polar form this gives
  \[
  \dot\theta = \omega \;-\; \beta\,r^2,
  \]
  so the instantaneous oscillator frequency is shifted by \(\beta\,\lvert z\rvert^2\).  On the limit cycle \(r = r_{\infty}=\sqrt{\mu/g}\), the asymptotic frequency is \(\omega - \beta\,(\mu/g)\).
\end{itemize}

Because the term \(\mu - g\,\lvert z\rvert^2\) vanishes exactly when \(\lvert z\rvert^2 = \mu/g\), it is immediately clear from this form that \(\lvert z\rvert^2\) must approach \(\mu/g\) in order for the radial growth \(\dot r\) to vanish.  Hence the limit‐cycle amplitude emerges naturally as \(\lvert z\rvert \to \sqrt{\mu/g}\).

\subsection{Stuart–Landau in Push/Pull Form}

Starting from
\[
\dot z \;=\; (\mu + i\,\omega)\,z \;-\; (g + i\,\beta)\,\lvert z\rvert^2\,z,
\qquad z = x + i\,y,\quad r^2 = x^2 + y^2,
\]
the Cartesian form is
\[
\begin{cases}
\dot x \;=\; \mu\,x \;-\;\omega\,y \;-\; g\,r^2\,x \;+\; \beta\,r^2\,y,\\[0.5em]
\dot y \;=\; \mu\,y \;+\;\omega\,x \;-\; g\,r^2\,y \;-\; \beta\,r^2\,x.
\end{cases}
\]
Define the amplitude‐dependent coefficient $\alpha$, a dynamic damping term, and $\Omega$, a dynamic instantaneous frequency,
\[
\alpha(r) \;=\; -\,\mu \;+\; g\,r^2,
\qquad
\Omega(r) \;=\; \omega \;-\; \beta\,r^2.
\]
Introduce the nonlinear differential operator
\[
L[f] \;:=\; \left[ \frac{d}{dt} \;+\; \alpha(r)\right] f 
\;=\; \dot f \;-(\;\mu - g\,r^2)\,f,
\]
where \(r^2 = x^2 + y^2\).  Then the Stuart–Landau equations can be compactly written as
\[
\boxed{
\begin{aligned}
L[x] &= -\,\Omega(r)\,y,\\
L[y] &= +\,\Omega(r)\,x
\end{aligned}
}
\quad
\]
In this form, all growth (\(\mu\)), saturation (\(g\,r^2\)), and nonlinear phase effects (\(\beta\,r^2\)) are absorbed into the operator \(L\) as a dynamic damping constant, while the right‐hand side retains a ``pure'' rotation at instantaneous frequency \(\Omega(r)\).

When the limit cycle is reached, $a=0$,  we are back at the undamped harmonic oscillator.


\subsection{DC-Shifted Formulation of the Oscillator} \label{app:shiftedoscillators}

In modeling oscillatory neural dynamics, one often needs to account for a tonic bias or constant drive, or more generally constant or very slowly changing (compared to the natural timescale of the system) ``forcing". In fact, the signals generated by the model oftentimes need to be positive quantities (e.g., firing rates), or at least not centered at zero (membrane potential perturbations). 
What is the natural way to do this in the HO and SL cases?

{\bf Harmonic Oscillator case.} Consider the damped, unforced oscillator in complex form:
\[
\dot z = (a + i\omega)\,z, \qquad a\in\mathbb{R},\ \omega>0.
\]
This equation gives solutions centered at zero, which is undesirable if we want to relate them with firing rate models or membrane potentials. E.g., a constant electric field produces a DC shift in membrane potential or firing rate in more realistic models. How can this simple model accommodate it? 

A simple additive term provides the desired solution behavior (a DC shift).
With a bias $c$, the equation becomes  
\begin{equation}\label{eq:HO_DCshifted}
\dot z = (a + i\omega)\,z +c.
\end{equation}

\begin{figure}
     \centering
          \includegraphics[width=0.48\linewidth]{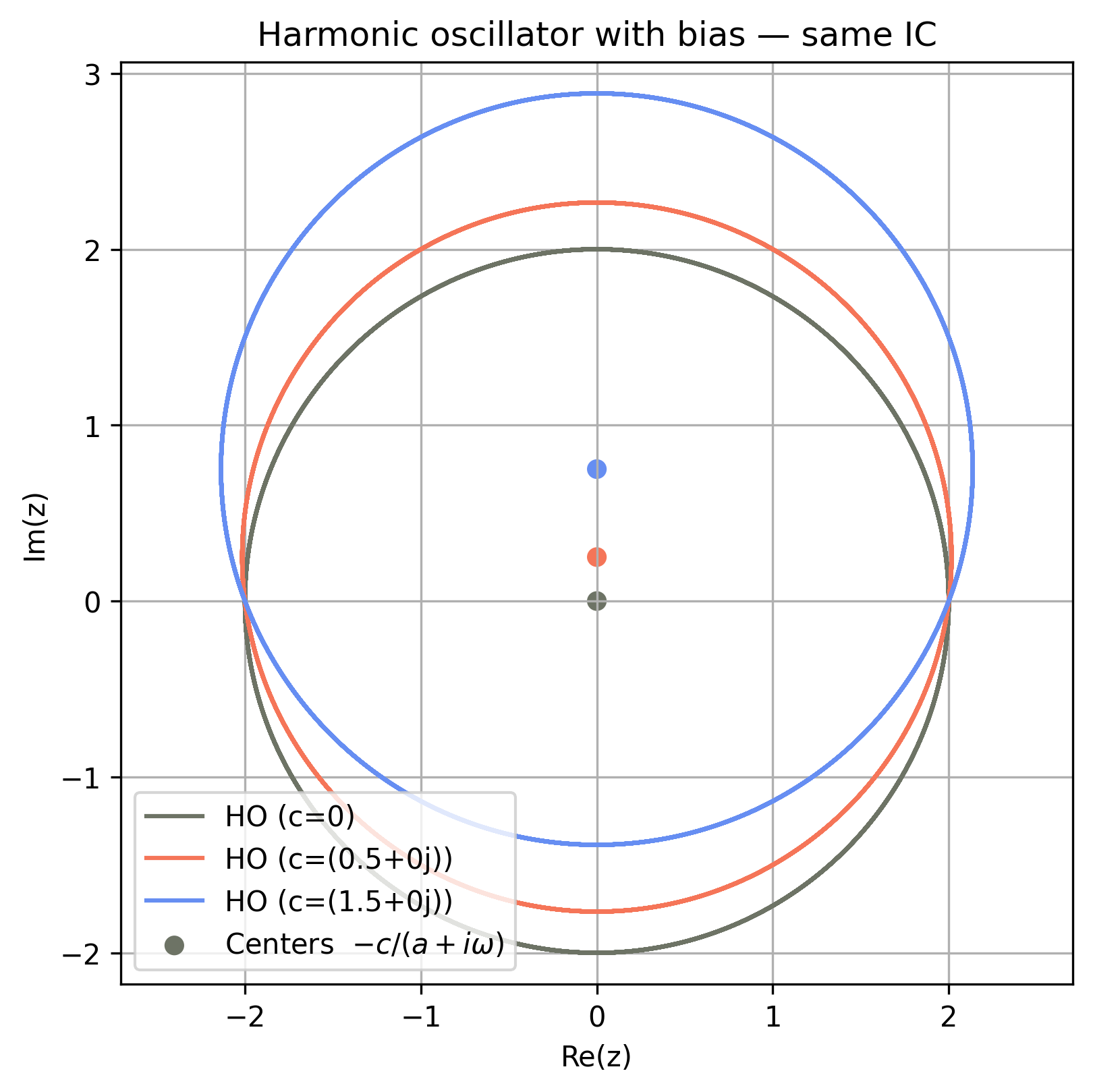}
         \includegraphics[width=0.48\linewidth]{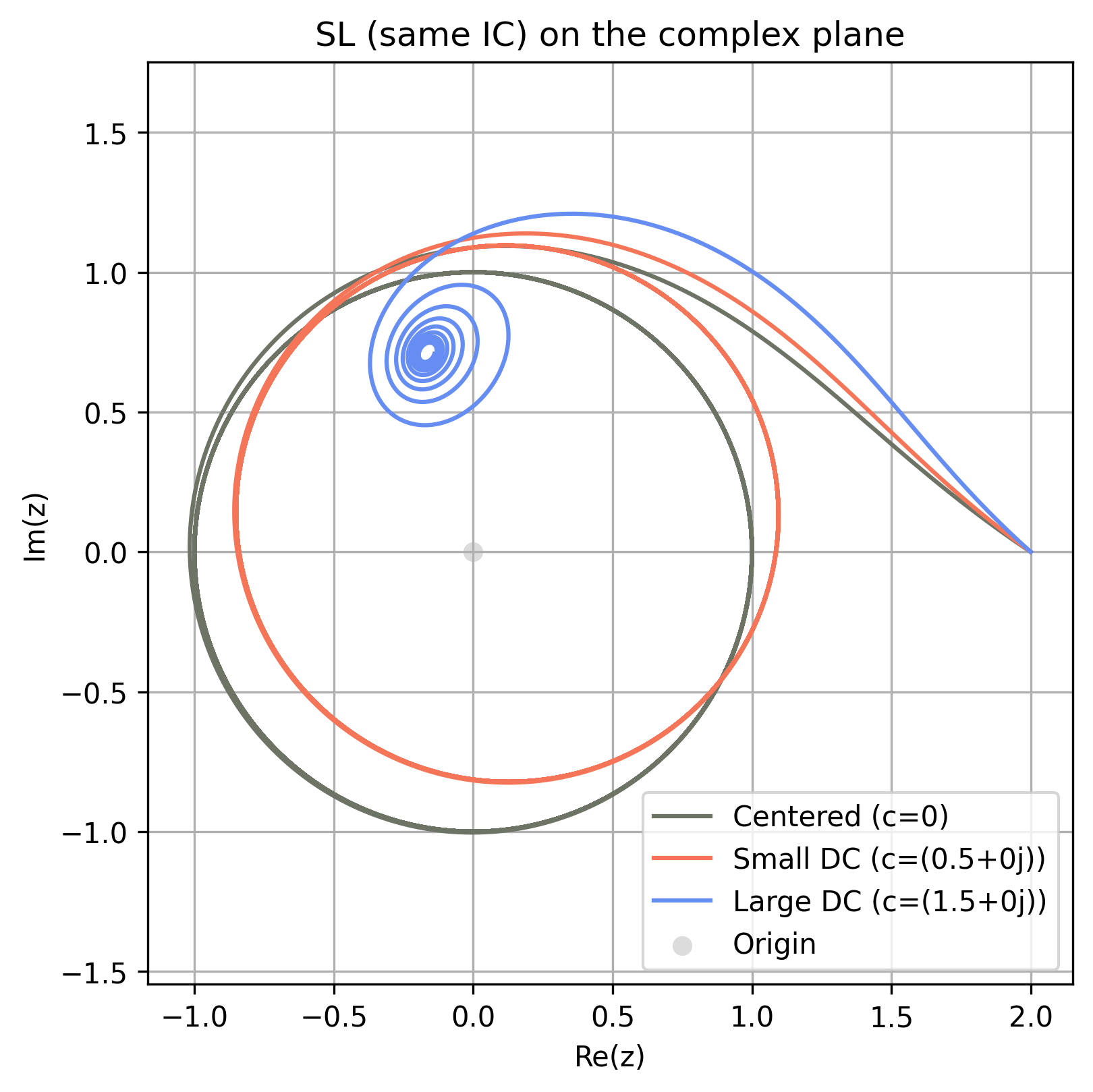}

     \includegraphics[width=1\linewidth]{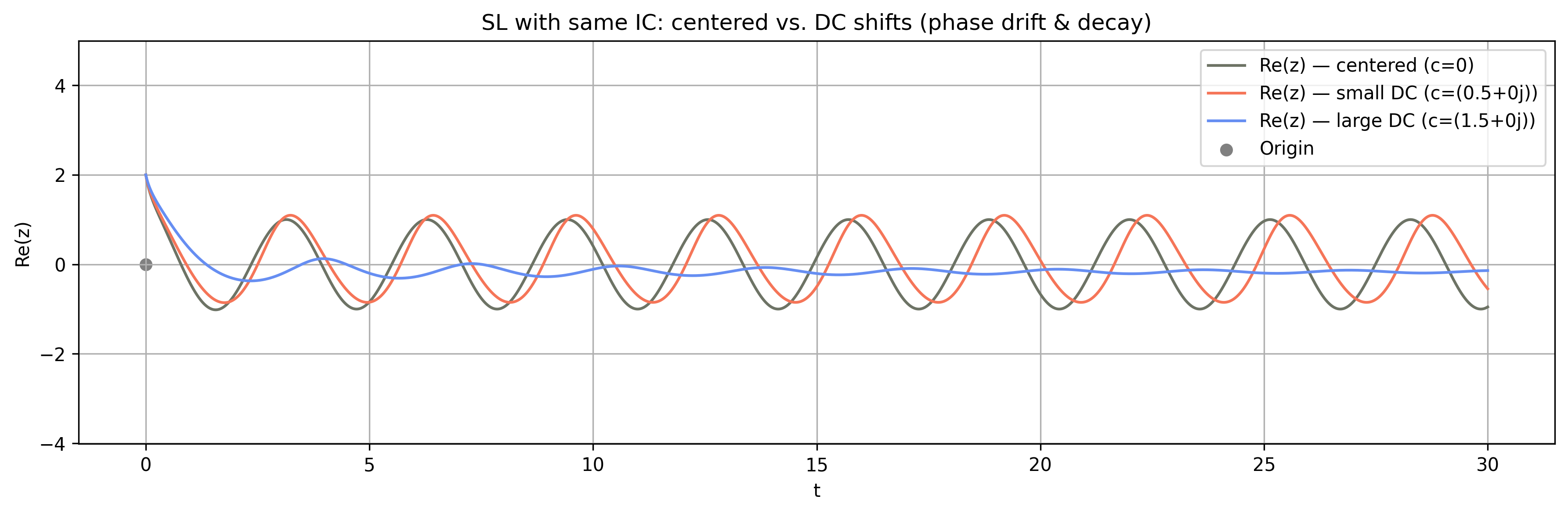}
     \caption{\textbf{Top: Harmonic Oscillator (HO) and Stuart–Landau (SL) with a bias: phase portraits.} Simulations display the effects of a small or larger DC bias offset, which induce frequency and amplitude changes, or destroy the limit cycle in SL while producing simple displacements in HO.
We integrate $\dot z=(\mu+i\omega)z-(g+i\beta)\lvert z\rvert^2 z + c$ with
$\mu=1$, $g=1$, $\omega=2$, $\beta=0$, using RK4 ($\Delta t=0.01$).
Cases: (i) centered SL ($c=0$), (ii) small DC shift ($c=0.5+0i$) yielding a displaced
limit cycle, (iii) large DC shift ($c=2+0i$) leading to convergence to the fixed point. Initial condition for all runs: $z_0=2R_0$ with $R_0=\sqrt{\mu/g}=1$.
Trajectories shown after transients to emphasize the displaced cycle and the decay. \textbf{Bottom: SL time-series overlay.} The three cases are shown, highlighting changes in frequency and amplitude as well as extinction. 
}

     \label{fig:SLtimeseries}
 \end{figure}

Introduce a constant shift \(C\in\mathbb{C}\) and define the change of variables (a translation)
\[
u = z + C.
\]
Then, the equation for $u$ is
\[
\dot u = \dot z
        = (a + i\omega)\,(u - C) +c
        = (a + i\omega)\,u \;-\; (a + i\omega)\,C +c.
\]
The last term is just a complex constant and we can set it to zero with
$$
C= \frac{c}{a+i\omega}
$$
so that $\dot u =(a+i\omega) u$  --- we recover the harmonic oscillator equation in the new coordinates. Hence, the generalized, shifted harmonic oscillator in Equation~\ref{eq:HO_DCshifted}
is a translation of the harmonic oscillator with a center at  $-C = -c/(a+i\omega)$ --- see Figure~\ref{fig:SLtimeseries} (top left) for examples. 
The explicit solution is
\[
z(t) = \Bigl(z_0 + \frac{c}{a+i\omega}\Bigr)e^{(a+i\omega)t}
       - \frac{c}{a+i\omega}.
\]
Hence, for \(a=0\), the motion is a uniform circular orbit around 
the displaced center \( -\,c/(i\omega)\); for \(a<0\), trajectories form logarithmic spirals 
converging to the point \( -\,c/(a+i\omega)\).

{\bf Stuart-Landau.}  Similarly to the HO case, a naive way to do so is to add a constant term \(c\in\mathbb{C}\) to the Stuart–Landau (SL) equation:
\[
\dot z = (\mu + i\omega)\,z \;-\;(g + i\beta)\,|z|^2z \;+\; c.
\]
This \emph{DC–shifted SL} now has
\begin{itemize}
  \item a limit cycle displaced away from the origin,
  \item broken $U(1)$ symmetry,
  \item and modified amplitude/frequency balance.
\end{itemize}

\smallskip
  To recover a zero‐mean description, one sets
\[
z = v + z_0,
\quad\text{where }z_0\text{ solves }(\mu + i\omega)z_0 - (g + i\beta)|z_0|^2z_0 + c = 0.
\]
Substitution gives
\[
\dot v
= (\mu + i\omega)\,v
- (g + i\beta)\,\bigl|v + z_0\bigr|^2\,(v + z_0)
\;+\;\underbrace{(\mu + i\omega)z_0 + c - (g + i\beta)|z_0|^2z_0}_{=0},
\]
so that the constant offset disappears.  However, the \emph{new} equation for \(v\) contains extra quadratic and linear terms arising from the expansion of \(\lvert v+z_0\rvert^2(v+z_0)\); it is no longer in the simple SL form.

\smallskip 
Despite this complication, \emph{normal form theory} guarantees that any smooth perturbation of a Hopf normal form can be brought back—through a further smooth, near‐identity change of variables—into the canonical Stuart–Landau structure, up to higher–order corrections. Le\'on and Nakao (2023) \cite{leonAnalyticalPhaseReduction2023} provide expressions for the frequency and amplitude shifts up to second order in DC shift.

More generally,
\[
v = u + H(u,\bar u),
\]
for a suitable cubic function \(H\), eliminates all non‐resonant quadratic and shifted cubic terms, leaving
\[
\boxed{
\dot u = (\tilde\mu + i\tilde\omega)\,u - (\tilde g + i\tilde\beta)\,\lvert u\rvert^2u
}
\]
to leading order.  The new parameters \(\tilde\mu,\tilde\omega,\tilde g,\tilde\beta\) absorb the effects of the original bias \(c\) and shift \(z_0\).  Thus, \emph{even with a DC offset}, the local oscillatory dynamics near the Hopf bifurcation remain of Stuart–Landau type: a self-saturating limit cycle with phase–neutral drift or fixed point (depending on parameters).

However, introducing a constant DC bias to the Stuart–Landau oscillator modifies its equilibrium position, breaks the symmetry, and shifts the critical Hopf bifurcation threshold, thus altering both amplitude and frequency of oscillations (see Figure~\ref{fig:SLtimeseries} for an example). Small biases result in second-order reductions in oscillation amplitude and frequency shifts, whereas sufficiently large DC offsets can completely eliminate the limit cycle. Analytically, this can be described as an imperfect Hopf bifurcation: the effective bifurcation parameter ($\mu$) is renormalized by the DC term, requiring a higher original parameter value to sustain oscillations. Hence, persistent oscillations occur only when the system’s gain overcomes the bias-induced suppression; otherwise, the oscillator settles into a stable equilibrium without oscillations.

\clearpage 

\section{Linear Operators, Green--Laplace Tools, and E-I Oscillations: a Pedagogical View}
\label{app:L-op}

Linear differential operators appear throughout neural modeling (synaptic kinetics, population filters, dendritic cable reductions). A compact way to see why they behave like filters is to write
\[
L[x](t)=u(t),\qquad L=\sum_{k=0}^n a_k\,\partial_t^k,
\]
and study two canonical objects. The \emph{homogeneous} solution \(L[x_h]=0\) reveals the system's natural modes, while the \emph{impulse response} \(h\) solves \(L[h]=\delta\) with \(h(t)=0\) for \(t<0\) and encodes the system's causal memory. If the characteristic polynomial \(p(\lambda)=\sum_{k=0}^n a_k\lambda^k\) factors as \(\prod_i(\lambda-\lambda_i)\) with distinct \(\lambda_i\), then \(x_h(t)=\sum_i c_i e^{\lambda_i t}\). The corresponding \(h(t)\) is also a linear combination of the same exponentials for \(t>0\), pinned by standard continuity conditions at \(t=0\) (all derivatives up to order \(n-2\) are continuous; \(x^{(n-1)}\) jumps by \(1/a_n\))  \cite{StakgoldHolst2011}.

\subsection{Two complementary tools: Laplace and Green}
\paragraph{Laplace viewpoint.}
With zero initial conditions, \(\mathcal{L}\{L[x]\}(s)=P(s)X(s)\) where \(P(s)=\sum_k a_k s^k\). The \emph{transfer function} is \(H(s)=X(s)/U(s)=1/P(s)\). Poles of \(H\) set decay rates, oscillation frequencies, and phase lag. Evaluating on the imaginary axis, \(H(j\omega)\), gives magnitude and phase; the \emph{group delay} is \(\tau_g(\omega)=-\frac{d}{d\omega}\arg H(j\omega)\).

\paragraph{Green (time‑domain) viewpoint.}
Equivalently, the causal Green's function \(G(t,t_0)\) satisfies \(L[G(\cdot,t_0)]=\delta(\cdot-t_0)\) with \(G=0\) for \(t<t_0\). For any input \(u\),
\[
x(t)=\int_{-\infty}^t G(t,t_0)\,u(t_0)\,dt_0=(h*u)(t),\qquad h(t)=G(t,0).
\]
Laplace uses exponentials \(e^{st}\) (an eigenbasis of \(\partial_t\)) to expose poles and phase directly; Green uses time‑localized impulses \(\delta(t-t_0)\) to show how past inputs are weighted and delayed  \cite{StakgoldHolst2011}. Both are the same mathematics seen from two bases.

\subsubsection*{A gentle taxonomy by order (with worked examples)}
For clarity we use \(L=a\,\partial_t+b\) for first order and \(L=m\,\partial_t^2+a\,\partial_t+b\) for second order, with \(m>0\), \(a\ge0\), \(b>0\). When convenient we switch to \(\omega_n\) and \(\zeta\) via \(b/m=\omega_n^2\) and \(a/m=2\zeta\omega_n\).

\paragraph{Zeroth order: memoryless gain.}
\(b\,y=f\Rightarrow h(t)=\frac{1}{b}\delta(t)\). There is no temporal memory.

\paragraph{First order: leaky integrator (and integrator limit).}
\(a\,\dot y+b\,y=f\) has \(h(t)=\frac{1}{a}e^{-(b/a)t}H(t)\). The time constant is \(\tau=a/b\). In Laplace, \(H(s)=1/(a s+b)=(1/b)\,1/(1+s\tau)\), so \(\tau_g(\omega)=\tau/(1+(\omega\tau)^2)\). If \(b=0\), \(h(t)=\frac{1}{a}H(t)\) and \(H(s)=1/(a s)\): an ideal integrator with perfect memory.

\paragraph{Second order: rise--decay, critical, underdamped, and negative damping.}
With \(L=m\,\ddot y+a\,\dot y+b\,y\), the roots
\[
r_{1,2}=\frac{-a\pm\sqrt{a^2-4mb}}{2m}
\]
organize all cases. Overdamped (\(a^2>4mb\)): \(h(t)=\dfrac{e^{r_1 t}-e^{r_2 t}}{m(r_1-r_2)}H(t)\), a difference of decays with a delayed peak. Critical (\(a^2=4mb\)): \(h(t)=(1/m)\,t\,e^{-\frac{a}{2m}t}H(t)\) (the alpha form). Underdamped (\(a^2<4mb\)): writing \(\alpha=\frac{a}{2m}\) and \(\omega_d=\frac{\sqrt{4mb-a^2}}{2m}\),
\[
h(t)=\frac{1}{m\omega_d}e^{-\alpha t}\sin(\omega_d t)\,H(t),
\]
an exponentially damped sinusoid. If \(a<0\) (negative friction) the real part of the poles is positive and the response grows.

\subsection{The synapse as a \emph{forced harmonic oscillator}}
It is perhaps more intuitive to understand synaptic delay and inertia by \emph{recognizing the operator} \(L=m\,\partial_t^2+a\,\partial_t+b\) as the mass-spring-damper driven by a force \(f(t)\):
\[
m\,\ddot y(t)+a\,\dot y(t)+b\,y(t)=f(t).
\]
Apply an impulse \(f=\delta\). The mass first acquires velocity, not displacement; energy shuttles between kinetic and spring energy, and damping removes energy. This automatically produces a \emph{delayed} peak in \(y\): the output must build up after the impulse. The same reasoning holds for the electrical RLC analog. In neural mass models, \(y\) is a postsynaptic potential and \(f\) is the presynaptic drive; the ``mass'' \(m\) summarizes effective inertial storage across coupled first‑order elements, while \(a\) and \(b\) summarize leak and restoring tendencies.

\paragraph{Worked example (critical alpha).}
Choose \(h(t)=A\,a\,t\,e^{-a t}H(t)\). Then \(\mathcal{L}\{h\}(s)=A a/(s+a)^2\) and
\[
(s+a)^2 Y(s)=A a\,\Sigma(s)\quad\Longleftrightarrow\quad (\partial_t+a)^2 y(t)=A a\,\sigma(t).
\]
Thus the alpha kernel \emph{is} the impulse response of a critically damped oscillator driven by the presynaptic input. Its delayed peak illustrates a causal, buffer‑like delay with smoothing, not a noncausal time shift.

\paragraph{Peak time and inertia: how mass slows the response.}
Factor \(L=m(\partial_t+a_r)(\partial_t+a_d)\) with \(a_r+a_d=a/m\) and \(a_r a_d=b/m\). In the overdamped case, \(h(t)=\frac{e^{-a_d t}-e^{-a_r t}}{m(a_r-a_d)}H(t)\) peaks at
\[
t_{\mathrm{peak}}=\frac{1}{a_r-a_d}\ln\!\frac{a_r}{a_d}.
\]
At the double pole \(a_r=a_d=a/(2m)\) one gets \(h(t)=(1/m)\,t\,e^{-\frac{a}{2m}t}\) with \(t_{\mathrm{peak}}=2m/a\). In the underdamped case,
\[
h(t)=\frac{1}{m\omega_d}e^{-\alpha t}\sin(\omega_d t)H(t),\qquad
t_{\mathrm{peak}}=\frac{1}{\omega_d}\arctan\!\frac{\omega_d}{\alpha},
\]
with \(\alpha=\tfrac{a}{2m}\) and \(\omega_d=\tfrac{\sqrt{4mb-a^2}}{2m}\). For fixed \(a,b>0\) and \(m\to\infty\), \(\alpha\to0\), \(\omega_d\sim \sqrt{b/m}\), and \(\arctan(\omega_d/\alpha)\to\pi/2\), hence
\[
t_{\mathrm{peak}}\sim \frac{\pi}{2}\sqrt{\frac{m}{b}}\to\infty.
\]
Inertia therefore \emph{increases} the causal delay. Importantly, the overdamped log formula must not be used once the poles become complex; the underdamped expression governs the peak.

\subsection{E--I motifs, Barkhausen conditions, and where the phase lag comes from} \label{app:barkhausen}
A pedagogical route to oscillation is to rewrite the undamped oscillator as a pair of coupled first‑order filters. Let \(z=x+i y\) and consider
\[
\dot z=(a+i\omega)z,\qquad a\ge0,\ \omega>0.
\]
Separating real and imaginary parts gives
\[
\dot x=a x-\omega y,\qquad \dot y=a y+\omega x.
\]
Each equation is a leaky integrator driven by the other in \(90^\circ\) phase. When \(a=0\) the loop produces sustained oscillations; when \(a>0\) the envelope decays as \(e^{-a t}\). This ``push--pull'' view is the simplest template to keep in mind as we turn to E-I populations.

\paragraph{Barkhausen as a phase-gain budget.}
For a feedback loop with transfer \(L(s)\), \textbf{necessary conditions for linear self‑oscillation} at \(\omega_0\) are \(|L(j\omega_0)|=1\) and \(\arg L(j\omega_0)=0^\circ \pmod{360^\circ}\) \cite{vonWangenheim2010}. The phase condition pins \(\omega_0\) by balancing element lags; the magnitude condition pins the product of gains, including the slope \(\kappa\) of the nonlinearity at the bias point. Nonlinear saturation then stabilizes amplitude \cite{AstromMurray2008}.

\paragraph{Wilson--Cowan with first‑order synapses.}
Linearize about a fixed point with slope \(\kappa\) and unit time constants:
\[
\dot{x}=-(1-w_{ee}\kappa)x - w_{ei}\kappa\,y + I_x,\qquad
\dot{y}=-y + w_{ie}\kappa\,x - w_{ii}\kappa\,y.
\]
The E\(\to\)I\(\to\)E loop has
\[
T_{EI}(s)=\frac{w_{ei}w_{ie}\kappa^2}{(s+1)^2},
\]
which can supply up to \(-180^\circ\) of phase---not enough alone to close the loop at a finite \(\omega\). Excitatory self‑coupling adds
\[
T_{EE}(s)=\frac{w_{ee}\kappa}{s+1},
\]
so the characteristic equation \(1-T_{EE}(s)-T_{EI}(s)=0\) can satisfy Barkhausen at some \(\omega_0\). A bias \(I_x\) ensuring \(\kappa>0\) is essential \cite{WilsonCowan1972}. This matches the intuition that two first‑order elements need additional phase (or an explicit transmission delay) to reach \(360^\circ\).

\paragraph{Jansen--Rit with second‑order synapses.}
For the cortical column with second‑order synapses,
\[
\ddot{x}+2a\dot{x}+a^2 x=A a\,\kappa(-w\,y+I_x),\qquad
\ddot{y}+2b\dot{y}+b^2 y=B b\,\kappa(w\,x),
\]
the linearized synapses are band‑pass filters
\[
H_{\mathrm{exc}}(s)=\frac{A a\,\kappa}{(s+a)^2},\qquad
H_{\mathrm{inh}}(s)=\frac{B b\,\kappa}{(s+b)^2}.
\]
The E\(\to\)I\(\to\)E loop transfer
\[
T(s)=-\,w^2\,H_{\mathrm{exc}}(s)\,H_{\mathrm{inh}}(s)
\]
can contribute \(-360^\circ\) of phase on its own, so self‑excitation is \emph{not} required to meet the phase condition. A bias \(I_x\) maintaining \(\kappa>0\) sets \(|T(j\omega_0)|=1\) at the selected frequency  \cite{jansen1995}. Empirically, \(\omega_0\) follows the loop's effective delay, which is controlled by synaptic poles \(a,b\) and any axonal conduction delay.

\paragraph{Information flow and effective loop delay.}
For narrowband loops, each element's phase behaves as \(\varphi_k(\omega)\approx -\omega\,\tau_k\) near resonance, so \(\sum_k \varphi_k(\omega_0)=0^\circ\) implies \(\omega_0 \approx 2\pi n/\tau_{\mathrm{loop}}\), where \(\tau_{\mathrm{loop}}\approx -\sum_k \varphi_k(\omega_0)/\omega_0\). Second‑order synapses provide larger group delay around their passband than single‑pole synapses. This is one reason gamma‑range E-I oscillations arise robustly once the synaptic dynamics are at least second order  \cite{BuzsakiWang2012}.

\paragraph{First-order worked example (frequency response to a sinusoid).}
To fix ideas, solve \(\dot x + a x = e^{j\omega t}\) with \(a>0\). In Laplace,
\[
(s+a)X(s)=\frac{1}{s-j\omega}\quad\Rightarrow\quad
X(s)=\frac{1}{(s+a)(s-j\omega)}.
\]
Partial fractions and inversion give
\[
x(t)=\frac{e^{-a t}}{-(a+j\omega)}+\frac{e^{j\omega t}}{a+j\omega}.
\]
The transient term dies as \(t\to\infty\); the steady state is \(x(t)=H(j\omega)\,e^{j\omega t}\) with \(H(j\omega)=1/(a+j\omega)\). Hence \(|H(j\omega)|=1/\sqrt{a^2+\omega^2}\) and \(\arg H(j\omega)=-\arctan(\omega/a)\), and \(\tau_g(\omega)=a/(a^2+\omega^2)\). This calculation makes explicit how a pole at \(-a\) sets both decay and phase lag.

\paragraph{Time-domain worked example (convolution with an alpha kernel).}
Consider \(h(t)=A a t e^{-a t}H(t)\) and an input \(\sigma(t)\). Then \(y(t)=(h*\sigma)(t)\) and \(\mathcal{L}\{h\}=A a/(s+a)^2\). Multiplication in \(s\)-space becomes
\[
Y(s)=\frac{A a}{(s+a)^2}\,\Sigma(s)\quad\Longleftrightarrow\quad
\left(\partial_t+a\right)^2 y(t)=A a\,\sigma(t),
\]
realizing the second‑order operator directly from the kernel. This is the synaptic analog of driving a critically damped mass-spring-damper.

\subsection*{Pointers for further reading}
A rigorous, operator‑theoretic treatment of \(L[h]=\delta\) and the jump conditions appears in  Stakgold \& Holst (2011)  \cite{StakgoldHolst2011}. For feedback, phase, and the Barkhausen criterion in context see {\AA str\"om \& Murray (2008)} \cite{AstromMurray2008} and the clarification in  vonWangenheim (2010) \cite{vonWangenheim2010}. For neural mass modeling with first‑ and second‑order synapses see  Wilson \& Cowan (1972)  \cite{WilsonCowan1972} and  Jansen \& Rit (1995) \cite{jansen1995}; for conductance‑based synaptic kinetics and canonical PSP shapes see  Destexhe\textit{ et al.} (1994) \cite{destexhe1994}; and for a broader review of E--I mechanisms of cortical rhythms see Buzs\'aki\&Wang (2012)  \cite{BuzsakiWang2012}.

\clearpage
\section{Oscillations, Topology and Simplicity}
\label{app:oscillation}
\noindent
The term \textit{oscillation} is used across mathematics, engineering and neurophysiology, yet each community defines it  differently  (see Table~\ref{tab:osc_types}).  In this section we show how
all views — dynamical systems, spectral tests, and Koopman eigenanalysis share the same core intuition: \textit{an oscillation is present when the data are better explained (or more concisely described) starting from the baseline of circular motion ($S^1$ topology) than by any simpler alternative.} We will formalize this through an algorithmic definition that more or less says "an oscillation is something that essentially goes around a circle".

An intuitive \textbf{classical definition} is the following.

\paragraph{Definition 1.}
\emph{A dynamical variable is said to oscillate when it exhibits sustained, approximately periodic departures around a reference value such that the system returns to a similar state after a characteristic approximate interval $T$ (its period), or, equivalently, at a dominant frequency $f=1/T$.  The repetition can be exact (strictly periodic) or approximate (quasi‑periodic, weakly modulated, or noise‑jittered); what matters is the recognisable cycle pattern.}

This phrasing captures the intuition of ``a signal that roughly repeats'' while making explicit (i) the presence of a characteristic timescale and (ii) tolerance for imperfect cycles.

This practical notion—a pattern that \emph{roughly} repeats—is compatible both with modern spectrum‑based detectors and with the dynamical‑systems idea of an attracting limit cycle.  But it can be generalized to include damped behavior:

\paragraph{Definition 2.}
\emph{An oscillation is a self‑sustained or externally driven fluctuation that revisits comparable states at quasi‑regular intervals, identifiable either as a closed trajectory in state space or as a narrow‑band spectral peak above the aperiodic background.}

This sentence bridges physics, signal processing, and neuroscience, preparing the ground for the algorithmic‑information view developed in the following sections. We revise first the definitions in different sub-fields.

\begin{table}[t!]\small
\centering
\caption{How major fields articulate the notion of oscillation.}
\label{tab:field_views}
\begin{tabular}{p{3.2cm}p{6.5cm}p{4.3cm}} 
\toprule
\textbf{Field} & \textbf{Typical wording} & \textbf{Key nuance} \\
\midrule
Classical physics \& engineering &
``Repetitive or \emph{periodic} variation, typically in time, of a quantity about a central value (often an equilibrium) or between two states.'' \cite{Kreyszig11}  &
Emphasizes small deviations from equilibrium and strictly periodic motion (e.g.\ undamped spring, AC current). \\[4pt]

Non‑linear dynamics / mathematics &
A \emph{periodic orbit} (or \emph{limit cycle}) satisfying $x(t+T)=x(t)$; at least one nearby trajectory spirals into it.\cite{Strogatz1994} &
Formal, coordinate‑free; includes self‑sustained oscillators (Van der Pol, Hodgkin–Huxley) and supports stability analysis. \\[4pt]

Neuroscience &
``Rhythmic or repetitive neural activity observed at all levels of the CNS.''  \cite{Basar13}&
Focus on multi‑scale biological generators; amplitude mainly indexes population synchrony, not a single source. \\[4pt]

Signal processing / spectral view &
``Narrow‑band peak that rises above the aperiodic 1/\emph{f} background in the power spectrum.'' \cite{Donoghue20}&
Detects oscillations without explicit time‑domain periodicity; robust for noisy or burst‑like data. \\
\bottomrule
\end{tabular}
\vspace{-6pt}
\end{table}

\paragraph{Oscillations and the Koopman Operator.}
A \textit{limit cycle} is a closed orbit $\gamma$ of an autonomous ODE to which at least one neighbouring trajectory spirals (stable Floquet multipliers) \cite{Strogatz94}. The system’s intrinsic period $T$ is exact, and, as we explain, next, its Koopman operator possesses an eigenpair $(\psi,\lambda=i\omega)$ whose argument $\theta=\arg\psi$ advances uniformly, embedding the dynamics on the circle $S^{1}$ \cite{Koopman31}.  

Thus,  periodic behavior can be precisely described from the  Koopman operator perspective \cite{Koopman31}. Relaxation oscillators (e.g.\ FitzHugh–Nagumo) fit the same definition but spend long intervals on slow manifolds, producing nonsinusoidal waveforms with rich harmonic content.

This approach fundamentally reframes the analysis. Instead of studying the nonlinear evolution of the system's state vector $\mathbf{x}(t) \in \mathbb{R}^n$ (governed by $\dot{\mathbf{x}} = \mathbf{f}(\mathbf{x})$), the Koopman operator $\mathcal{K}^t$ describes the \textit{linear} evolution of   ``observables'' $g(\mathbf{x})$ (cleverly chosen functions of the state). The operator is defined by how it advances any such function along the system's trajectories: $(\mathcal{K}^t g)(\mathbf{x}_0) = g(\mathbf{x}(t))$, where $\mathbf{x}(t)$ is the flow starting from $\mathbf{x}_0$. The crucial insight is that $\mathcal{K}^t$ is a \textbf{linear operator} acting on the (infinite-dimensional) space of observables, even when the underlying dynamics $\mathbf{f}(\mathbf{x})$ are nonlinear.

Because $\mathcal{K}^t$ is linear, we can use spectral methods. For a limit cycle, the operator's \textit{infinitesimal generator} $L$ (where $\mathcal{K}^t = e^{Lt}$) possesses an eigenpair $(\psi, \lambda=i\omega)$ corresponding to the system's fundamental frequency $\omega = 2\pi/T$. This special observable $\psi$, the \textbf{Koopman eigenfunction}, evolves simply in time: $\psi(\mathbf{x}(t)) = (\mathcal{K}^t \psi)(\mathbf{x}_0) = e^{\lambda t} \psi(\mathbf{x}_0) = e^{i\omega t} \psi(\mathbf{x}_0)$. Consequently, its argument $\theta=\arg\psi$ advances uniformly ($\theta(t) = \theta_0 + \omega t$), embedding the complex, multi-dimensional dynamics onto a simple rotation on the circle $S^{1}$. Relaxation oscillators (e.g.\ FitzHugh–Nagumo) fit the same definition, but their corresponding eigenfunctions $\psi$ are more complex (capturing all the harmonics), resulting in nonsinusoidal waveforms.

It is important to clarify what the eigenfunction $\psi$ represents. It is a special \textbf{observable}, meaning it is a function of the state vector, $\psi(\mathbf{x})$, that maps the $n$-dimensional state space (where the dynamics are nonlinear) to the complex plane $\mathbb{C}$ (where the dynamics are linear). Its special property is that when evaluated \textit{along} a trajectory $\mathbf{x}(t)$, its value evolves with perfect simplicity according to $\psi(\mathbf{x}(t)) = e^{i\omega t}\psi(\mathbf{x}_0)$.


In essence, $\psi$ acts as a ``magic'' coordinate transformation. While the state $\mathbf{x}(t)$ traces a complex orbit, the scalar observable $\psi(\mathbf{x}(t))$ simply rotates in the complex plane at a constant frequency $\omega$. The level sets of its phase, $\theta = \arg \psi(\mathbf{x})$, are the system's \textbf{isochrons}: surfaces of points in the state space that all share the same asymptotic phase on the limit cycle.

\paragraph{Koopman perspective as compression.}
Finding an eigenfunction $\psi$ with generator eigenvalue $ \lambda=i\omega$ represents a form of compression. Once this function $\psi$ is known, the full $n$‑dimensional trajectory $\mathbf x(t)$ can be encoded (losslessly near the attractor) by a single complex phase variable $\psi(\mathbf{x}(t))$, or separated into its phase $\theta(t)=\arg\psi(\mathbf x(t))$ and amplitude $r(t)=|\psi(\mathbf x(t))|$ \cite{Koopman31}. In effect, the Koopman transform finds a ``magic'' coordinate system $\psi$ in which the nonlinear dynamics become simple linear rotation. It replaces a complex waveform with uniform rotation on $S^{1}$, turning geometry into a one‑line program: \texttt{output $r(t) \cdot \cos(\theta(t))$}.

\paragraph{Signal‑processing criteria}.
In experimental neurophysiology, oscillations are said to  be  \textit{detected} according to some criteria.  
Two widely used operational rules are  
(i) the BOSC power\,+\,duration test \cite{Whitten11} and  
(ii) spectral parameterisation (``FOOOF'') that labels any narrow‑band bump lying above the aperiodic 1/\emph{f} background as oscillatory \cite{Donoghue20}.  
Both are statistical surrogates for asking whether a periodic template explains the data substantially better than a broadband model. Spectral peaks are suggestive of reduced entropy or increased compressibility.

Table \ref{tab:osc_types} aligns the main modeling traditions—from classical limit–cycle theory to modern information–theoretic views—while the text links them through the common intuition that  \textit{circular motion in an abstract coordinate revealed by compression}.

Next, we discuss the formalization of this intuition and  generalization of the above definitions using the language of algorithmic information theory and compression (AIT)\cite{cover_elements_2006}.

\begin{table}[t!]
\centering
\caption{Representative oscillator types and the lenses through which they are defined or detected.}
\label{tab:osc_types}
\begin{tabular}{p{3.3cm}p{4.5cm}p{3.5cm}p{3.3cm}}
\toprule
\textbf{Oscillation class} & \textbf{Canonical models} & \textbf{Koopman lens } & \textbf{Spectral lens}\\
\midrule
\textbf{Limit cycle} & Harmonic oscillator; Van der Pol \cite{VanDerPol26}; Stuart–Landau \cite{Stuart58} & $\lambda = i\omega$ (purely imaginary) & Dirac comb (infinitely sharp harmonics)  \\
\addlinespace[2pt]
\textbf{Relaxation oscillator} & FitzHugh–Nagumo \cite{FitzHugh61}; Morris–Lecar \cite{Morris81} & $\lambda  = i\omega$ plus strong higher harmonics &   ($\uparrow$ bandwidth) \\
\addlinespace[2pt]
\textbf{Damped spiral / ring‑down} & Linear focus; LCR circuits & $\lambda = \alpha + i\omega,\; \alpha < 0$ & Lorentzian PSD peak (width $\sim|\alpha|$)  \\
\addlinespace[2pt]
\textbf{Noise‑sustained quasi‑cycle} & Linear focus $+$ stochastic drive \cite{McKane05} & Same $\lambda$ as above, noise keeps $| \psi|>0$ & Finite‑width peak  \\
\bottomrule
\end{tabular}\\[2pt]
\end{table}
\subsection*{Algorithmic‑information definition}

Algorithmic Information Theory (AIT) takes a computational perspective and quantifies the information content of \emph{individual} objects via computation rather than probability. 
For a binary string $x$, the (prefix) Kolmogorov complexity $K_U(x)$ is defined as the length (in bits) of the shortest program $p$ that makes a fixed universal prefix-free Turing machine $U$ output $x$ and halt,
\begin{equation}
K_U(x) = \min_{p \in \{0,1\}^*} \{\, |p| \; : \; U(p) = x \,\}.
\end{equation}
By the \emph{Kolmogorov invariance theorem}, $K_U(x)$ depends on the choice of $U$ only up to an additive constant independent of $x$, and one typically writes $K(x)$. 
The conditional version $K(x\,|\,y)$ is defined analogously (what is the complexity of $x$ if we know $y$?). 
Kolmogorov complexity is uncomputable (though upper-semicomputable) and provides the foundation for formal notions of randomness, structure, and the ultimate limits of lossless compression.
For a detailed treatment, see classical textbooks on algorithmic information theory \cite{cover_elements_2006,li_applications_2007}.

\paragraph{Compressing data from an oscillator.} Let $x_{1:N}$ be the measured signal.  
Our compressor stores the following program: 
(i) a periodic template $u_{1:T}$ with some frequency (U1 limit‑cycle model with nonlinear mapping / Koopman operator, $K_{\text{LC}}$ bits) and  
(ii) a residual code $e$ (modulation, burst gaps,  remaining error / noise) of length $K_e$.  
If
\[
K_{\text{LC}} + K_{\text{e}} \;\; \ll \;\; L_{\text{raw}} \; ,
\]
where $L_{\text{raw}}$ is the length of a generic lossless code (e.g.\ LZ‑77 or Huffman on the empirical alphabet), we say the sequence \emph{contains an oscillation}.  
No reference model is required—the gain is measured against the unstructured data description.

\paragraph{Relation to Koopman.}
The ideal template $u_{1:T}$ corresponds to one traversal of the Koopman phase; storing $\theta_0$ and $\omega$ plus a small update map for $r$ therefore yields the shortest program in the Kolmogorov sense.  
Thus, the ``kernel'' of every oscillator is circular motion, and oscillation = substantial description‑length reduction via an $S^{1}$ code.

\paragraph{Generative view.} We say that a dataset contains an oscillation if it can be Lie-generated by the U(1) group. In other words, the latent space coordinate of the generative model is $\theta \in S^1$. The generative  (compressive) model is of the form data = $M(\theta)$ + noise.
 In AIT, an \textit{algorithmic agent} would declare, \textit{``An oscillation is a detected pattern: a signal that approximately repeats."}  A bit more precisely, a dynamical dataset is said to contain an oscillation when a periodic‑template model (program) compresses the data significantly better than any model that lacks periodic structure.  But we can be more precise using the notion of Lie-generated model \cite{ruffiniAITFoundationsStructured2022,ruffiniStructuredDynamicsAlgorithmic2023}:

\begin{tcolorbox}[colback=gray!12,colframe=gray!12,enlarge left by=0mm,
  boxrule=0pt,sharp corners]
  \textbf{Propoed definition:}  A dataset is said to represent an oscillation when it can be most succinctly Lie-generated from a representation of U1 plus noise.
\end{tcolorbox}
 Topologically, every limit cycle is a circle. Hence, a compression algorithm will naturally start from the specification of the circle (the U1 group) plus a mapping and corrections. Or, more generally, from the specification of a U1 invariant object.  We analyze this further in the next section.

\subsection*{U1 and the topology of the Stuart-Landau equation}
\label{app:limitcycles}
The origin of U1 and the topology of $S^1$ can be unearthed cleanly in the case of the Hopf bifurcation.

Oscillatory dynamics---limit cycles in the phase space of a dynamical system---play a central role in modeling neural population activity and other biological rhythms.

We begin this journey with the simplest possible oscillator: a \emph{phase clock}, defined by a variable $\theta(t)$ advancing at constant rate $\dot{\theta} = \omega$. This system has no amplitude, only phase, and its trajectory lies on a circle. Importantly, it enjoys continuous phase-shift invariance: the dynamics are unchanged under $\theta \to \theta + \phi$, for any fixed phase offset $\phi$. This is the defining symmetry of the circle group $U(1)$---the group of rotations in the plane.

The next natural step introduces amplitude: the \emph{harmonic oscillator} (HO). It describes circular motion in two dimensions at a fixed radius, and takes the form
\[
\dot z = i\omega z,
\]
in complex notation. The solution is $z(t) = z_0 e^{i\omega t}$, and the motion traces a perfect circle in phase space. Again, we see the same $U(1)$ phase symmetry: multiplying a solution by $e^{i\phi}$ simply rotates the initial condition, leaving the dynamics invariant. However, the amplitude $|z|$ remains constant---there is no mechanism for growth, decay, or saturation. Any initial amplitude persists indefinitely, and the system is neutrally stable.

Real-world oscillators behave differently. In practice, oscillations may grow or decay, but often they settle onto a stable limit cycle: a rhythm of fixed amplitude that persists after transients decay. To capture this, we must go beyond the linear HO and introduce nonlinearities that regulate amplitude. Consider adding smooth perturbations to the harmonic oscillator:
\[
\dot z = (\mu + i\omega)\,z + F(z,\bar z),
\]
where $F$ contains higher-order nonlinear terms. The goal is to find the simplest nonlinear correction that leads to amplitude saturation---i.e., to a self-limiting oscillator.

To identify the relevant terms, we use tools from normal form theory: center manifold reduction followed by a near-identity change of coordinates \cite{guckenheimer2013nonlinear,kuznecovElementsAppliedBifurcation1995}. These procedures systematically eliminate all \emph{non-resonant} terms---i.e., terms whose angular dependence does not match that of the linear part and thus oscillate out of phase. At third order, the only resonant term that survives the coordinate transformation process is
\[
F(z,\bar z) = - (g + i\beta)\,|z|^2 z.
\]
All other cubic combinations, such as $z^2$ or $\bar z^2$, are non-resonant and can be removed by choosing proper coordinates. The resulting simplified system is the \emph{Stuart-Landau equation}:
\[
\dot z = (\mu + i\omega)\,z - (g + i\beta)\,|z|^2 z,
\]
which describes a self-excited oscillator: for $\mu > 0$, the amplitude $r = |z|$ grows until it stabilizes at $r = \sqrt{\mu/g}$; for $\mu < 0$, oscillations decay to rest. The equation also governs the phase evolution $\theta = \arg(z)$, with $\dot\theta = \omega - \beta r^2$.

\paragraph{The emergence of \texorpdfstring{$U(1)$}{U(1)} symmetry.}
At no point did we assume that the nonlinear system was $U(1)$-symmetric. We began with a generic perturbation of the harmonic oscillator. However, after applying normal form reduction, we find that only \emph{resonant} terms cannot be removed—those that transform under rotation in the same way as the linear term. At third order, the only such term is $|z|^2 z$, which happens to be $U(1)$-covariant. All non-covariant terms are eliminated by a coordinate transformation. Thus, the $U(1)$ symmetry of the Stuart-Landau equation emerges \emph{as a consequence} of the reduction process.  But the ultimate origin of this is topological: limit cycles are topological circles, and the ``simplest" description of a cycle, the ``platonic" cycle is a circle.

\paragraph{Cohomology and emergence of resonant terms.} The emergence of the Stuart-Landau normal form from generic nonlinear oscillators can thus be understood by appealing to topology and cohomology. Near a Hopf bifurcation, the system dynamics collapse onto a limit cycle—a smooth, closed loop in phase space that is topologically equivalent to the circle, $S^1$.   The circle $S^1$ is a simple but topologically rich space. It admits a special type of 1-form, $d\theta$, which is \emph{closed}, meaning that its exterior derivative vanishes ($d(d\theta)=0$), but it is not \emph{exact}, meaning it cannot be expressed globally as the differential of any smooth scalar function. Exact forms represent trivial cohomological classes because they can be integrated globally, whereas closed but non-exact forms represent fundamental, nontrivial topological features. This distinction is captured by the nontrivial first de Rham cohomology of the circle, $
H^1_{\mathrm{dR}}(S^1)\cong \mathbb{R}.
$
To eliminate nonlinear terms near the bifurcation, we perform a near-identity coordinate transformation, attempting to remove as many nonlinear perturbations as possible. Specifically, we consider coordinate changes of the form
\[
z \mapsto z + H(z,\bar{z}),
\]
and ask whether a given nonlinear perturbation $F(z,\bar z)$ can be eliminated. Under the linearized dynamics, characterized by pure rotations with frequency $\omega$, the infinitesimal rotation operator naturally arises as the Lie derivative,
\[
\mathcal{L}_0 = \omega \frac{\partial}{\partial \theta},
\]
which describes how functions and vector fields vary as we rotate around the limit cycle. Formally, solving the normal-form reduction involves repeatedly solving equations of the form
\[
\mathcal{L}_0 H = F - N,
\]
where $F$ is the nonlinear perturbation we start with, $N$ is the simplified normal form we desire, and $H$ is the coordinate change we seek to perform.

The crucial point is that certain nonlinear terms are \emph{resonant}: their angular frequency exactly matches the natural rotation frequency $\omega$. Such resonant terms belong to the kernel (null space) of the Lie derivative $\mathcal{L}_0$. Geometrically, these terms correspond precisely to closed but non-exact forms on $S^1$. Cohomologically, resonant terms represent nontrivial elements of the first cohomology group associated with $\mathcal{L}_0$:
\[
H^1(\mathcal{L}_0)=\frac{\ker \mathcal{L}_0}{\mathrm{im}\,\mathcal{L}_0}\cong H^1_{\mathrm{dR}}(S^1)\cong \mathbb{R}.
\]
Therefore, no smooth local coordinate transformation (which can only add exact forms) can eliminate these resonant terms, and thus they constitute genuine cohomological obstructions.

At cubic order, this resonance condition selects uniquely the term $|z|^2 z$, ensuring its survival after normal-form reduction. Similarly, higher-order resonant terms take the general form $|z|^{2m} z$, while non-resonant terms oscillate out of synchrony with the natural rotation and lie within the image of $\mathcal{L}_0$. Thus, these non-resonant terms correspond to exact forms and can always be removed by appropriate coordinate changes.

This topological argument generalizes straightforwardly to higher-order nonlinear terms, ensuring that at every odd order only terms of the form $|z|^{2m} z$ survive. Hence, the general normal form near a Hopf bifurcation is universally structured as
\[
\dot z = (\mu+i\omega)z + z\,g(|z|^2),
\]
a direct consequence of the underlying circle geometry and its associated cohomological constraints.


\end{document}